\documentclass[12pt]{iopart}
\usepackage{array}
\usepackage{amssymb}
\usepackage{amsfonts}
\usepackage{amstext}
\usepackage{graphicx}
\usepackage{amscd}
\usepackage{enumitem}
\usepackage{blindtext}

\expandafter\let\csname equation*\endcsname\relax
\expandafter\let\csname endequation*\endcsname\relax
\usepackage{amsmath}

\begin{document}

\title[]{Information transmission and control in a chaotically kicked spin chain}

\author{Lucile Aubourg and David Viennot}
\address{Institut UTINAM (CNRS UMR 6213, Universit\'e de Bourgogne-Franche-Comt\'e, Observatoire de Besan\c con), 41bis Avenue de l'Observatoire, BP1615, 25010 Besan\c con cedex, France.}
\ead{lucile.aubourg@utinam.cnrs.fr}
\maketitle

\begin{abstract}
We study spin chains submitted to disturbed kick trains described by classical dynamical processes. The spin chains are coupled by Heisenberg and Ising-Z models. We consider chaotic processes by using the kick irregularity in the multipartite system (the spin chain). We show that the both couplings transmit differently the chaos disorder along the spin chain but conserve the horizon of coherence (when the disorder into the kick bath is transmitted to the spin chain). An example of information transmission between the spins of the chain coupled by a Heisenberg interaction shows the interest of the horizon of coherence. The use of some chosen stationary kicks disturbed by a chaotic environment allows to modify the information transmission between the spins and to perform a free control during the horizon of coherence.
\end{abstract}
\pacs{03.65.Yz, 05.45.Mt, 75.10.Jm, 75.10.Pq}

\section{Introduction}
The emergence of quantum information protocols (to perform logic gates and for the transport and the teleportation of information) and the nanosciences have given an interest in dynamics and in control of multipartite quantum systems. A key problem is the understanding of the effects of dynamical processes on the whole multipartite quantum system. They can have consequences on each component of the system and could induce decoherence, relaxation and chaotic processes (\cite{breuer,lages,gedik,lages2,rossini,zhou,castanino,xu,brox}). In order to understand these problems we consider a spin chain, i.e. a set of $N$ $\frac{1}{2}$-spins two by two coupled to form a line chain. The coupling is modelled by the Heisenberg or the Ising-Z interaction which allows a ``cohesion'' into the spin chain. Each spin of the chain is submitted to a train of ultra-short kicks which are disturbed by a chaotic dynamical process.

The subjects concerning decoherence and chaotic processes of regularly kicked spin chains have been studied by some authors \cite{prosen,prosen2,prosen3,prosen4,lakshmin,boness,pineda}. In a previous paper \cite{viennot2013} we have extended the analyses to irregular kicks on spin ensembles without any coupling between the spins. Some interesting behaviours of the density matrix of the spin ensemble have been observed as for example an ``horizon of coherence'' for chaotic dynamics (it corresponds to the time from which the disorder of the kick bath is transmitted to the spins). However, this ensemble cannot be considered as a multipartite system (no information is exchanged between the spins) but only as a set of independent systems dephased during the evolution. A goal of the present paper is to see the behaviours and to understand the state modifications of a kicked spin chain coupled by the Heisenberg or the Ising-Z interaction when the dynamics of the ultra-short kick trains is chaotically disturbed. A main question is to know if the horizon of coherence remains in spite of the coupling between the spins and if it is possible to control the spins before this horizon using appropriate stationary kicks. In the paper \cite{aubourg} we have already seen the general behaviour of a ten spin chain coupled by a nearest-neighbour Heisenberg, Ising-Z and Ising-X interaction and submitted to various ultra-short kick dynamics (stationary, drift, microcanonical and Markovian). It results from this model that the coupling between the spins of the chain allows a better transmission of the disorder (the disorder into a spin chain being defined as a large difference between the states of the different spins) whatever the coupling. An initial dispersion of the kicks induces a disorder and an entanglement between the spins. The entanglement between two spins increases with the increase of the disorder, and so of the decoherence. The Ising-X coupling always induces decoherence, even if there is no kick. It is the worst model to realize quantum controls. So, the spin chain coupled by an Ising-X interaction will not be studied here. The behaviour of a spin chain coupled by an Ising-Z interaction is nearly identical to the one of a chain coupled by the Ising-X interaction except that there is an initial ``plateau'' of coherence. It allows a conservation of the coherence (which is not maximum) during a little number of kicks (see section \ref{consplateau}). The Heisenberg coupling seems to be the most efficient to realize quantum controls. This coupling is isotropic. Two neighbour spins tend to be in the same state due to the coupling if there is no kick. For this coupling, all spins follow the behaviour of the average spin of the chain. 

The dynamical processes describing the trains of ultra-short kicks can be considered as being induced by an environment which disturbs a primary train of kicks. The disturbance can attenuate the kick strengths and/or delay the arrival kicks. Since each kick train can be irregular, the spins can feel different trains. The set of kick trains is called a kick bath since we can assimilate the model to a spin chain in contact with a kind of classical bath. For a chaotic kick bath, the chaotic process is defined by continuous automorphisms of the torus, i.e a dynamics characterised by its matrix and by its modulo $2 \pi$ (this process have a lot of interesting properties, it is chaotic, ergodic, Anosov...). One of the advantages of the chaotic process is the property of sensitivity to initial conditions. This notion means that two points initially really close, do not remain close during the dynamics. They separate each other after a time called the horizon of predictability. This horizon in our model is the time from which the similar kick trains on different spins become different. In a spin ensemble, we have seen that this irregularity of the kicks induces an irregularity of the spin states from the horizon of coherence. This last horizon is larger than the horizon of predictability and corresponds to an initial conservation of the coherence. 

A spin can be assimilated to a qubit. The up state is supposed to be the value 1 and the down state the value 0. So a variation of the spin population can be identified as a variation of the quantum information and a fall of the coherence can be a lost of information. During the horizon of coherence, the coherence is conserved and so all the information. But after it, the coherence falls to 0, the information is completely lost. But we have only seen this phenomenon for a spin ensemble, the spins (or the qubits) cannot exchange any information. This paper studies the use of the interaction between the spins to show the possibility to realize information transports from one spin to another one and to control the spins (using stationary kicks) during the horizon of coherence.\\

This paper is organized as follows. Section II presents the model of the disturbed kicked spin chain. Section III is devoted to the behaviours of the spin chain submitted to chaotic kicks according to the kind of the coupling : Heisenberg or Ising-Z. The last section talks about the information transmission and control along the chain. The use of chaotic kick processes with the Heisenberg coupling, allows to give an interesting example of information transmission. We can extend the previous example to a control of a closed spin chain using stationary kicks disturbed by a chaotic dynamical process.

\section{Dynamics of kicked spin chain}
\begin{figure}
\begin{center}
\includegraphics[width=9.cm]{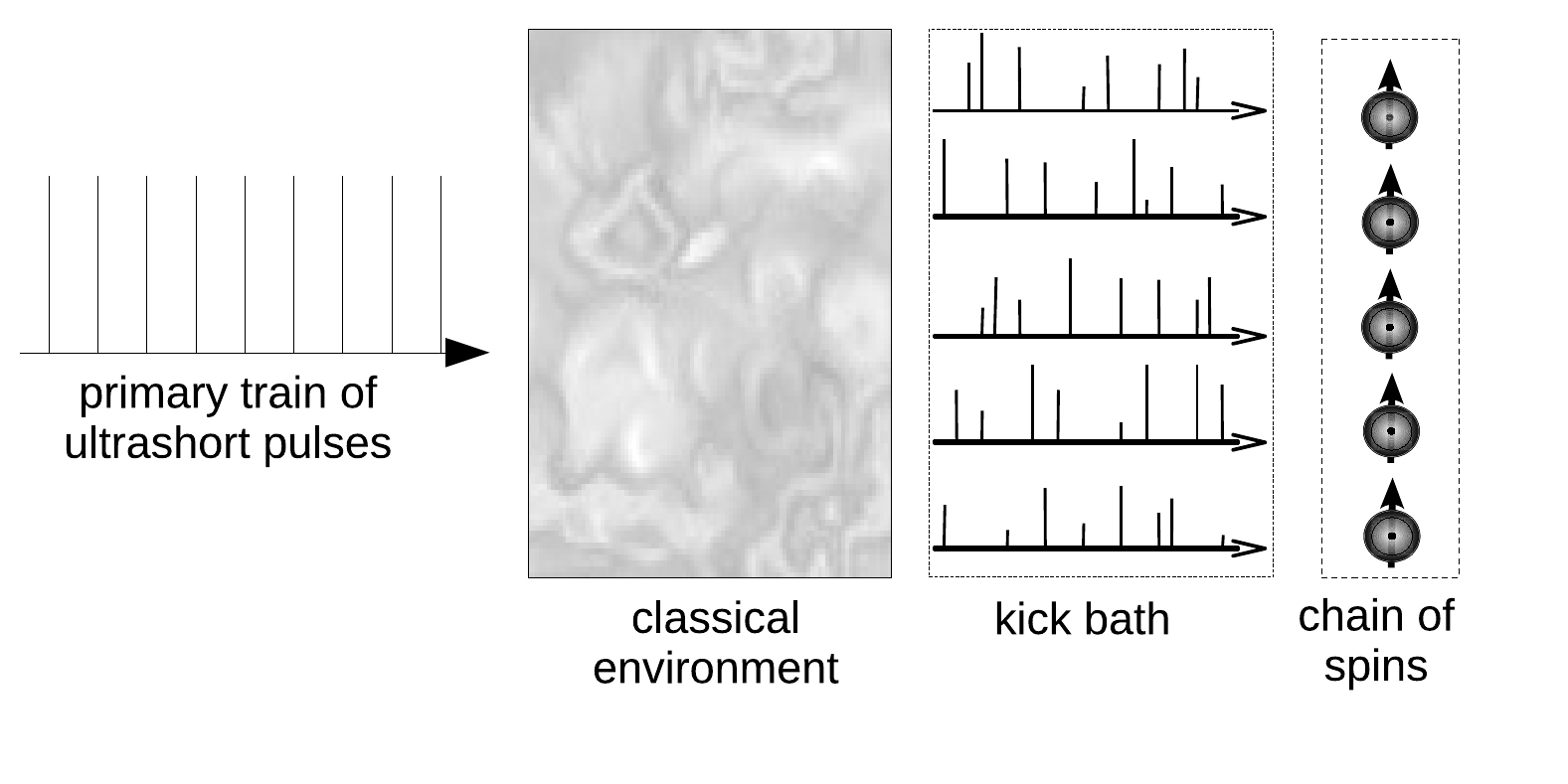}
\end{center}
\caption{\label{kickedspinbath2} Schematic representation of a quantum spin chain controlled by a disturbed train of ultrashort pulses. The set of kick trains issued from the disturbance constitutes a kind of ``classical kick bath''.}
\end{figure}

We consider an open chain of $N$ spins coupled by nearest-neighbour interactions. A constant and uniform magnetic field $\vec B$ is applied on the spin chain inducing an energy level splitting by Zeeman effect. We denote by $\frac{\hbar \omega_1}{2}$ the energy splitting. At the initial time $t=0$, the chain can be coherent or incoherent. In a coherent case, the spins are in the same quantum state $|\psi_0\rangle = \alpha |\uparrow \rangle+ \beta |\downarrow \rangle$ ($|\alpha|^2+|\beta|^2=1$ with $\alpha,\beta\not=0$ -- $|\psi_0\rangle$ is a ``Schr\"odinger's cat state'' -- ). For $t>0$ the chain is submitted to a train of ultrashort pulses kicking the spins. Each pulse can be disturbed by a classical environment such that each spin ``views'' a different train (fig.~\ref{kickedspinbath2}).
Let $\omega_0 = \frac{2\pi}{T}$ be the kick frequency of the primary train. We suppose that the classical environment can attenuate kick strengths and can delay kicks. We denote by $\lambda_n^{(i)}$ and by $\tau_n^{(i)}$ the strength and the delay of the $i$-th kick on the $n$-th spin of the chain. Let ${H_{0_n} = \mathrm{id}^{\otimes (n -1)} \otimes \frac{\hbar \omega_1}{2} |\downarrow \rangle \langle \downarrow|} \otimes \mathrm{id}^{\otimes (N-n)}$ be the quantum Hamiltonian of the $n$-th spin with the Zeeman effect (where we have removed a constant value without significance) and $H_I$ be the nearest-neighbour interaction Hamiltonian which can be for the $n$-th spin of the chain one of the following operators
\begin{enumerate}
\item Heisenberg coupling 
\begin{equation}
\label{heisenberg}
H_{I_n} =  -J \mathrm{id}^{\otimes (n -1)} \otimes (S_x \otimes S_x + S_y \otimes S_y + S_z \otimes S_z) \otimes \mathrm{id}^{\otimes (N-n-1)}
\end{equation}
where $S_i= \frac{\hbar}{2}\sigma_i$, $\{\sigma_{i}\}_{i=x,y,z}$ are the Pauli matrices and $\mathrm{id^{\otimes n}}$ is the tensor product of ``$n$'' identity matrices of order two.
\item Ising-Z coupling 
\begin{equation}
\label{isingz}
H_{I_n} = -J \mathrm{id}^{\otimes (n-1)} \otimes S_z \otimes S_z
\otimes \mathrm{id}^{\otimes (N-n-1)}
\end{equation}
\end{enumerate}

Let $\theta = \frac{2\pi t}{T} = \omega_0 t$ be the reduced time. The quantum Hamiltonian of a kicked spin chain is
\begin{equation}
\label{dynamics}
H(\theta) = \sum_{n=1}^N \Big( H_{0_n} + H_{I_n} + \mathrm{id}^{\otimes (n -1)}
\otimes \hbar W \sum_{i \in \mathbb N} \lambda_n^{(i)} \delta\left(\theta - 2i \pi +\varphi_n^{(i)} \right)\otimes \mathrm{id}^{\otimes (N-n)} \Big)
\end{equation}
where $\delta(t)$ is the Dirac distribution and where the kick operator $W$ is a rank one projection : $W = |w \rangle \langle w|$ with the kick direction ${|w \rangle = \cos \vartheta |\uparrow \rangle + \sin \vartheta |\downarrow \rangle}$ (for the sake of simplicity we do not consider a relative phase between the two components of $|w \rangle$). $\varphi_n^{(i)} = \omega_0 \tau_n^{(i)}$ is the angular delay. The $i$-th monodromy operator (the evolution operator from $t=\frac{2 i \pi}{\omega_0}$ to $\frac{2(i+1)\pi}{\omega_0}$) \cite{viennot}, for the spins organized from the smallest delay (for $n=1$) to the greatest one (for $n=N$) is
\begin{multline}
\label{monodromy}
U^{(i)} = e^{-\frac{\imath H_{0,I}}{\hbar \omega_0} (2\pi - \varphi_N^{(i)})} \prod_{n=1}^{N-1} \left[ \mathrm{id}^{\otimes (N-n)}\right.
 \otimes (id+(e^{-\imath \lambda_{N-n+1}^{(i)}}-1)W) \otimes \mathrm{id}^{\otimes (n-1)}\\
\left. \times e^{-\frac{\imath H_{0,I}}{\hbar \omega_0} (\varphi_{N-n+1}^{(i)} - \varphi_{N-n}^{(i)})}\right] \mathrm{id}^{\otimes (N-1)}
 \otimes (\mathrm{id}+(e^{-\imath \lambda_{
 1}^{(i)}}-1)W) e^{-\frac{\imath H_{0,I}}{\hbar \omega_0} \varphi_1^{(i)}}
\end{multline}
with $H_{0,I} = \sum_{n=1}^N H_{0_n} + \sum_{n=1}^{N-1} H_{I_n}$.
We see that the monodromy operator is $2\pi$-periodic with respect to the kick strength. $\lambda_n^{(i)}$ is then defined modulo $2\pi$ from the viewpoint of the quantum system. Thus the strength-delay pair $(\lambda,\varphi)$ defines a point on a torus $\mathbb T^2$ which plays the role of a classical phase space for the kick train. So, we can consider the kick dynamics as being continuous automorphisms of the torus $\mathbb T^2$ like the Arnold's cat map (in \cite{viennot2013}).

Let $|\psi^{(i)} \rangle \in \mathbb{C}^{2N}$ be the state of the chain at time $t=iT$ ($|\psi^{(i)} \rangle$ represents the ``stroboscopic'' evolution of the chain). By definition of the monodromy operator we have
\begin{equation}
|\psi^{(i+1)} \rangle = U^{(i)} |\psi^{(i)} \rangle
\end{equation}
The density matrix of the chain is then
\begin{equation}
\rho^{(i)} = \frac{1}{N} |\psi^{(i)}\rangle \langle \psi^{(i)}|
\end{equation}
and the density matrix of the $n$-th spin is
\begin{equation}
\rho_{n}^{(i)} = \Tr_{i=1,...,n-1,n+1,...,N}(\rho^{(i)})
\end{equation}
$ \Tr_{i=1,...,n-1,n+1,...,N}$ is the partial trace on all the spin Hilbert spaces except the $n$-th one. It encodes two fundamental informations. The first information concerns the diagonal elements of the density matrix. They represent the occupation probabilities of the state $|\uparrow\rangle$ and $|\downarrow \rangle$ for the $n$-th spin and are called the populations ($\langle \uparrow|\rho_n^{(i)}|\uparrow \rangle$ and $\langle \downarrow|\rho_n^{(i)}|\downarrow \rangle$). The second one is associated with the non-diagonal terms. It is a measure of the entanglement of the $n$-th spin with the others of the chain \cite{breuer,bengtsson} and is called the coherence ($|\langle \uparrow| \rho_n^{(i)} |\downarrow \rangle|$).

We deduce from $\rho_{n}^{(i)}$ the density matrix of the average spin of the chain for the $i$-th kick
\begin{equation}
\rho_{tot}^{(i)} = \frac{1}{N} \sum_{n=1}^N \rho_{n}^{(i)}
\end{equation}

The kick baths are also defined by the initial distribution of the first kicks $\{ (\lambda_n^{(0)},\varphi_n^{(0)}) \}_{n=1,...,N}$. Since the dynamical processes are considered as being chaotic, the kicks are characterised by the sensitivity to initial conditions. The first kicks are randomly chosen in $[\lambda_*,\lambda_*+d_0] \times [\varphi_*,\varphi_*+d_0]$ (with uniform probabilities) with a small $d_0$. $(\lambda_*,\varphi_*)$ can be viewed as the parameters of the primary kick train. The length of the support of the initial distribution (the initial dispersion) $d_0$ is the magnitude of the disturbance on the first kick.\\

Using the model described in this section, we want to know the effects of chaotic kick trains on a spin chain coupled by an Ising-Z or a Heisenberg interaction. Especially we are interested in controlling the informations of the system in spite of the kicks and of the coupling. But for a control, it is necessary that the spins of the chain remain coherent. For a sake of simplicity, in the following analyses, we consider that $\hbar=1$.

\section{The chaos}
We have shown in \cite{viennot2013} with the model of kicked spins without interaction, that a large coherence plateau appears when the classical kick dynamics is chaotic. In the same paper, we have found an empirical expression of the length of the plateau (corresponding to a kick number) which determines the horizon of coherence
\begin{equation}
\label{horizon}
n_* = n_{\square} + \frac{1}{2} \sqrt{1 + \frac{8 S_{max}}{\ln |\lambda_+|}} - \frac{1}{2}
\end{equation}
$S_{max}$ is the maximum entropy, $\ln |\lambda_+|$ is the Lyapunov exponent of the dynamical system in the instable direction and $n_{\square}$ is the horizon of predictability of the kick bath and is given by the sensitivity to initial conditions of the chaotic dynamics. For a continuous automorphism of the torus, the horizon of predictability is given by
\begin{equation}
n_{\square} = \frac{\ln d_{\square} - \ln \frac{d_0}{\sin \gamma}}{\ln |\lambda_+ |}
\end{equation}
where $\gamma = \arctan \frac{e_+^{(\phi)}}{e_+^{(\lambda)}}$ is the angle between $e_+$ the instable direction of the automorphism matrix of the torus $\mathbb{T}^2$ ($\lambda_+$) and the strength axis ($\lambda$) of $\mathbb{T}^2$. The dispersion ($d_0$) of the projection of the initial distribution on the unstable axis is approximately $\frac{d_0 }{\sin \gamma}$. $d_{\square}$ is the microstate length of an equipartition of $\mathbb{T}^2$ ($\mathbb{T}^2$ is covered by a set of disjoint cells of dimensions $d_{\square} \times d_{\square} $ which constitute the classical microstates).

It is important to note that with a spin ensemble, the horizon of coherence does not correspond to the horizon of predictability. It is larger and allows a conservation of the coherence. \\

This section studies the robustness of this horizon of coherence regarding to the interactions between the spins and the validity of eq.~\ref{horizon} in this context. This is a very important question, because the horizon of coherence is a time during which it could be possible to control the spins before the decoherence disturbs their quantum behaviours.

\subsection{An almost destruction of the plateau}
\label{sectionisingz}

\begin{figure}
\begin{center}
\includegraphics[width=10cm]{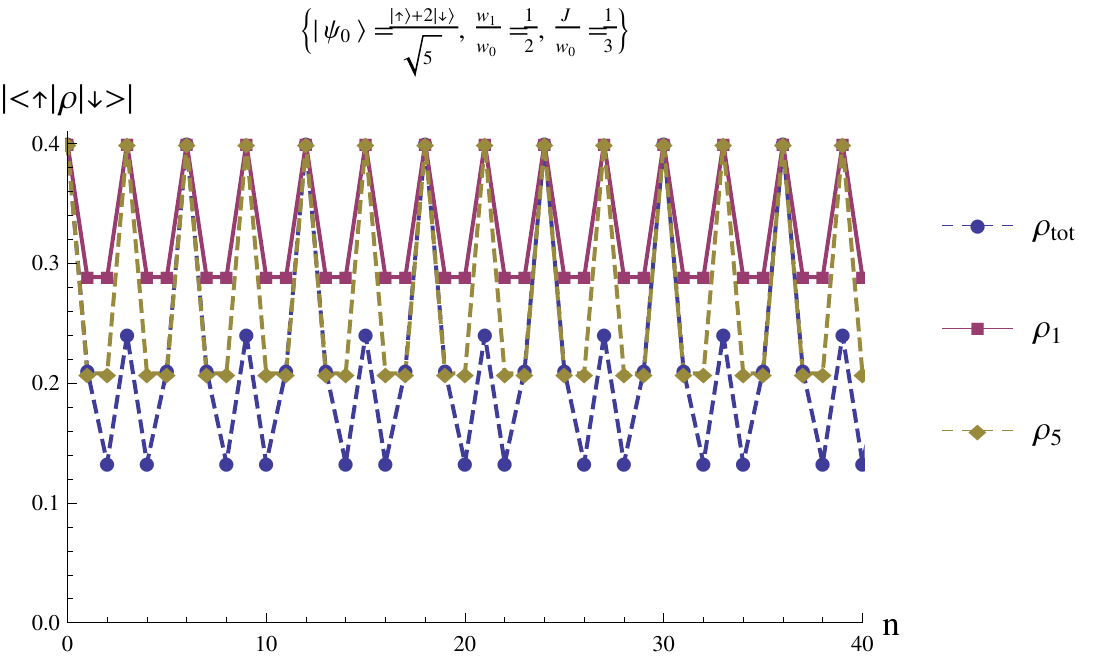}
\end{center}
\caption{\label{isingzfreeevolution} Coherence evolution of the average spin ($\rho_{ tot}$) and of the first and the fifth spin of the chain ($\rho_1$  and $\rho_{5}$). Each spin is coupled with its nearest neighbours by an Ising-Z interaction and is in the initial state $\psi_0$.}
\end{figure}

\begin{figure}
\begin{center}
\includegraphics[width=15cm]{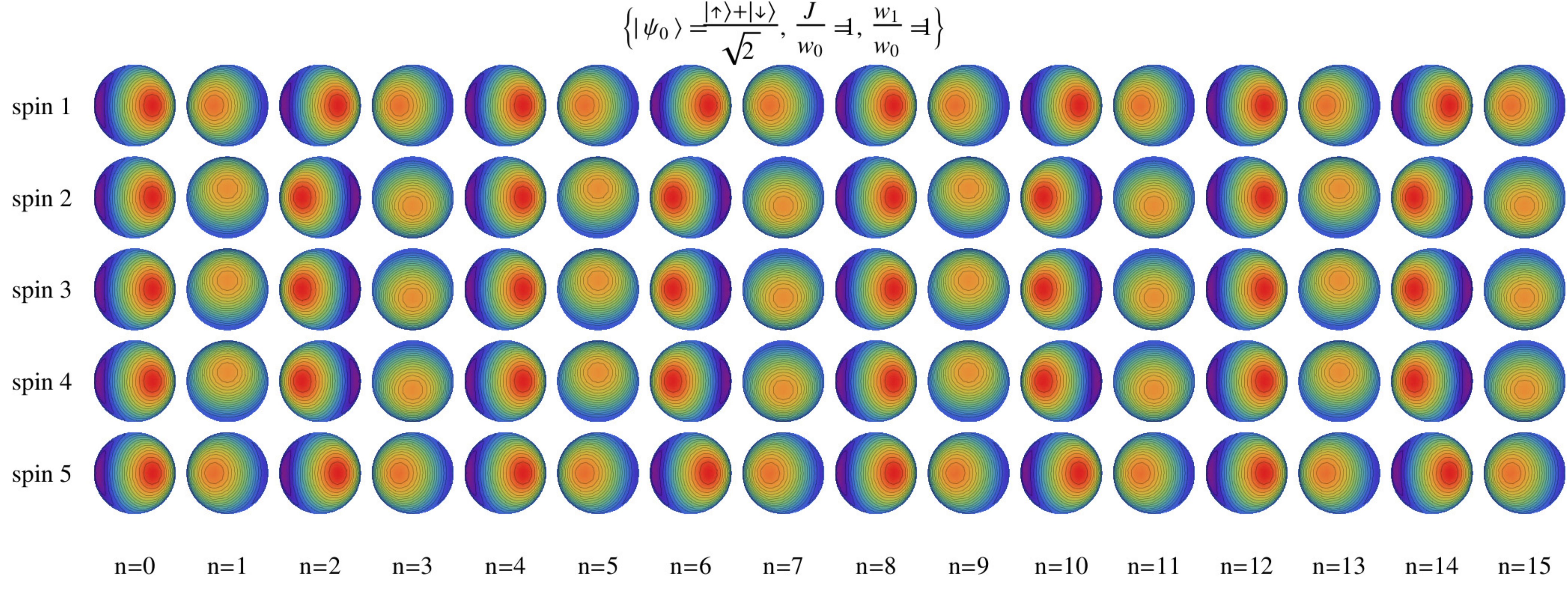}
\end{center}
\caption{\label{husimiIsingZ} Evolution of the Husimi distribution of the five spins of a chain coupled by the Ising-Z interaction. There is a free evolution of the spin chain and each spin is in the same initial state $\psi_0$. The highest probability is represented in red and the smallest one in blue. The entanglement process is also shown by the Husimi distribution. In this case, the disk goes to the green colour.}
\end{figure}

We consider a spin chain coupled by a nearest-neighbour Ising-Z interaction. Without any kind of kick, there is an oscillation of the coherence below the initial coherence value which is generally less important for the edge spin than for the others (because an edge spin has only one neighbour and so is ``less coupled'') as we can see on fig.~\ref{isingzfreeevolution}. 
There is no modification of the population.  Each one remains at its initial value even if the states of the spins are not the same. This is due to the fact that the coupling is completely diagonal and only induces a dephasing in the absence of kicks (see \ref{isingzdemo}). Thus, there is no information transmission between the spins. In order to see more precisely what happens, we consider a semi-classical analysis \footnote{A spin could be viewed as a classical magnetic moment vector, inducing a local magnetic field $\vec B_{loc} \propto \langle \vec S \rangle = \tr(\rho \vec S)$ (where $\vec S$ are the spin operators and $\rho$ is spin density matrix) which is felt by their neighbours. We talk about the (classical) spin orientation in place of the (quantum) spin state (a quantum spin state $\alpha |\uparrow \rangle + \beta |\downarrow \rangle$ being equivalent to the classical spin orientation $\theta = 2 \arctan \left| \frac{\beta}{\alpha} \right|$ and $\varphi = \arg \beta - \arg \alpha$, or in other words we identify the Bloch sphere (the space of the spin states without global phase) with a sphere of classical vector directions)} of the spin chain by the use of the Husimi distribution \cite{husimi}. This distribution is defined by :
\begin{equation}
H_n^{(i)}(\theta,\varphi) = | \langle \theta, \varphi|\rho_n^{(i)} |\theta, \varphi\rangle |^2
\end{equation}
where $|\theta, \phi \rangle= \cos( \frac{\theta}{2})| \uparrow \rangle + e^{\i \varphi} \sin(\frac{\theta}{2})|\downarrow \rangle $ is the spin coherent state. The Husimi distribution measures the quasiprobability distribution of a quantum state onto the classical phase space (here, the sphere of the classical spin direction). This sphere will be represented by an azimuthal projection map (north pole at the center and south pole as being the limit circle). The entanglement processes are also shown by the Husimi distribution. The distribution becomes uniform for a maximal entanglement state. Figure \ref{husimiIsingZ} represents the evolution of the Husimi distribution with respect to the spin and to the kick number. We see that periodically, the spins become entangled (the distribution goes to the green colour). In \ref{isingzdemo}, we have obtained the value of the coherence of two spins coupled by the Ising-Z interaction. This term is $2\pi$ periodic (which explains the oscillation) and is inherent to the quantum aspect. Every time that there is a system where the coupling is completely diagonal, these oscillations appear. They are due to the interferences between the phases of the energies of each spin. \\

\begin{figure}
\begin{center}
\includegraphics[width=9cm]{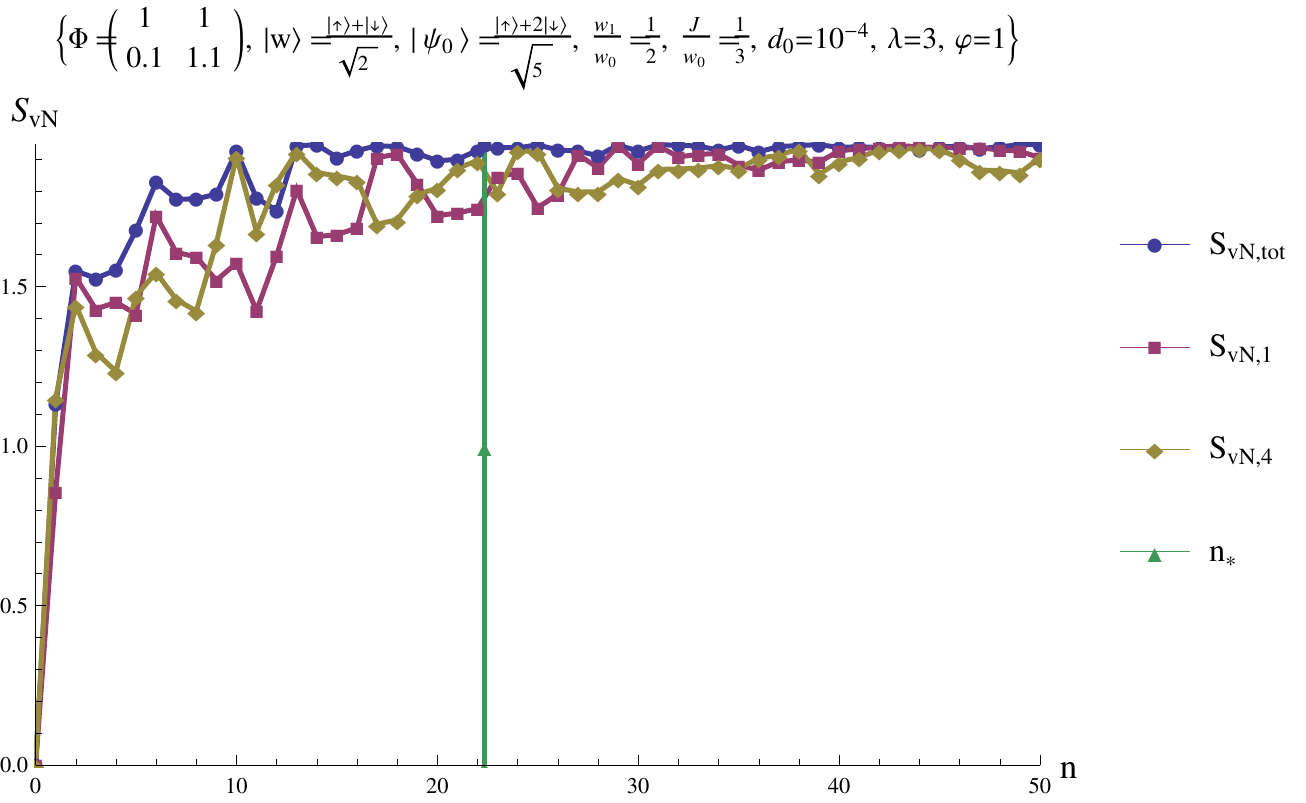}
\includegraphics[width=15cm]{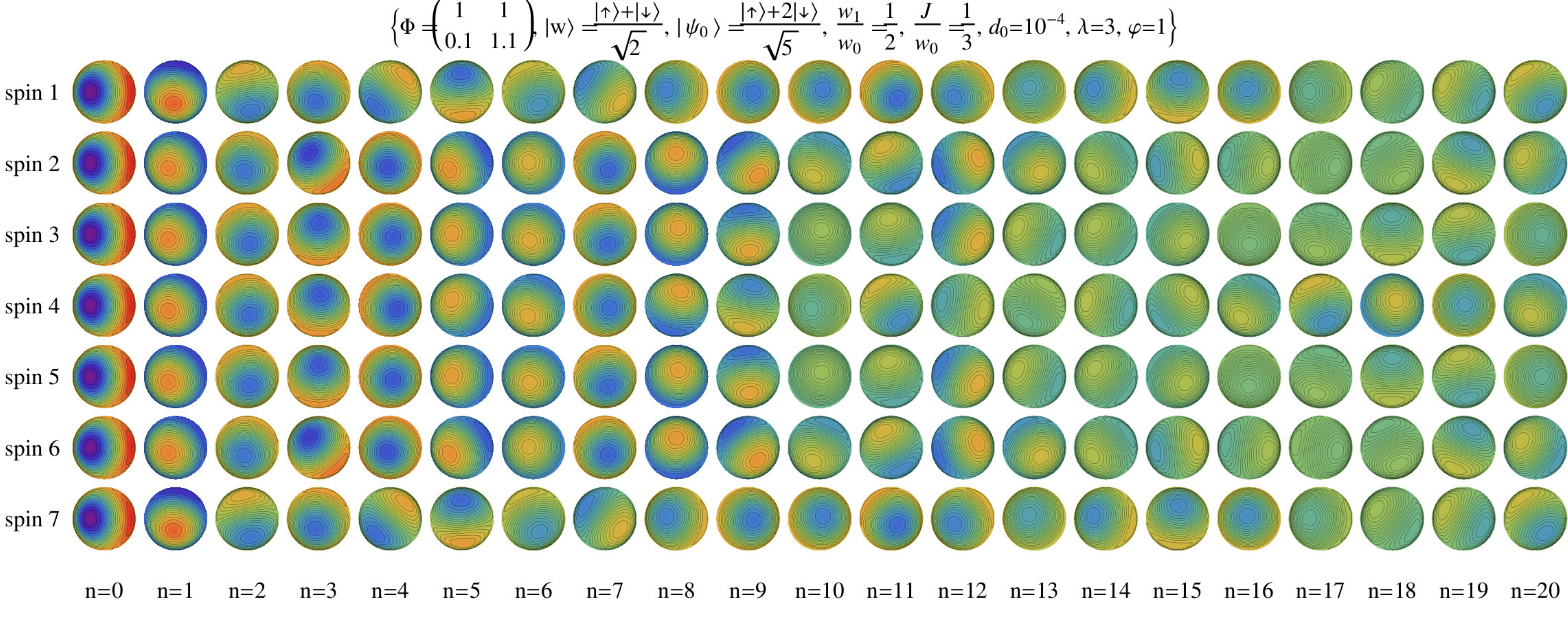}
\end{center}
\caption{\label{Husimisingzentanglement} Evolution of the entropy (up) and the Husimi distribution (down) for a seven spin chain coupled by an Ising-Z interaction. The chain is submitted to several kicks evolving according to a chaotic dynamics on the torus. Each spin is in the same initial state $\psi_0$. $\Phi$ is the matrix defining the automorphism of the torus. On the up graphic, the vertical green line corresponds to the horizon of coherence. On the down one, the highest probability is represented in red and the smallest one in blue. The entanglement process is also shown by the Husimi distribution. In this case, the disk goes to the green colour.}
\end{figure}

For a spin chain coupled by an Ising-Z interaction and submitted to a kick bath disturbed by a chaotic process (a continuous automorphism of the torus), the coupling induces disorder and entanglement. The coherence and the populations go toward a microcanonical distribution (relaxation of the population toward $\frac{1}{2}$ and fall of the coherence to 0) and the entropy increases a lot. Figure \ref{Husimisingzentanglement} presents the evolution of the entropy (up) and of the Husimi distribution (down) of a seven spin chain chaotically kicked in a direction different of the one of the eigenvectors. We see that the entropy increases rapidly and that the Husimi distribution tends to become entirely green which is a sign of the entanglement. The value of the maximum entropy corresponds to when the Husimi distribution is the closest to the green color (about 13 kicks). 

\begin{figure}
\begin{center}
\includegraphics[width=10cm]{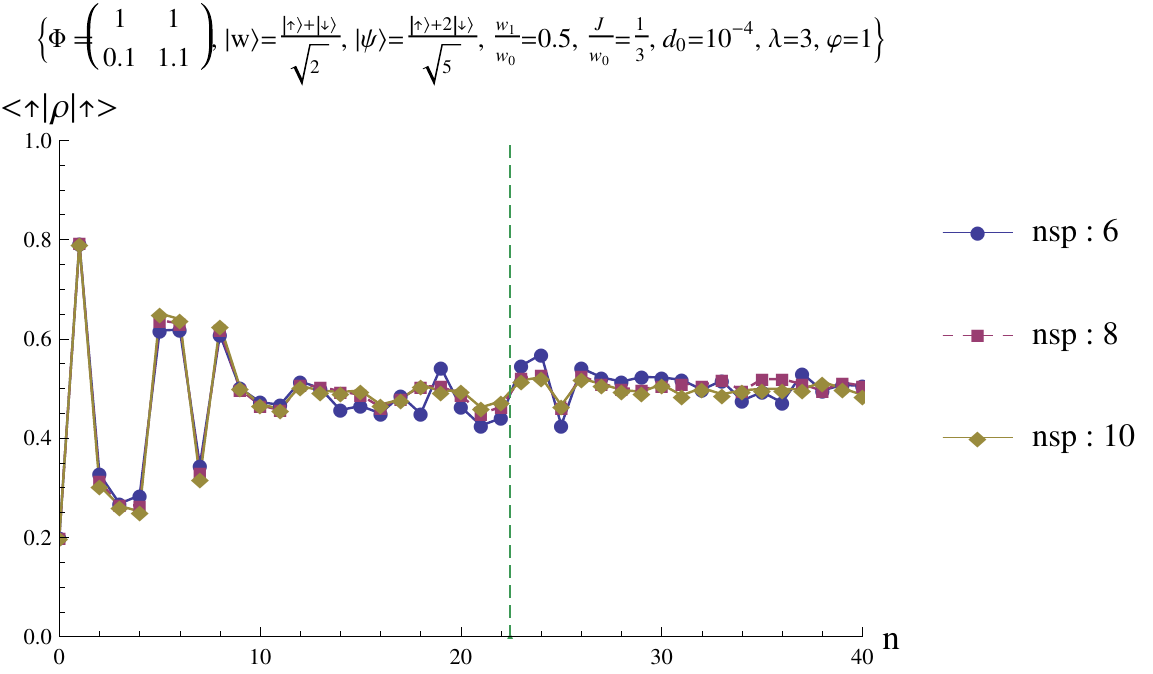}
\includegraphics[width=10cm]{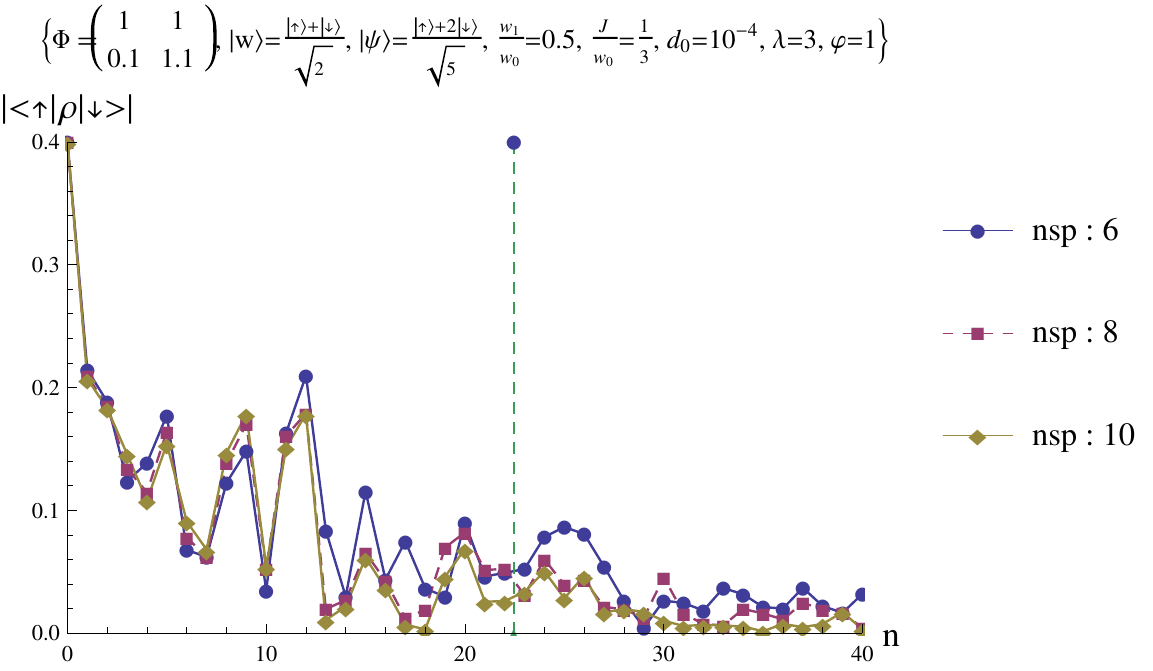}
\end{center}
\caption{\label{plateau} Evolution of the population (up) and of the coherence (down) for different numbers of spins of a chain coupled by an Ising-Z interaction. The chain is submitted to several kicks evolving according to a chaotic dynamics on the torus. Each spin is in the same initial state $\psi_0$. $\Phi$ is the matrix defining the automorphism of the torus. The vertical green line corresponds to the horizon of coherence.}
\end{figure}

However, an interesting phenomenon appears for the coherence which can be seen on fig.~\ref{plateau} : a little initial coherence conservation. This coherence plateau is described by the presence of some oscillations of the coherence before going to the microcanonical distribution (the coherence falls near to 0). This low coherence conservation is more visible for the individual spin coherence than for the coherence of the average spin of the chain because of the oscillation addition of each spin. During this plateau, there is some oscillations of the population before it relaxes toward $\frac{1}{2}$, the microcanonical distribution, when the coherence goes to zero. So, before that the coherence goes to $0$, there is a little conservation of the spin information. The coherence plateau does not depend on the dynamics, on the initial dispersion and apparently on the number of spins. It does not correspond to a maximal coherence and its value is about 0.2-0.3, it only depends on the coupling value. The larger the coupling is, the less large the plateau is and the less it can be viewed for the average spin of chain (the plateau always appears on the population and the coherence for an individual spin of the chain). This is not the plateau due to the chaotic process, because it ends before 23 kicks (obtained using eq.~\ref{horizon}). This is a result of the Ising-Z coupling.\\

\begin{figure}
\begin{center}
\includegraphics[width=10cm]{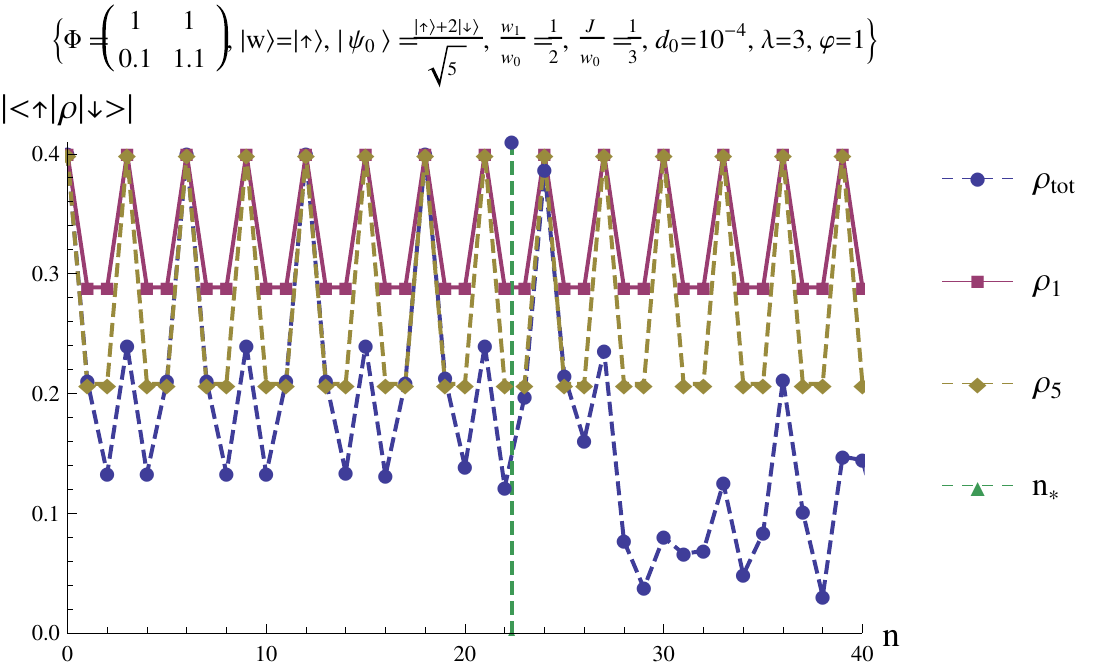}
\end{center}
\caption{\label{isingzgreatplate} Coherence evolution of the average spin ($\rho_{ tot}$) and of the first and the fifth spin of a ten spin chain ($\rho_{5}$) coupled by the Ising-Z interaction. The chain is submitted to several kicks evolving according to a chaotic dynamics on the torus. Each spin is in the same initial state $\psi_0$. $\Phi$ is the matrix defining the automorphism of the torus. The green vertical line corresponds to the horizon of coherence. This graphic is the same than fig.~\ref{isingzfreeevolution} but with kicks disturbed by a chaotic dynamics}
\end{figure}

The plateau linked to the horizon of coherence is only seen in one case for a spin chain coupled by an Ising-Z interaction : when the kick direction is $|\uparrow \rangle$ or $|\downarrow \rangle$, i.e. when the kick is in the direction of an eigenvector. Figure \ref{isingzgreatplate} is realized in this condition. It shows that each spin conserves a coherence with strong down oscillations whatever the kick number. There is no modification of the population. The behaviour of the coherence of the average spin is a little different. The coherence is conserved with large down oscillations only before the horizon of coherence delimited by the green vertical line on fig.~\ref{isingzgreatplate}. After it, the average of these oscillations falls to zero. The comparison between fig.~\ref{isingzfreeevolution} and \ref{isingzgreatplate} shows that before the horizon of coherence, the kicked chain has the same behaviour than a free chain. It is as if the spins do not feel the kicks. If we consider a spin without any interaction with its neighbours, we see in \ref{withoutcoupling} that the strength and/or the delay do not influence the population when the kick is in the direction of an eigenvector. The strength only induces a pure dephasing. But here, we have in addition a coupling between the spins. We have demonstrated on \ref{isingzdemo2} that two coupled and kicked spins never feel the delay (it does not appear in the evolution operator). For the strengths, two cases appear. If the kick strengths on two spins are the same, there is no effect on the coherence, but, if the kick strengths are different, the coherence is modified. This can be easily extended to a larger number of spins. The coupling induces a ``cohesion'' between the spins. If the cohesion is complete (same strength and delay) the system has a free behaviour. But if the cohesion is lost, when the kick bath disorder (different strengths) is transmitted to the spin chain, there is some coherence interferences which can induce a lost of the quantum property. Before the horizon of coherence, the spins are quantum and in a state superposition. After it, the fall of the coherence means than the spins become classical, they are either on the up or in the down state with a probability given by the up and the down population.\\

\begin{figure}
\begin{center}
\includegraphics[width=10cm]{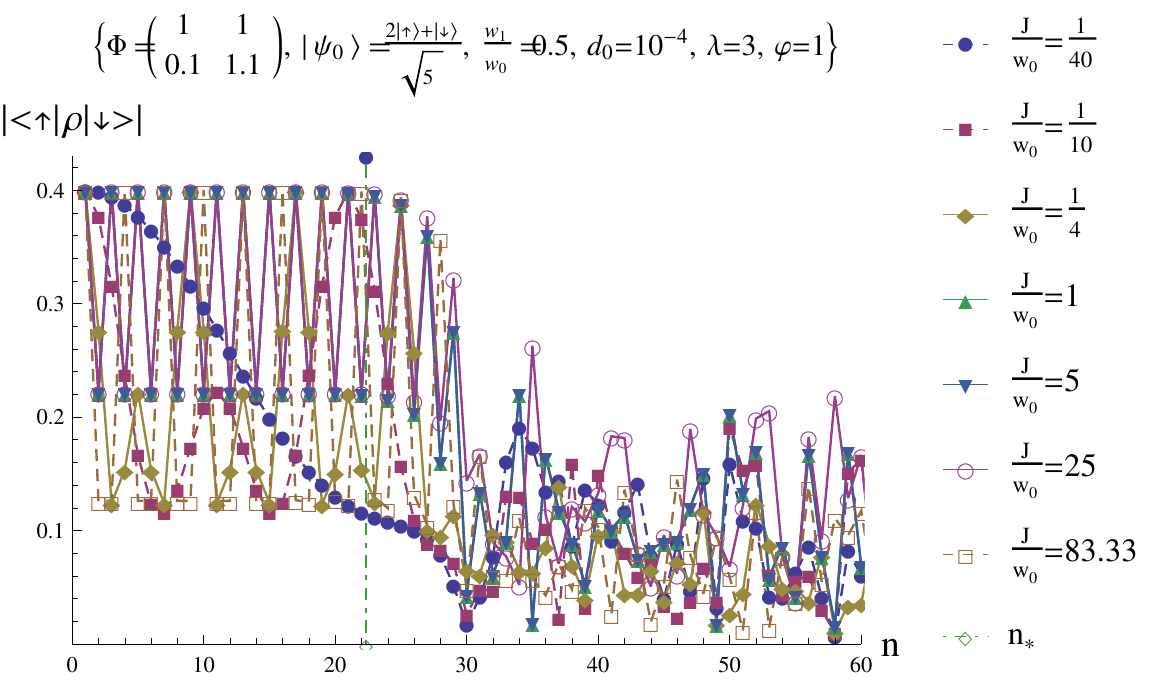}
\includegraphics[width=10cm]{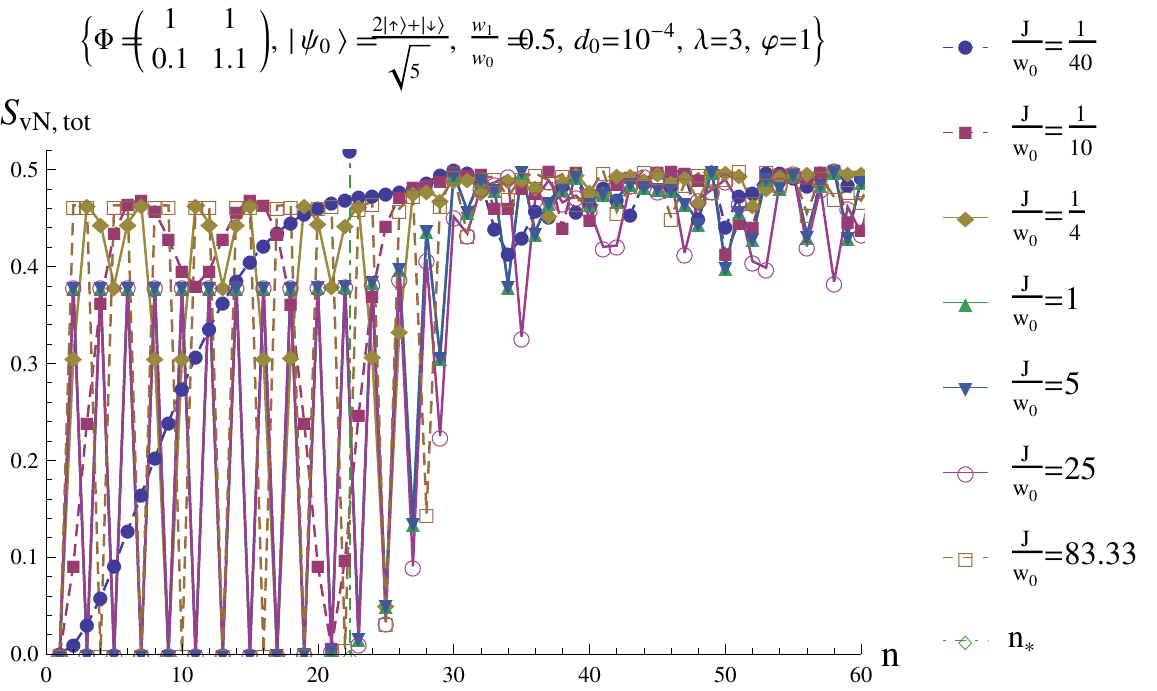}
\end{center}
\caption{\label{isingzplateauinteraction} Evolution of the coherence (up) and of the entropy (down) of the average spin of a ten spin chain coupled by the Ising-Z interaction. The chain is submitted to several kicks evolving according to a chaotic dynamics on the torus. Each spin is in the same initial state $\psi_0$. $\Phi$ is the matrix defining the automorphism of the torus. The green vertical line corresponds to the horizon of coherence.}
\end{figure}

The comparison with the results obtained without interaction (the yellow point curve fig.~\ref{heisenbergplate} for fifty spins), gives a same coherence plateau and a same fall of the average coherence after the horizon of coherence but without the down oscillations. In order to know if eq.~\ref{horizon} is still correct (the green vertical line fig.~\ref{isingzgreatplate} and \ref{isingzplateauinteraction}) for the average spin of a chain coupled by an Ising-Z interaction we have to see the evolution of the coherence plateau with the interaction parameter. Figure \ref{isingzplateauinteraction} shows the coherence (up) and the entropy (down) of the average spin of a ten spin chain with respect to the kick number and to the interaction parameter. If the interaction parameter is too small, the plateau disappear. We are nearly in the case of ten spins without interaction. This spin number is not sufficient to see the coherence plateau when the spins are not coupled. If $\frac{J}{w_0}$ is large enough, we see coherence and entropy oscillations before the horizon of coherence. After it, they fall to 0. In the entropy graphic, we see that the horizon of coherence always corresponds to the kick number for which the entropy oscillations begin to decrease and so the average oscillations begin to increase. For an Ising-Z coupling, the empirical formula also corresponds to the kick number for which the entropy begins to increase.  \\

If we chaotically kick the spins in a direction which do not correspond to an eigenvector, there is a lost of the information and the coherence goes to $0$. In this condition it is impossible to realize a control of the information even during the initial little plateau. The kicks in the direction of an eigenvector allow, before the horizon of coherence, a kind of conservation of the coherence with large down oscillations. In addition, whatever the strengths and the delays of the kicks, there is no modification of the populations, so nothing can be controlled. Thus this model is not efficient to realize quantum control and information transmission. This coupling could eventually be interested if we want to conserve the spin state and if we can force the environment to be in an eigenvector direction. In this case and only in this one there is a conservation of the spin state.

\subsection{Conservation of the plateau}
\label{consplateau}
\begin{figure}
\begin{center}
\includegraphics[width=10cm]{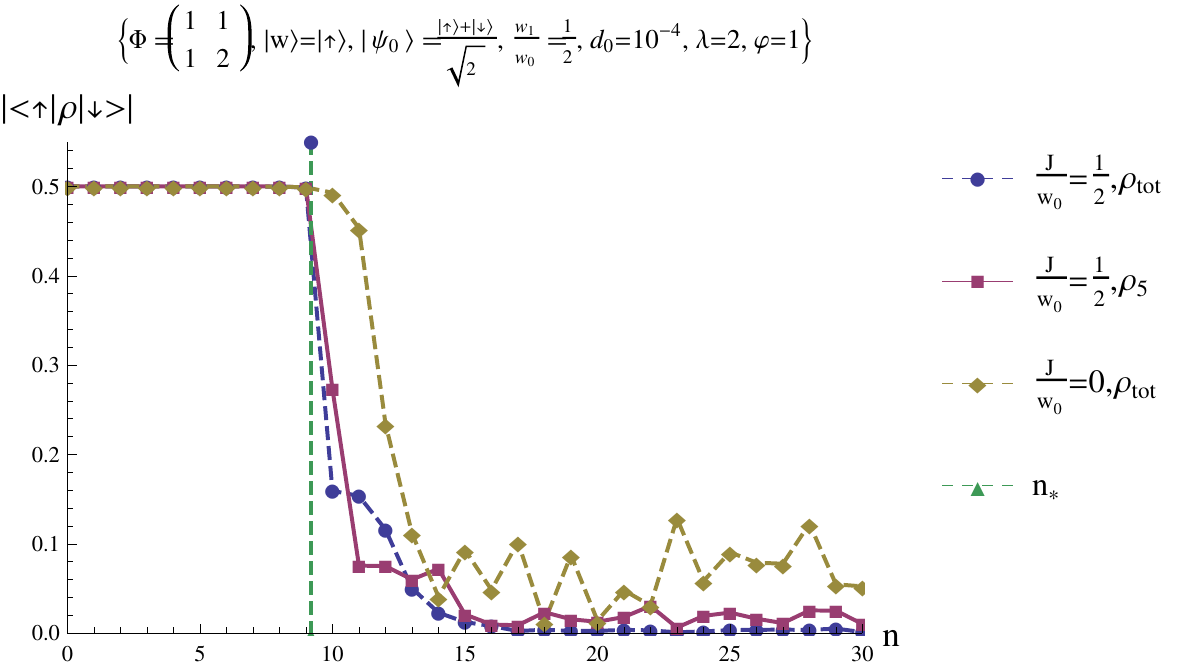}
\includegraphics[width=10cm]{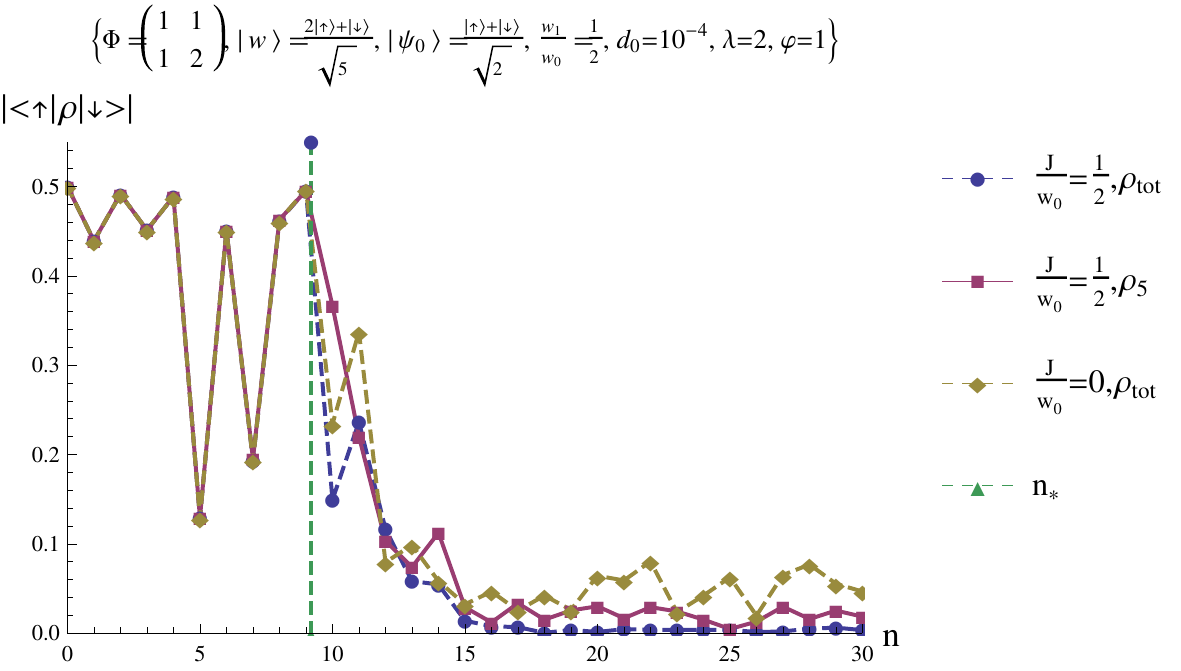}
\end{center}
\caption{\label{heisenbergplate} Evolution of the coherence, for 50 spins without interaction ($J=0, \rho_{tot}$), for 5 chains of 10 spins ($\frac{J}{w_0}=\frac{1}{2}, \rho_{tot}$) and for the fifth spin of the first chain ($\frac{J}{w_0}=\frac{1}{2},\rho_{5}$) coupled by the Heisenberg interaction. The spins are submitted to several kicks evolving according to the Arnold's cat map (a chaotic dynamics). $\Phi$ is the matrix defining the automorphism of the torus. Each spin is in the same initial state $\psi_0$. For the up graphic, the kicks are in the direction of an eigenvector ($|\uparrow>$) whereas for the down one, the kicks are in the superposition of the both states of a spin. The green vertical axis corresponds to the horizon of coherence ($n_*$).}
\end{figure}

Consider now a spin chain coupled by a Heisenberg interaction. We have reminded in the introduction that, this interaction is isotropic (two coupled spins tend to be in the same state or to become entanglement if they cannot), and each spin follows the average evolution. Figure \ref{heisenbergplate} shows the evolution of the coherence of the average spin of five chains of ten spins, of one spin of one chain and of an ensemble of fifty spins. The spins of the chains are submitted to the Heisenberg interaction and the classical dynamics is chosen to be the Arnold's cat map. The up graphic is for a kick in the direction of an eigenvector ($|\uparrow \rangle$ or $|\downarrow \rangle$) and the down one is when the kick direction is a superposition of the both eigenvectors of a spin. For these both cases, all the coherence curves are merged. We have the same behaviour for a spin chain than for a spin ensemble with the particularity that each spin exhibits the coherence plateau. The Heisenberg coupling allows to conserve for spin chains and for each spin of the chains, the interesting result obtained for a spin ensemble. 

\begin{figure}
\begin{center}
\includegraphics[width=10cm]{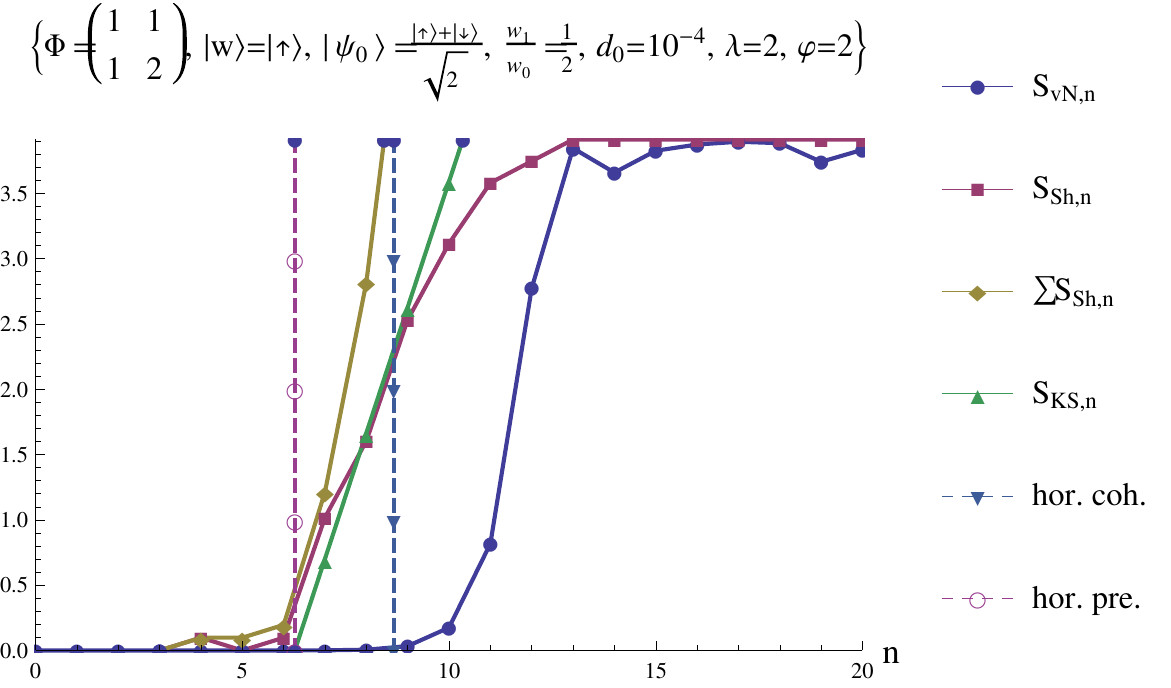}
\includegraphics[width=10cm]{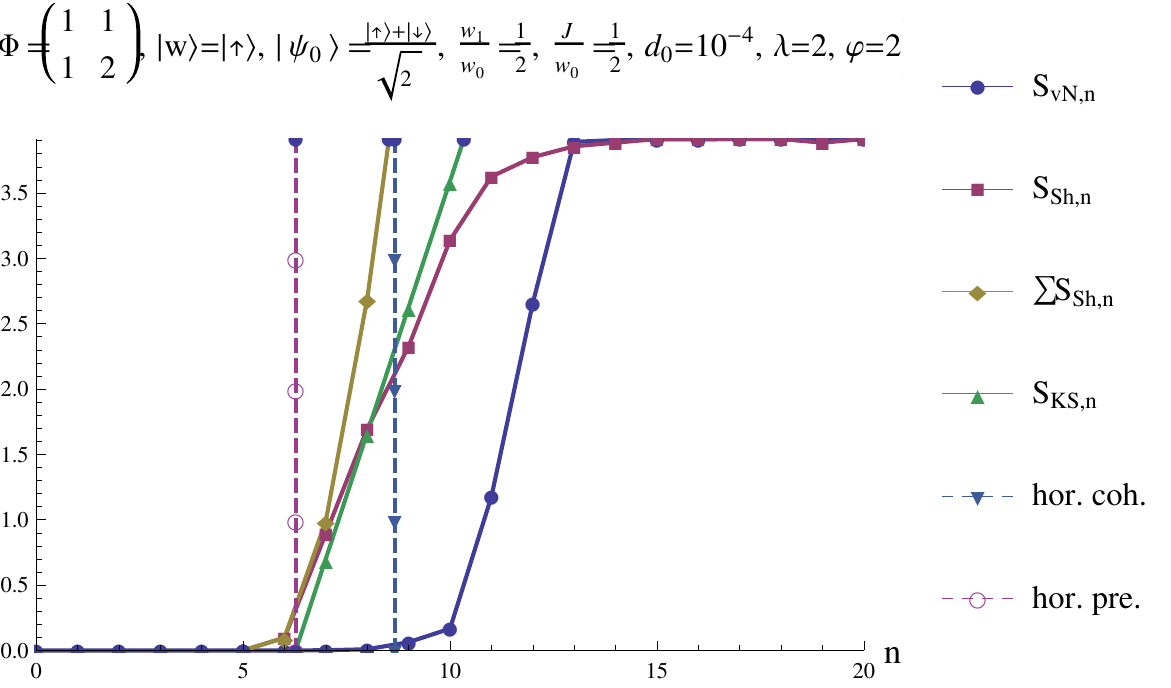}
\end{center}
\caption{\label{entropy} For the Arnold's cat flow : von Neumann entropy (a measure of the disorder into the spin chain) of the spin ensemble, Shannon entropy (a measure of the disorder and of the entanglement into the kick bath) of the kick bath, cumulated Shannon entropy of the kick bath and entropy of the kick bath predicted by the Kolmogorov-Sina\"i analysis (a measure of the production of the disorder by the flow predicts by the dynamical system theory \cite{benoist}). The horizon of predictability of the kick bath and the horizon of coherence of the spin ensemble are indicated by vertical dashed lines. The up graphic is for 50 spins without interaction and the down one is for 10 chains of 5 spins coupled by the Heisenberg interaction. Each spin is in the initial state $\psi_0= \frac{1}{\sqrt{2}}(|\uparrow \rangle + |\downarrow \rangle)$}
\end{figure}

The length of the coherence plateau does not change for a coupled spin chain and corresponds to the kick number given by the eq.~\ref{horizon}. This can be better seen using the entropy, fig.~\ref{entropy}. The quantum entropy, the entropy into the spin chain is measured by the von Neumann entropy
\begin{equation}
S_{vN,n}=- \gamma tr(\rho_n \log \rho_n)
\end{equation} 
The factor $\gamma$ is arbitrary. To define the classical entropy it is necessary to introduce a partition of the phase space $\mathbb{T}^2$. Let X be this partition. The dimension of the phase space is $2 \pi \times 2 \pi$ and the partition is chosen to be $\{i\frac{\pi}{64}\}_{i = 0, ..., 128} \times \{j\frac{\pi}{64}\}_{j = 0, ..., 128}$. A cell of $X$ constitutes one of the classical microstates for one kick train. The classical entropy, the entropy into the kick bath, is defined by the Shannon entropy
\begin{equation}
S_{Sh,n}=\theta \sum_{i,j} - p_{ij,n} \ln p_{ij,n}
\end{equation} 
where $p_{ij,n}$ is the fraction of kick trains which are in the microstate $(i,j)$ at the $n$-th iteration and $\theta$ is another arbitrary factor. The arbitrary factor in the von Neumann and in the Shanon entropy is chosen in order to have a same maximum for the classical and the quantum entropy.

\begin{figure}
\begin{center}
\includegraphics[width=10cm]{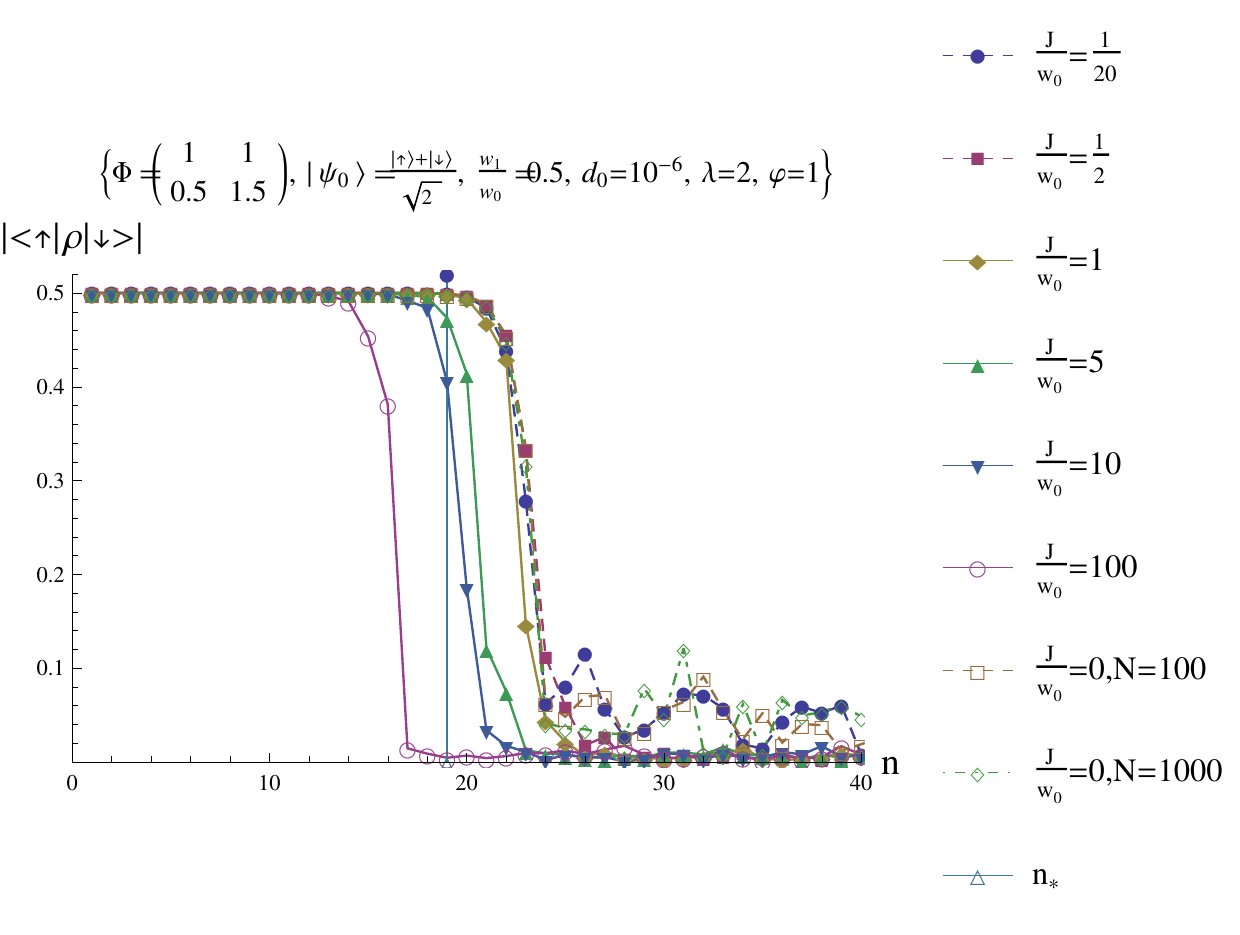}
\includegraphics[width=10cm]{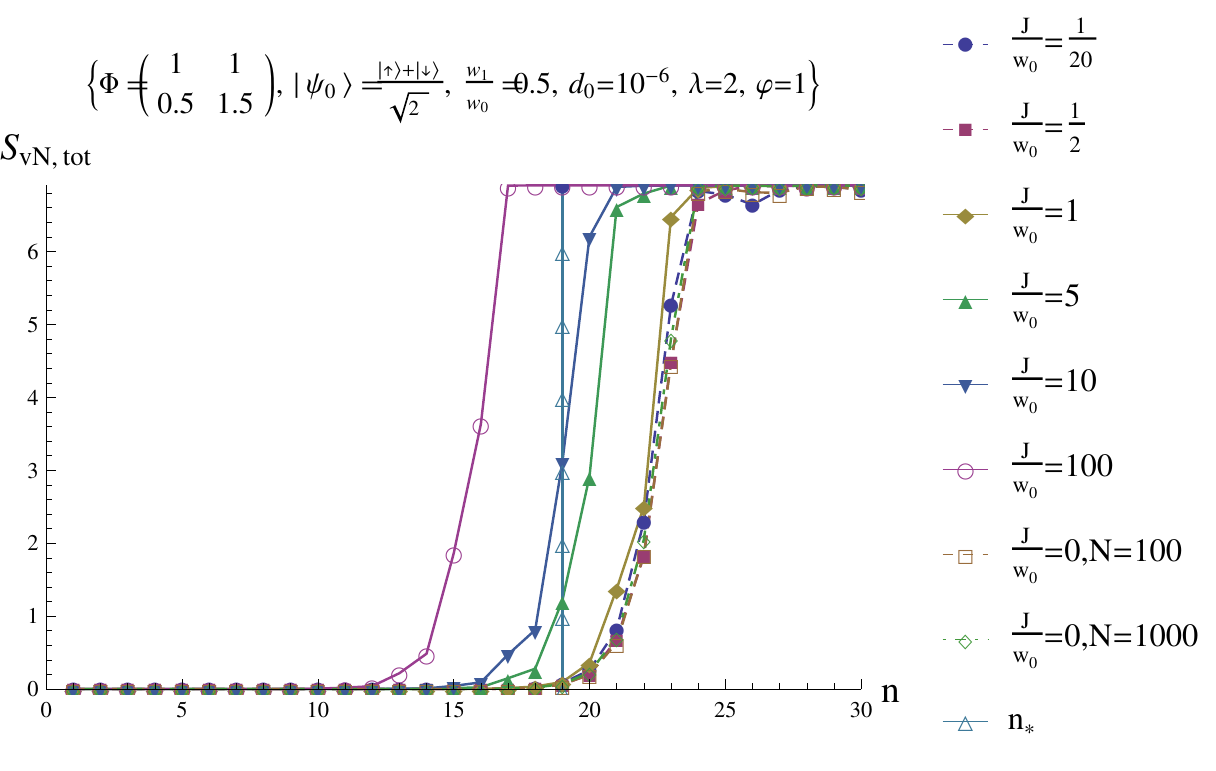}
\end{center}
\caption{\label{entropieheisenberg} Evolution of the coherence (up) and of the entropy (down), for a hundred and a thousand spins without interaction ($\frac{J}{w_0} =0$) and for the average spin of a ten spin chain ($\frac{J}{w_0} \neq 0$) coupled by the Heisenberg interaction. The spins are submitted to several kicks evolving according to a continuous automorphism of the torus (the phase space). Each spin is in the same initial state $\psi_0$. $\Phi$ is the matrix defining the automorphism of the torus. $N$ correspond to the number of spins without any coupling between them. The blue vertical axis correspond to the horizon of coherence.}
\end{figure}

The entropy is a measure of disorder. In the kick bath, the disorder is given by a variation of the kick strengths and delays received by the spins. In the spin chain, the quantum entropy corresponds to a large difference between the states of the spins into the chain and/or to a large entanglement and comes from the kicks (since we choose all spins initially in the same state). The classical entropy (the disorder of the kick bath) begins from the horizon of predictability and the quantum entropy (the disorder of the spin chain), begins from the horizon of coherence. Even if the interaction and the entanglement allow a better transmission of the disorder into the spin chain (see \cite{aubourg}), the time required for the transmission of the disorder from the classical bath to the quantum one is the same than without interaction. We also see that the entropy of the spin chain increases only if the cumulated classical entropy exceeds a threshold value, $S_{max}$. \\

For only five or seven coupled spins, the quantum entropy follows the evolution obtained for fifty spins without interaction, which is not the case for the classical entropy. The classical entropy can be modeled by the Kolmogorov-Sina\"i entropy \cite{benoist} which requires a large number of kick trains. From fifty kick trains (and so fifty spins), the evolution of the classical entropy corresponds to the Kolmogorov-Sina\"i prediction. So, the notion of ``a large number of spins'' is different according to the disorder is quantum or classical. The disorder into the kicks, which is a classical disorder, requires a large number of kick trains to be in conformity with the prediction (Kolmogorov-Sina\"i) whereas a lower number of spins is sufficient to see the disorder into the spin chain. \\

In order to know the modifications of the horizon of coherence when the spins are coupled by a Heisenberg interaction, we have to see the evolution of the coherence and of the entropy with respect to $\frac{J}{w_0}$. Figure \ref{entropieheisenberg} shows the coherence (up) and the quantum entropy (down) evolution with respect to the kick number. When $J<w_0$ the entropy and the coherence behave as a spin ensemble and the empirical formula given by eq.~\ref{horizon} can be used here. The dynamics induced by the kicks dominates the internal dynamics of the chain. In this condition, the results can be compared with the one of a spin ensemble. But, if $J>w_0$ the entropy increase and the fall of the coherence begin earlier and earlier. There is still a horizon of coherence but it cannot be predicted by eq.~\ref{horizon}. The interne dynamics induces more disorder and other phenomena which are not taken into account in eq.~\ref{horizon}. \\

Contrary to the results obtained for a spin chain coupled by an Ising-Z interaction, the Heisenberg coupling allows to conserve for each spin and for the average spin of the chain, the coherence during the horizon of coherence. The analysis of the entropy allows to confirm the length of the plateau given by eq.~\ref{horizon} only when $J\leq w_0$. In this way, we can think that the information transmission between the spins can be conserved and well performed and maybe that some controls can be realized during this horizon. In the next section, we choose to stay in the case where $J\leq w_0$ in order to know the value of the horizon of coherence.  

\section{The Heisenberg coupling, an appropriate interaction to realize information transmission and control during the horizon of coherence}
We have just seen that for a spin chain coupled by a Heisenberg interaction and submitted to a chaotic kick bath, there is a time during which the spins conserve their coherence. The conservation of the coherence is linked to a conservation of the information. For kicks no in a direction on an eigenvector, there is an oscillation of the population and of the coherence due to the kicks. So, before the horizon of coherence, the coherence is conserved with down oscillations. But, if the kicks are in the direction of an eigenvector, during the horizon, even if the spins are kicked, there is a complete conservation of the coherence without any oscillation.

Since the spin could represent a qubit, we will consider the wave observed on the density graphics as an information transmission along the spin chain. The first subsection is devoted to the means to conserve the information during the horizon of coherence. The second one talks about the manner of transmitting an information. The last one uses the information obtained in the second subsection in order to realize a control during the horizon of coherence using stationary kicks on a closed spin chain.

For all the following analysis, we consider that $J\leq w_0$. In this case, we can predict the value of the horizon of coherence and the information transmissions in the density graphic are visible. We also choose this condition in order that the control of the dynamics by the kicks dominates compared to the internal dynamics of the chain.

\subsection{Information conservation}

\begin{figure}
\begin{center}
\includegraphics[width=7.7cm]{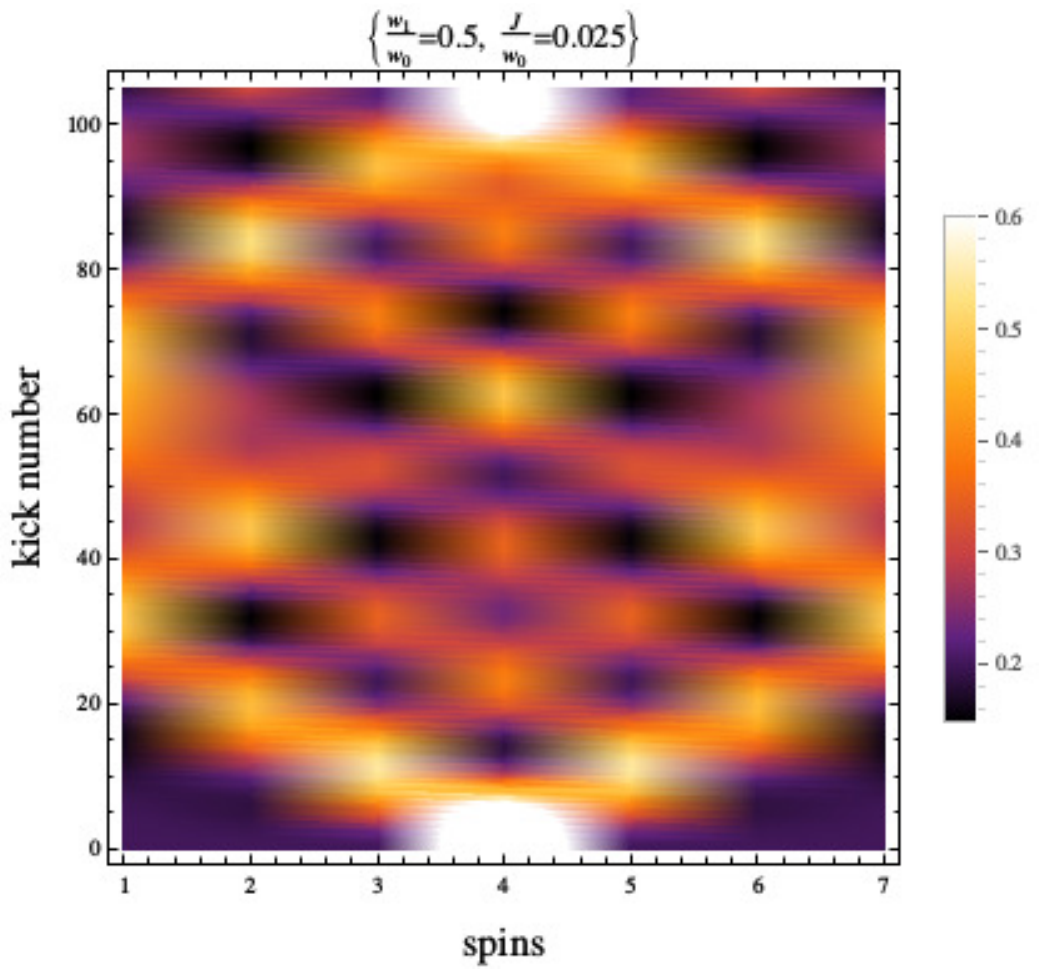}
\includegraphics[width=7.7cm]{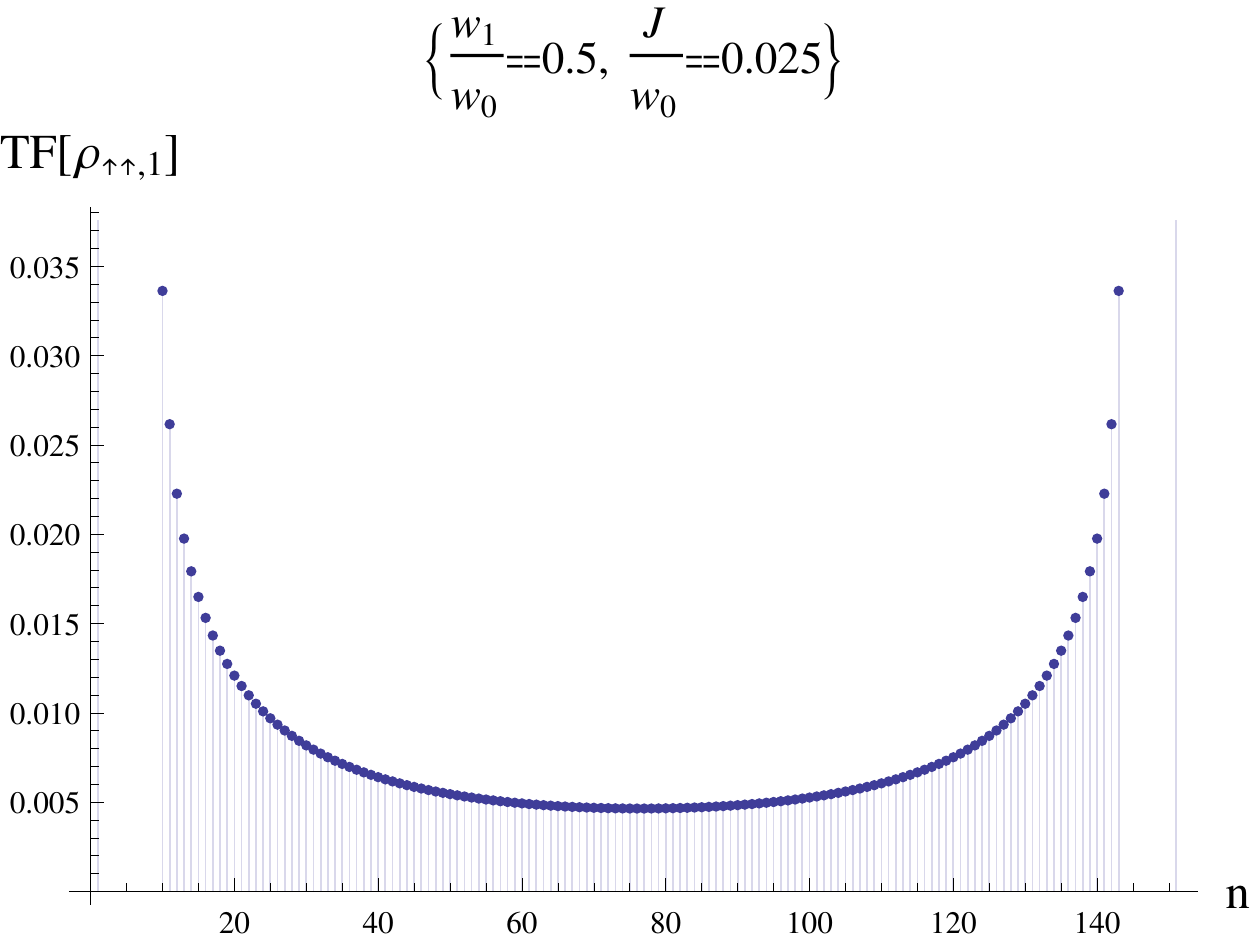}
\end{center}
\caption{\label{heisenbergdensiteonde} Left density of the populations $\langle \uparrow | \rho_n | \uparrow \rangle$  of 7 spins coupled by the Heisenberg interaction with respect to the spins and to the kick number when the spins are not kicked. The right graphic shows the Fourier transform of the first spin population associated with the one on the density graphic. Each spin is in the initial state $\frac{1}{\sqrt{5}} (|\uparrow \rangle + 2|\downarrow \rangle)$ except the fourth one which is $|\uparrow \rangle$.}
\end{figure}

We consider a spin chain coupled by a nearest-neighbour Heisenberg interaction. Figure \ref{heisenbergdensiteonde} represents the evolution of the up population of each spin with respect to the time. All spins are in the initial state $\frac{1}{\sqrt{5}} (|\uparrow \rangle + 2|\downarrow \rangle)$ except the center one (here the fourth) which is in the up state. Since there is no kick, we see an information transmission between the spins due to the Heisenberg coupling represented by the orange-yellow colour (the Heisenberg coupling is isotropic and induces a same state for the coupled spins). This figure presents density peaks (yellow, orange and white colour) which result from the interferences between the various waves. The more yellow point at the end of the graphic seems to be a revival of the initial wave (at $t=0$) and at the middle, the peak looks like an inverse revival of the information (the populations are inverted) : this graphic looks like a wave revival. However the Fourier transform of the population with respect to the time, right graphic on fig.~\ref{heisenbergdensiteonde}, presents a broadband which is a signature of chaotic oscillations \cite{broad}. Thus, the wave packet does not have a complete revival. We called this phenomenon an almost-revival.

\begin{figure}
\begin{center}
\includegraphics[width=8cm]{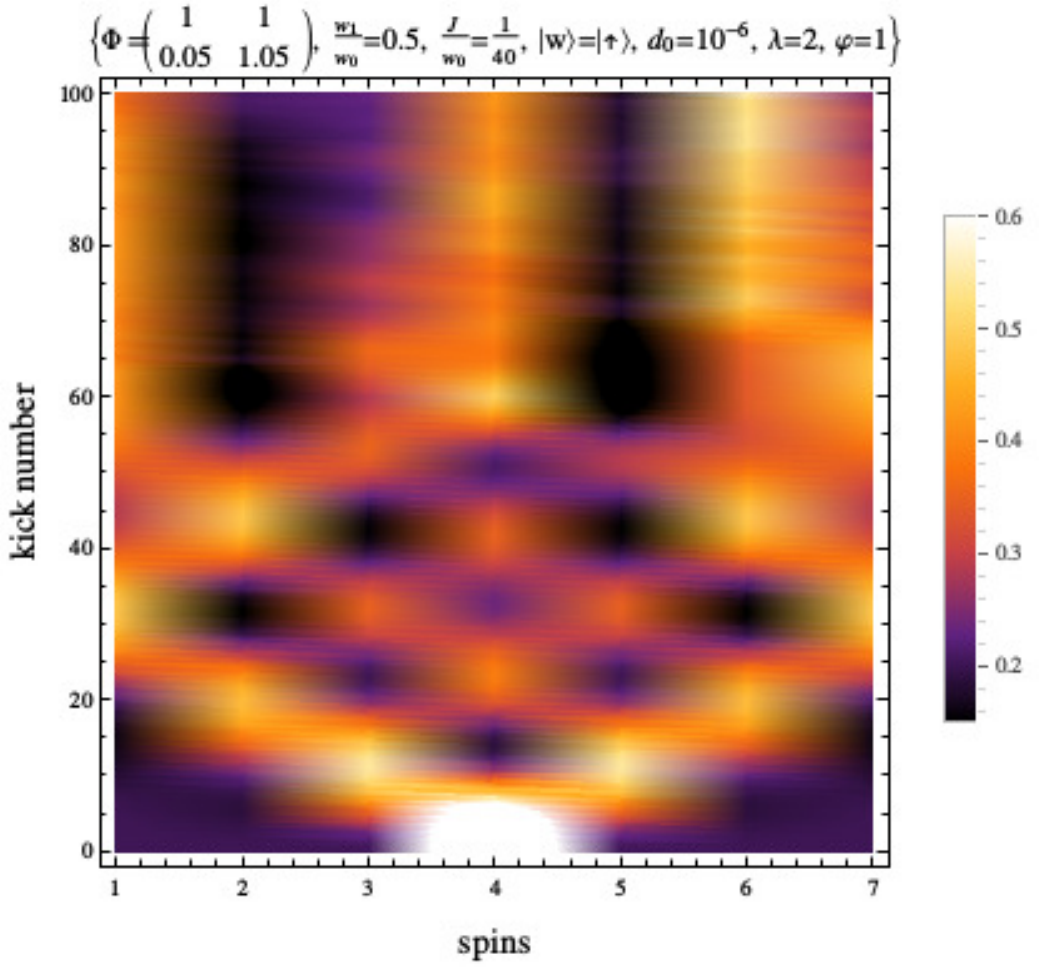}
\includegraphics[width=9cm]{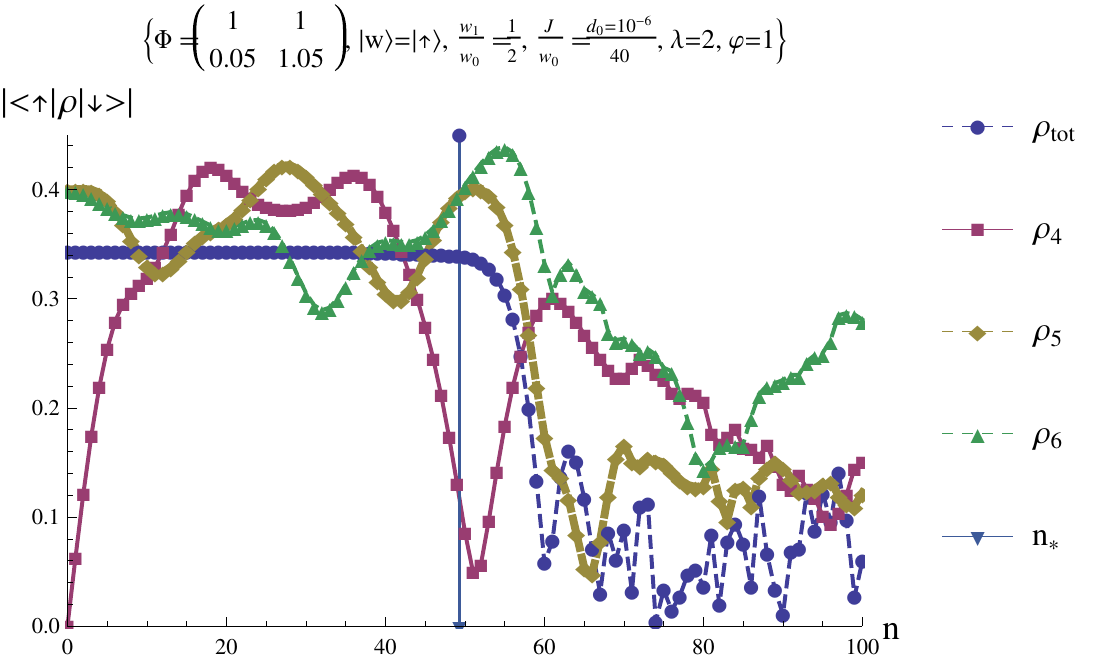}
\includegraphics[width=9cm]{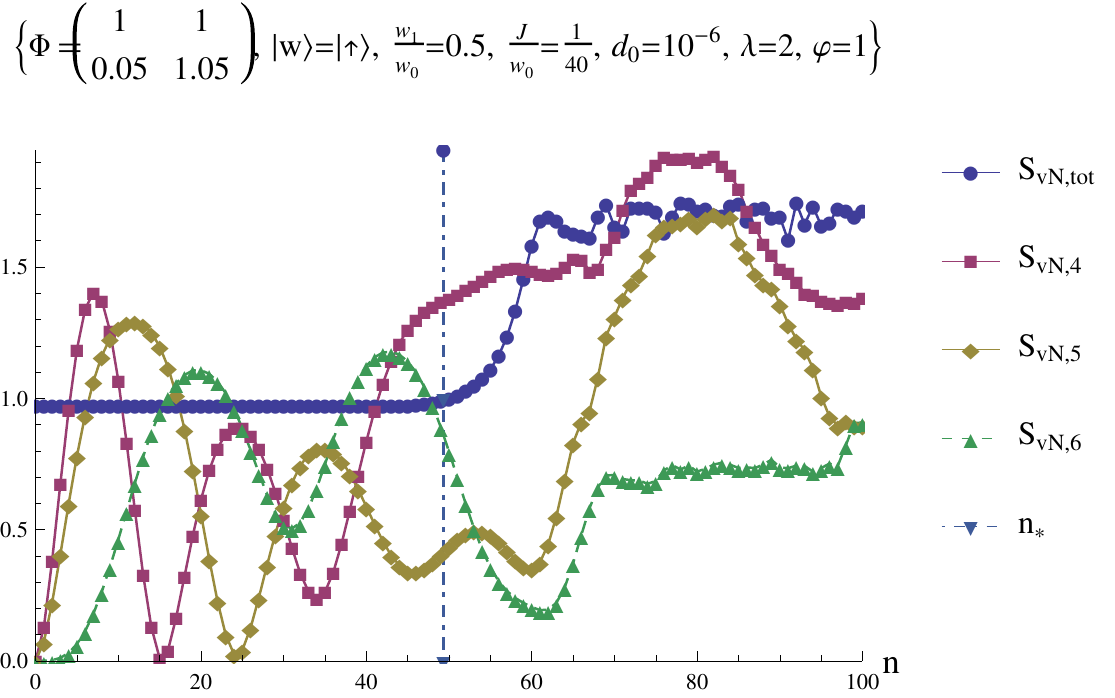}
\end{center}
\caption{\label{heisenbergdensitechaotiqueonde} Density of the populations (up) $\langle \uparrow | \rho_n | \uparrow \rangle$  of 7 spin chain, and evolution of the coherence (second) and of the entropy (down) of the average spin of the chain ($\rho_{tot}$ and $S_{vN,tot}$) of the fourth ($\rho_4$, $S_{vN,4}$), the fifth ($\rho_5$, $S_{vN,5}$) and the sixth ($\rho_6$, $S_{vN,6}$) spin of the chain. Each spin of the chain is coupled by the Heisenberg interaction  and in the initial state $\frac{1}{\sqrt{5}} (|\uparrow \rangle + 2|\downarrow \rangle)$ except the fourth one which is in the state $|\uparrow\rangle$. The spins are submitted to a chaotic kick bath where each kick is in the direction of an eigenvector. $\Phi$ is the matrix defining the automorphism of the torus. The vertical line on the coherence and on the entropy graphics corresponds to the horizon of coherence.}
\end{figure}

\begin{figure}
\begin{center}
\includegraphics[width=8cm]{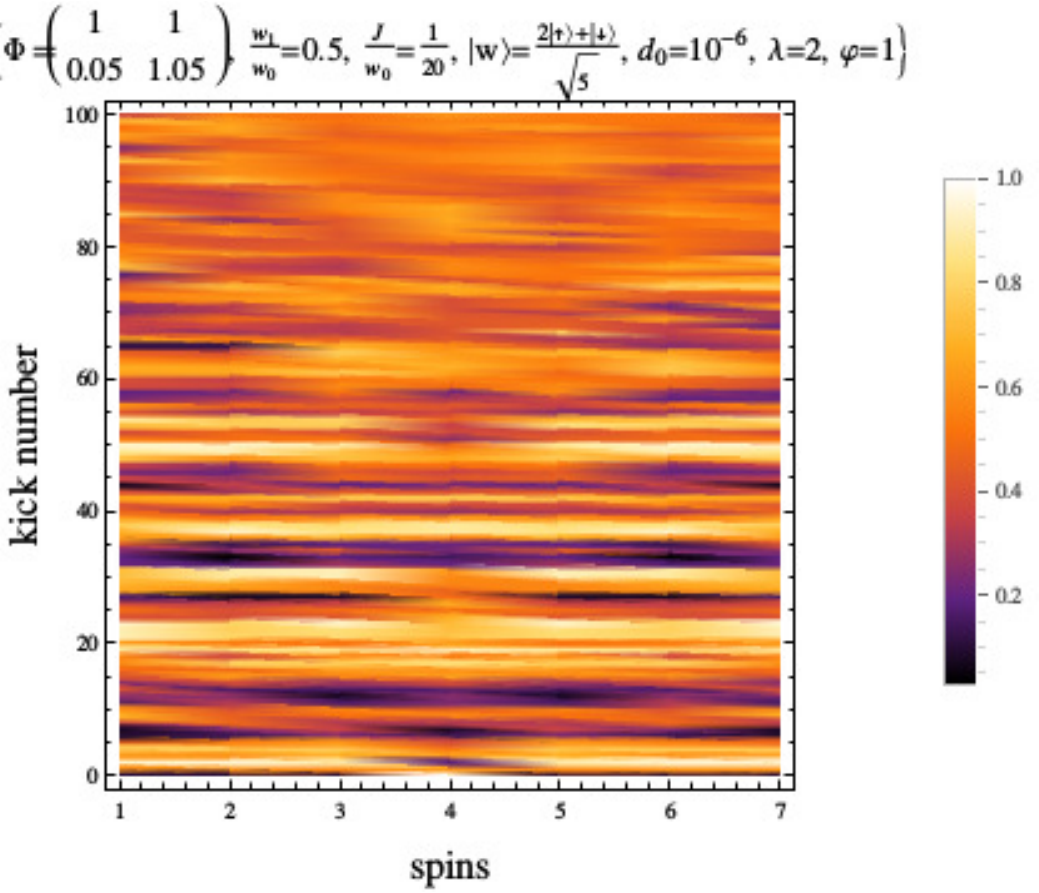}
\includegraphics[width=9cm]{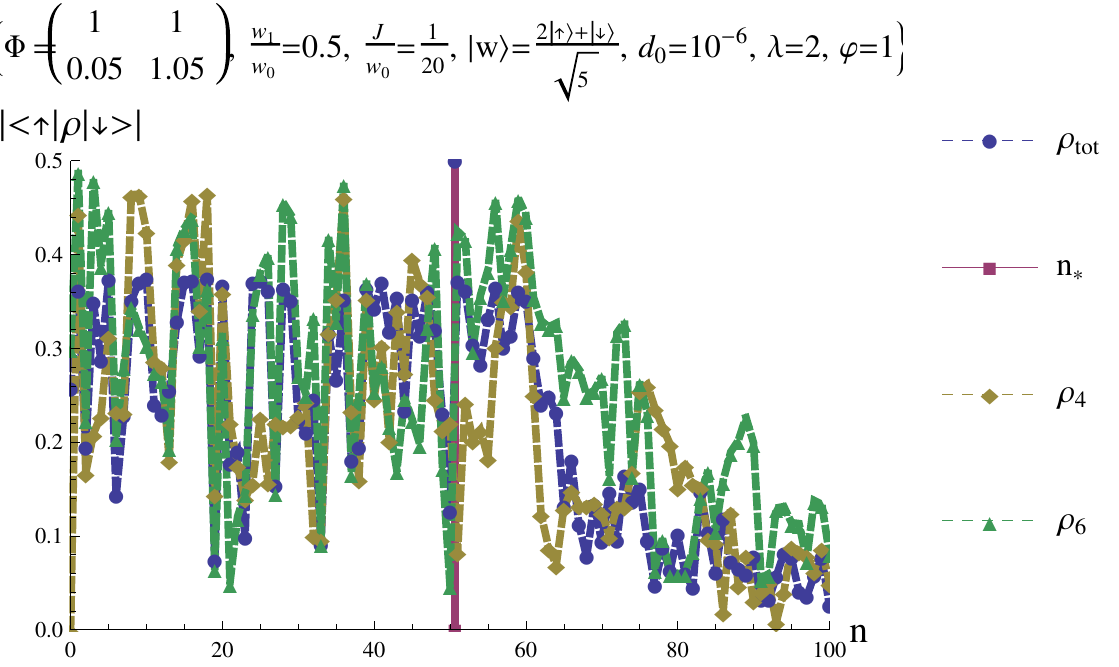}
\end{center}
\caption{\label{heisenbergdensitechaotiquew} Up density of the population $\langle \uparrow | \rho_n | \uparrow \rangle$ of 7 spins coupled by the Heisenberg interaction. The down graphic represents the evolution of the coherence of two spins of the chain and of the average spin of the chain with the kick number. The chain is submitted to a chaotic kick bath where the kick direction is not the one of an eigenvector. At time $t=0$, all spins are in the state  $\frac{1}{\sqrt{5}} (|\uparrow \rangle + 2|\downarrow \rangle)$ except the fourth one which is in the state $|\uparrow\rangle$. $\Phi$ is the matrix defining the automorphism of the torus. The horizontal axis on the coherence graphic (down) corresponds to the horizon of coherence.}
\end{figure}

The up graphic on fig.~\ref{heisenbergdensitechaotiqueonde} is the same than the left one on fig.~\ref{heisenbergdensiteonde} except that all spins are kicked in the direction of an eigenvector. The kicks are disturbed by a chaotic dynamics. We clearly see that the information is completely transmitted along the spin chain until a certain number of kicks, in exactly the same manner than when there is no kick. The kick number for which the information transmission is stopped corresponds to the duration of the horizon of coherence. The second graphic of fig.~\ref{heisenbergdensitechaotiqueonde} shows that the average spin of the chain has a coherence which falls after the horizon of coherence. It is the same thing for one spin of the chain but with large oscillations. These oscillations appear because the coupling is chosen to be not too large in order to see the information transmission and to have a prediction of the value of the horizon of coherence (for a large coupling, the oscillations are really fast). The spins are kicked differently from the horizon of coherence, which explains the lost of information transmission. However, if the kick direction does not correspond to an eigenvector, the information wave cannot be seen, as in fig.~\ref{heisenbergdensitechaotiquew}.

A kick in the direction of an eigenvector allows a transmission of information before the horizon of coherence as if there is no kick. The demonstration is made on \ref{force} and shows that if the strengths are the same for all spins, they do not affect the population. In the case of a chaotically kicked spin chain, all trains of kicks are almost similar until the horizon of predictability. The spins only feel the difference at the horizon of coherence. So before the horizon, the population is not modified by the kicks and we only conserve the coupling variations. Inversely, if there is a modification of the strength kicks between two kick trains or more, the spin populations are not modified in the same manner. This induces a lost of coherence. Since the coupling induces a ``cohesion'' between the spins, if the kicks are all the same the cohesion remains, so the population and the coherence do not change. But if the kicks are modified, when the spins feel this modification, the cohesion into the chain is disturbed and some interferences between the coherence wave appears. If the kicks are not in an eigenvector direction, they modify the spin states and so the spin populations. The interaction can also add some population modifications because it induces a same state for the coupled spin. It produces a modification of the states and same an entanglement between the spins if their states are too different.

For a kick not in the direction of an eigenvector (down graphic of
fig.~\ref{heisenbergplate}), some oscillations appear during the coherence plateau. At the beginning of the dynamics, all spins are in the same state. The kicks on the spins are approximately the same (the initial dispersion of the initial strengths and delays of the kicks is small). So, no disorder is transmitted from the kick bath to the spin chain. But, the kick direction (for a superposition) modifies the spin states and disturbs the transmission of information (up graphic of fig.~\ref{heisenbergdensitechaotiqueonde}). The states of the spins can be more or less close to a classical state and then can lose or gain some coherence. This explains the presence of some coherence oscillations before the horizon of coherence and the lost of information in fig.~\ref{heisenbergdensitechaotiquew}. But, if the kick direction is the one of the eigenvectors, we only conserve the modification due to the interaction and not the one due to the kicks. The up graphic of fig.~\ref{heisenbergplate} shows that kicks in the direction of an eigenvector do not modify the coherence before the horizon of coherence, i.e before the dispersion induced by the sensitivity to initial conditions.\\

If the kicks are not in a direction of an eigenvector, the states of the spins are completely modified. Since the automorphism of the torus induces all the time a variation of the strength and of the delay, sometimes the strength is larger than other times, and so sometimes the spins are more in the direction of the kicks than other times. It is really complicated to realize a control in this condition. However a spin chain coupled by a Heisenberg interaction and kicked in an eigenvector direction transmits all the information (until the horizon of coherence) like a no kicked chain. So we can realize some information transmission during the horizon of coherence.

\subsection{Information transmission}
We consider a spin chain coupled by a nearest-neighbour Heisenberg interaction where $J \leq w_0$. Because of the Heisenberg coupling, two neighbour spins tend to be aligned in the same direction. This allows to obtain an information transmission if two neighbour spins are not in the same initial state. Let the spins be submitted to a chaotic kick bath and be initially in the state $\frac{1}{\sqrt{17}} (|\uparrow> + 4|\downarrow \rangle$ except the first one which is in the up state. We have just seen that for a spin chain chaotically kicked in the direction of an eigenvector, before the horizon of coherence, the spin state evolution is only due to the coupling. But after this horizon there is a modification of the population, a fall of the coherence and an increase of the entropy. If the kick is not in an eigenvector direction, there exists two kinds of oscillations of the population. The first oscillation is due to the kick and the spin frequency and corresponds to the carrier wave. The second one results from the coupling and is the envelope. For a kick not in a direction of an eigenvector, these both oscillations describe the population behaviours. But, for a kick in a direction of an eigenvector, there is only the oscillations due to the coupling if all spins are initially similarly kicked.

\begin{figure}
\begin{center}
\includegraphics[width=7.7cm]{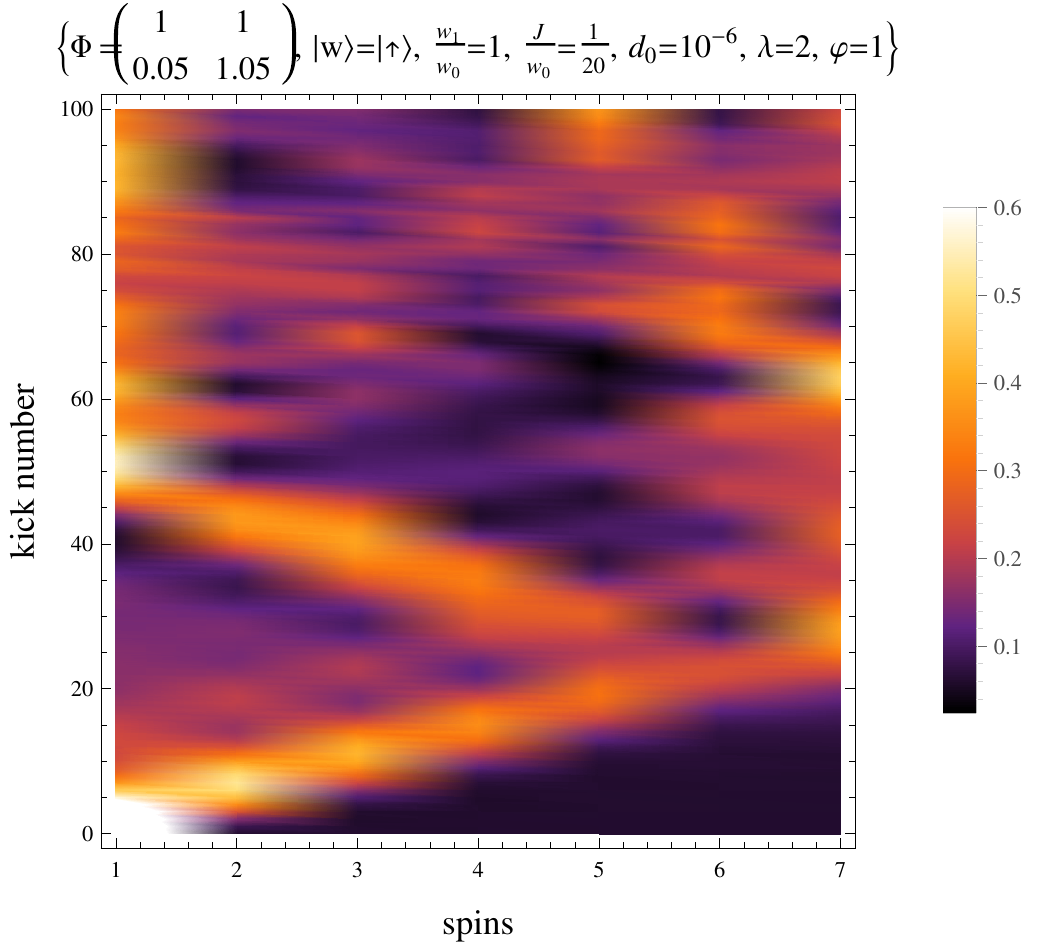}
\includegraphics[width=7.7cm]{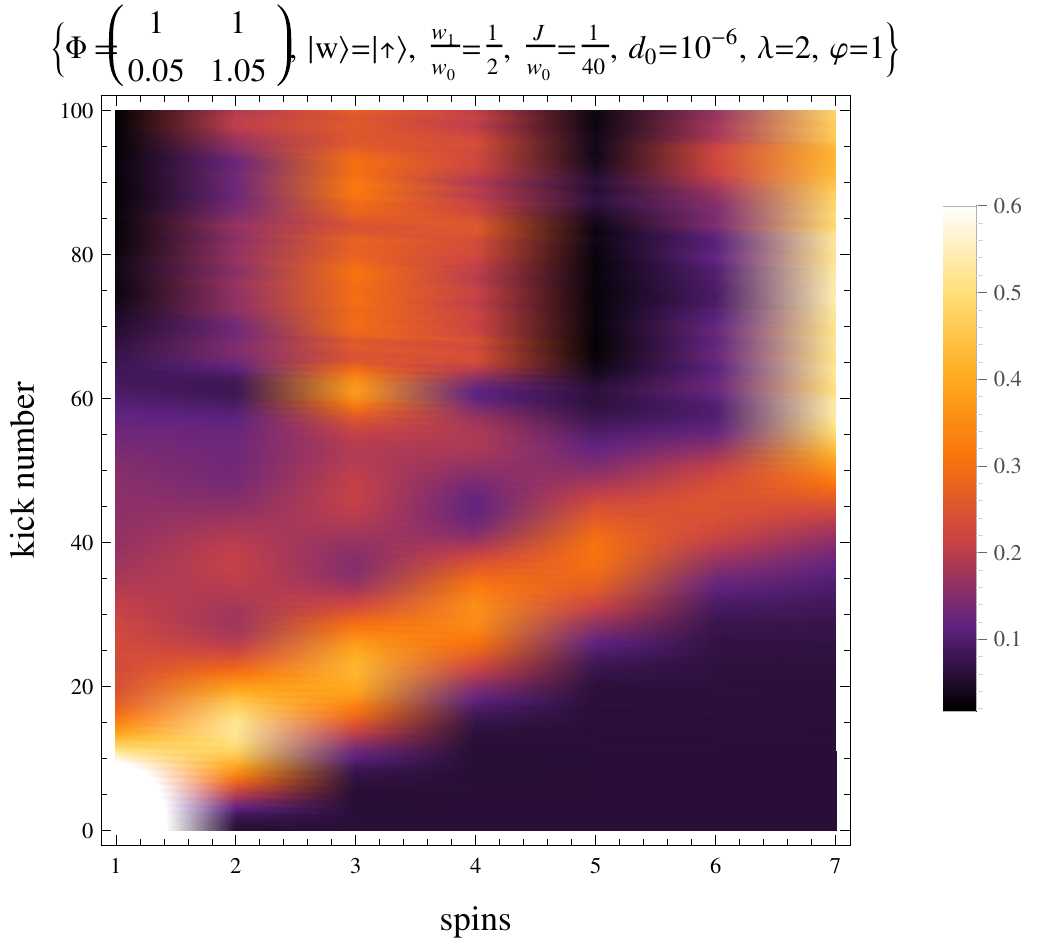}
\includegraphics[width=7.7cm]{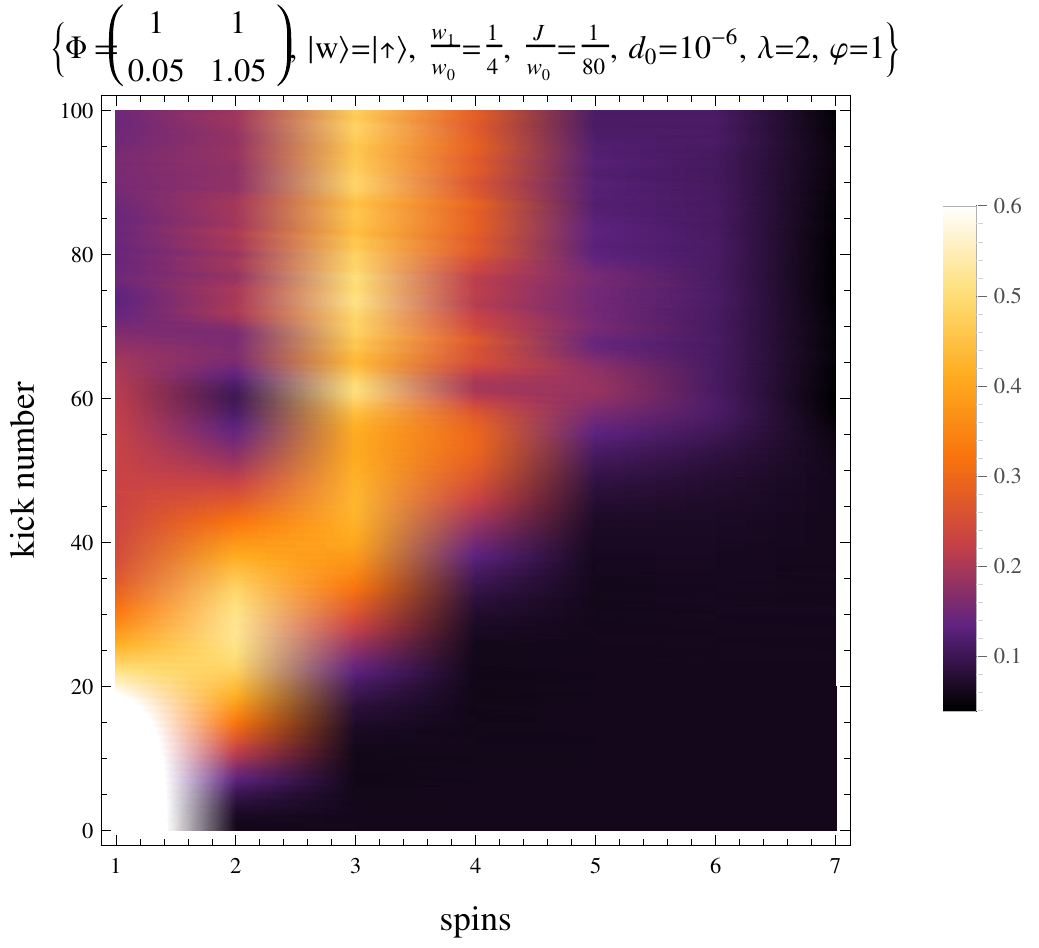}
\includegraphics[width=9cm]{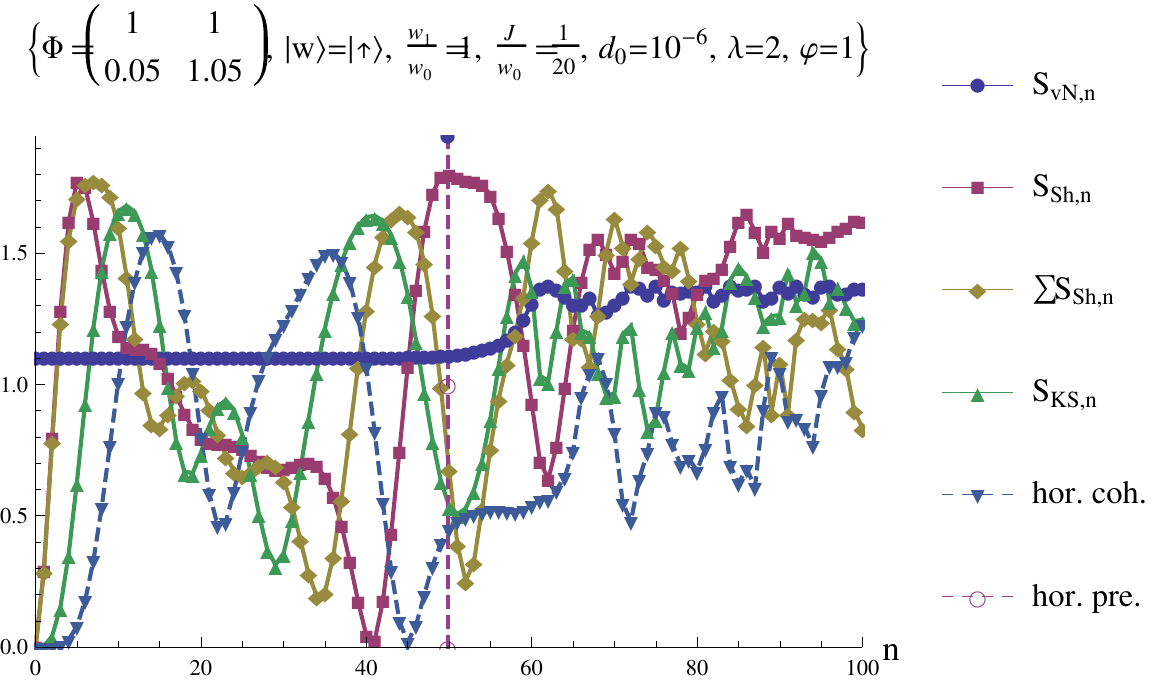}
\end{center}
\caption{\label{informationtransmission} Density of the populations $\langle \uparrow | \rho_n | \uparrow \rangle$  of 7 spins coupled by the Heisenberg interaction with respect to the kick number and to the spins. The down graphic represents the evolution of the coherence with respect to the kicks and is associated with the first density graphic. The chain is submitted to a chaotic kick bath where the kick direction is the one of an eigenvector. At time $t=0$, all spins are in the state $\frac{1}{\sqrt{17}} (|\uparrow> + 4|\downarrow \rangle)$ except the first one which is in the state $|\uparrow\rangle$. $J$ (the interaction parameter) increases from the first graphic to the third one. $\Phi$ is the matrix defining the automorphism of the torus.}
\end{figure}

We want to know the number of spins through which the information passes during the horizon of coherence, with respect to $w_0$ (the kick frequency). We remind that the horizon of coherence can be predicted using eq.~\ref{horizon} (because we are in the condition $J\leq w_0$). Consider fig.~\ref{informationtransmission}. We see a variation of the number of  spins reached by the information before the horizon of coherence with respect to $w_0$. Especially, more $w_0$ increases, less the number of reached spins is large. In the monodromy operator eq.~\ref{monodromy}, $w_0$ is only included in $e^{-\imath \frac{H_{0, I}}{\hbar w_0}}$ . When $w_0$ tends to zero, this exponential presenting a lot of fast oscillations which behaves as if it is equal to zero (Riemann-Lebesgue lemma), and if $w_0$ is large it tends to one. If the exponential tends to one, the impact of this factor on the spin states is lower than if this factor tends to 0, which is in agreement with our observations. We can also make this analysis by considering the variation of the interaction parameter. The results will be the same. Physically, larger the interaction parameter is, faster the spins tend to be in the same state and so faster they transmit their information.

\begin{figure}
\begin{center}
\includegraphics[width=10cm]{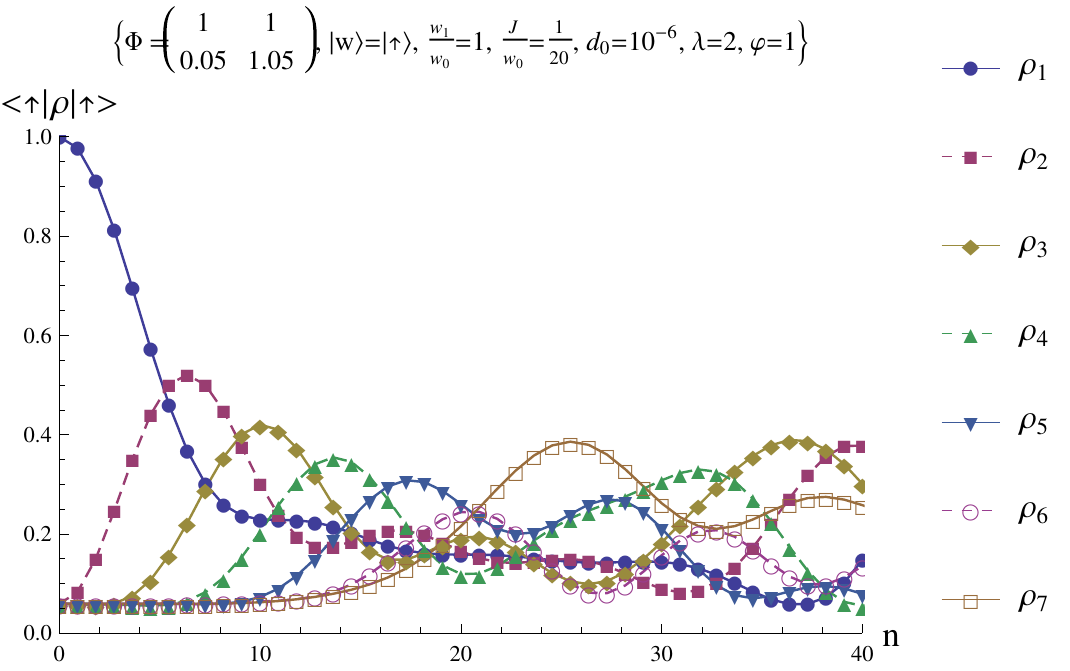}
\end{center}
\caption{\label{spintransmission} Evolution of the up population of all the spins of a seven spin chain with respect to the kick number. Each spin is coupled to its nearest-neighbours by a Heisenberg interaction and submitted to chaotic kicks where the kicks are in the direction of an eigenvector. All spins are initially in the state  $\frac{1}{\sqrt{17}} (|\uparrow> + 4|\downarrow \rangle)$ except the first one which is in the up state. This evolution correspond to the first density graphic on fig.~\ref{informationtransmission}. $\Phi$ is the matrix defining the automorphism of the torus.}
\end{figure}

In order to obtain the number of spins reached by the information before the horizon of coherence, we need to know the transmission velocity. Figure \ref{spintransmission} presents the up population evolution of the seven spins of a chain coupled by a Heisenberg interaction with respect to the kick number and corresponds to the up left graphic of fig.~\ref{informationtransmission}. Each spin transmits its information to its neighbours. The state of the first spin is up. It transmits its information to the second spin. The state of the second spin depends on the state of its two neighbours and it tends to be a superposition of them. It is the same thing for the other spins. In addition, the up populations of the spins do not decrease to $0$ but to a value upper than $0$ at the end of an oscillation. So each spin conserves a little information which explains the decrease of the peak height of the up state from one spin to the following one with the kick number. The last spin has a single neighbour, it is only influenced by it. This is like a wave in a box, it has an increase of the information of the previous spins (a kind of concentration of the wave). We observe the classical phenomenon of signal scattering during its propagation (the spreading of the signal with an attenuation of its maximal intensity). Here the signal corresponds to the population with the maximal up state which spreads along the chain.\\

We want to obtain the oscillation period of one spin coupled with only one spin (so an edge spin in our case). The interaction Hamiltonian of two coupled spins is given by the following matrix
\begin{equation}
\begin{pmatrix}
-\frac{J \hbar}{4 w_0} & 0 & 0 & 0 \\
0 & \frac{J \hbar}{4 w_0} & -\frac{J \hbar}{2 w_0}& 0 \\
0 & -\frac{J \hbar}{2 w_0} & \frac{J \hbar}{4w_0}& 0 \\
0 & 0 & 0 & -\frac{J \hbar}{4 w_0}
\end{pmatrix}
\end{equation}
The coupling part is in the middle of this matrix with the non-diagonal terms. We consider the matrix block associated with the states $(|\uparrow \downarrow \rangle, |\downarrow \uparrow \rangle)$,
$\begin{pmatrix}
\frac{J \hbar}{4 w_0} & -\frac{J \hbar}{2 w_0} \\
-\frac{J \hbar}{2 w_0} & \frac{J \hbar}{4w_0}
\end{pmatrix}$ 
for which the eigenvalues are $\lambda_{\pm}= \frac{J \hbar}{4 w_0} \pm \frac{J \hbar}{2 w_0}$. Then the frequency of the Rabi oscillations for a spin which has only one neighbour corresponds to $\lambda_{+}-\lambda_{-} = \frac{J \hbar}{ w_0}$. An edge spin has an oscillation period of
\begin{equation}
T^{eff}_{edge}=\frac{w_0}{J \hbar}.
\end{equation} 
A spin with two neighbours has its frequency multiplicated by two and so its period divided by two
\begin{equation}
 T^{eff}_{mid}=\frac{w_0}{2 J \hbar}
\end{equation}

\begin{figure}
\begin{center}
\includegraphics[width=10cm]{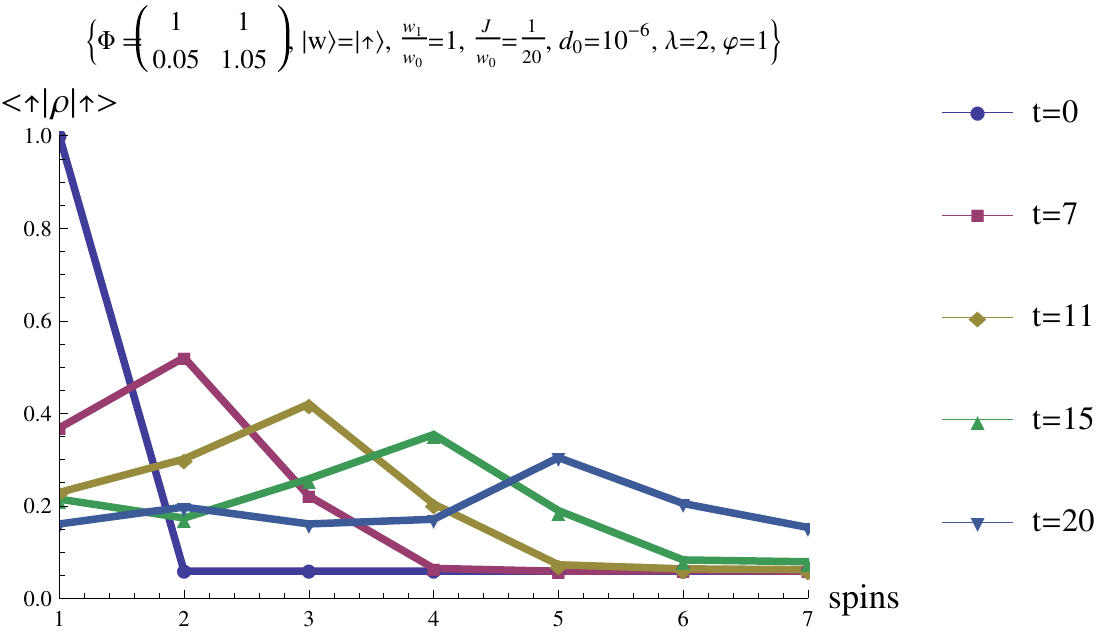}
\end{center}
\caption{\label{dispersiononde} Evolution of the population of a seven spin chain with respect to the spin position into the chain. This information is represented for five times. They correspond to the time predictions of when the first, the second, the third, the fourth and the fifth spin respectively are reached by the maximal information. Each spin is coupled to its nearest-neighbours by a Heisenberg interaction and submitted to chaotic kicks. All spins are initially in the state  $\frac{1}{\sqrt{17}} (|\uparrow> + 4|\downarrow \rangle)$ except the first one which is in the up state. This evolution corresponds to the one observed on the first density graphic on fig.~\ref{informationtransmission}. $\Phi$ is the matrix defining the automorphism of the torus.}
\end{figure}

During the information propagation, there is a wave packet spreading. So, the oscillation period of each spin increases during the propagation of the information. This phenomenon can be seen fig.~\ref{dispersiononde}. More the time increases, more the wave packet is spread on a larger number of spins.

The oscillation period of the first spin is $T^{eff}_{edge}$. But the second one, which has two neighbours, receives the information from a spin which only has one neighbour and so does not have the same oscillation period than itself. The oscillation period of the second spin is then the average between the one of one spin with two neighbours and the one for one spin with only one neighbour
\begin{equation}
T^{eff,2}_{average,}=\frac{1}{2} (\frac{w_0}{J \hbar} +\frac{w_0}{2 J \hbar})
\end{equation}
In the same way, for the other spins, we obtain the oscillation period 
\begin{equation}
T^{eff^n}_{average}=\frac{1}{2} (T^{eff,n-1}_{average}+T^n)
\end{equation}
with $T^n= T^{eff}_{edge}$ or $T^{eff}_{mid}$ the oscillation period of the $n$-th spin only induced by the nearest-neighbours.

Now, we know the oscillation period of all spins of the chain. If we obtain the time to transmit the information from the maximal up population of one spin to the maximal up population to the following spin, we have all data that we need. Consider a spin called ``$sp$'' which has two nearest-neighbours. This spin has its maximal information when the one before it and the one after it cross each other what is well seen fig.~\ref{spintransmission} and \ref{dispersiononde}. This is only seen for the first transmission from the first spin to the last one, i.e only for a one-way transmission of the information and not for the return way because of the scattering and the interferences. On fig.~\ref{spintransmission}, at $t=0$ only the first spin has the information. When $t$ increases, the number of spins reached by the information increases, but also, the wave spreads on a larger number of spins. The up population of the spin before the spin $sp$ decreases whereas the one after it increases. The maximal information that the spin $sp$ can obtain is when its neighbours have the same information and so when they cross each other. Since the shape of the wave packet is symmetric with respect to the maximal up population, the spin $sp+1$ has the maximal information when the spin $sp$ is at a quarter of its oscillation. The dispersion and the interferences of the wave packet induces that it is hard to obtain the value of the up population of all spins with the time. The dispersion is not only between three spins but more. One spin population has its maximum at half of its oscillation period and transmits it at three-quarter of it. Then, 
\begin{equation}
T_{Trans}^n = \frac{1}{4} T^{eff,n}_{average}
\end{equation}
This does not concern the last spin of the chain in the transmission direction. The last spin has twice the period of a middle spin. So $T                     ^N_{Trans}$ has to be multiplicated by two. Finally a complete period of information transmission from the first spin to the last one is (a one-way)
\begin{equation}
P= \frac{3}{4} T^{eff,2}_{average} + \sum_{n=3}^{N-2} T_{Trans}^n  + 2T_{Trans}^{N} 
\end{equation}
where $N$ corresponds to the number of spins. The first term gives the time to obtain the maximum information of the third spin, the second gives the quarter of an oscillation period of the spins from the third to the second to last one, and the last term is linked to the maximal information of the last spin of the chain. For the model chosen,
\begin{multline}
P=\frac{3}{4}\left[\frac{1}{2}\left(\frac{w_0}{\hbar J} +\frac{w_0}{2\hbar J} \right) \right] + \frac{1}{4}\left[\frac{1}{2}\left(\left[\frac{1}{2}\left(\frac{w_0}{\hbar J} +\frac{w_0}{2\hbar J} \right) \right] +\frac{w_0}{2\hbar J}\right) \right]\\
 + \frac{1}{4}\left[\frac{1}{2} \left(\left[\frac{1}{2}\left(\left[\frac{1}{2}\left(\frac{w_0}{\hbar J} +\frac{w_0}{2\hbar J} \right) \right] +\frac{w_0}{2\hbar J}\right) \right] +\frac{w_0}{2\hbar J} \right)\right]\\
 + \frac{1}{4}\left[\frac{1}{2}\left( \left[\frac{1}{2} \left(\left[\frac{1}{2}\left(\left[\frac{1}{2}\left(\frac{w_0}{\hbar J} +\frac{w_0}{2\hbar J} \right) \right] +\frac{w_0}{2\hbar J}\right) \right] +\frac{w_0}{2\hbar J} \right)\right] +\frac{w_0}{2\hbar J} \right) \right]\\
 + \frac{1}{2}\left[\frac{1}{2}\left(\left[\frac{1}{2}\left( \left[\frac{1}{2} \left(\left[\frac{1}{2}\left(\left[\frac{1}{2}\left(\frac{w_0}{\hbar J} +\frac{w_0}{2\hbar J} \right) \right] +\frac{w_0}{2\hbar J}\right) \right] +\frac{w_0}{2\hbar J} \right)\right] +\frac{w_0}{2\hbar J} \right) \right] + \frac{w_0}{\hbar J} \right)\right]
\end{multline}

Finally, to obtain the number of spins ($nsp$) reached by the information, we calculate 
\begin{equation}
nsp=\frac{n_*}{P} \times N -NTurn
\end{equation}
$NTurn$ is the number of one-way transmissions from the first spin of the chain to the last one in the direction of the transmission. This number has to be removed in order to not add the last spin or the first one two times. In order to know the value of the horizon of coherence ($n_*$), we use eq.~\ref{horizon}. To obtain it, we realize a simulation with 700 classical systems (700 trains of kicks). For this study we need a large number of classical systems in order that the Kolmogorov Sina\"i analyses would be efficient. With this number of spins, the horizon of coherence of fig.~\ref{informationtransmission} is approximately 50. On the entropy graphic (the down one) of fig.\ref{informationtransmission}, we see that the entropy begins to increase at 50 kicks. However, the increase is relatively low. The large increase begins approximatively at 55 kicks. This value is in accordance to when the disorder becomes to be visible on the density graphics of the same figure. So let $n_*$ = 55. The prediction of the number of spins reached by the information, for the down graphic of fig.~\ref{informationtransmission} is 3.5, the prediction for the second one is 7 and for the upper one, the prediction is 13. These values correspond to what we obtain on the graphics. But it is necessary to watch out. We use a nearest-neighbour interaction. So the information of one spin is transmitted to two spins. To simplify the calculation the possibility to have a revival information by the wave interferences is not taken into account. We only consider the transmission of the information of the first spin.\\

\begin{figure}
\begin{center}
\includegraphics[width=8cm]{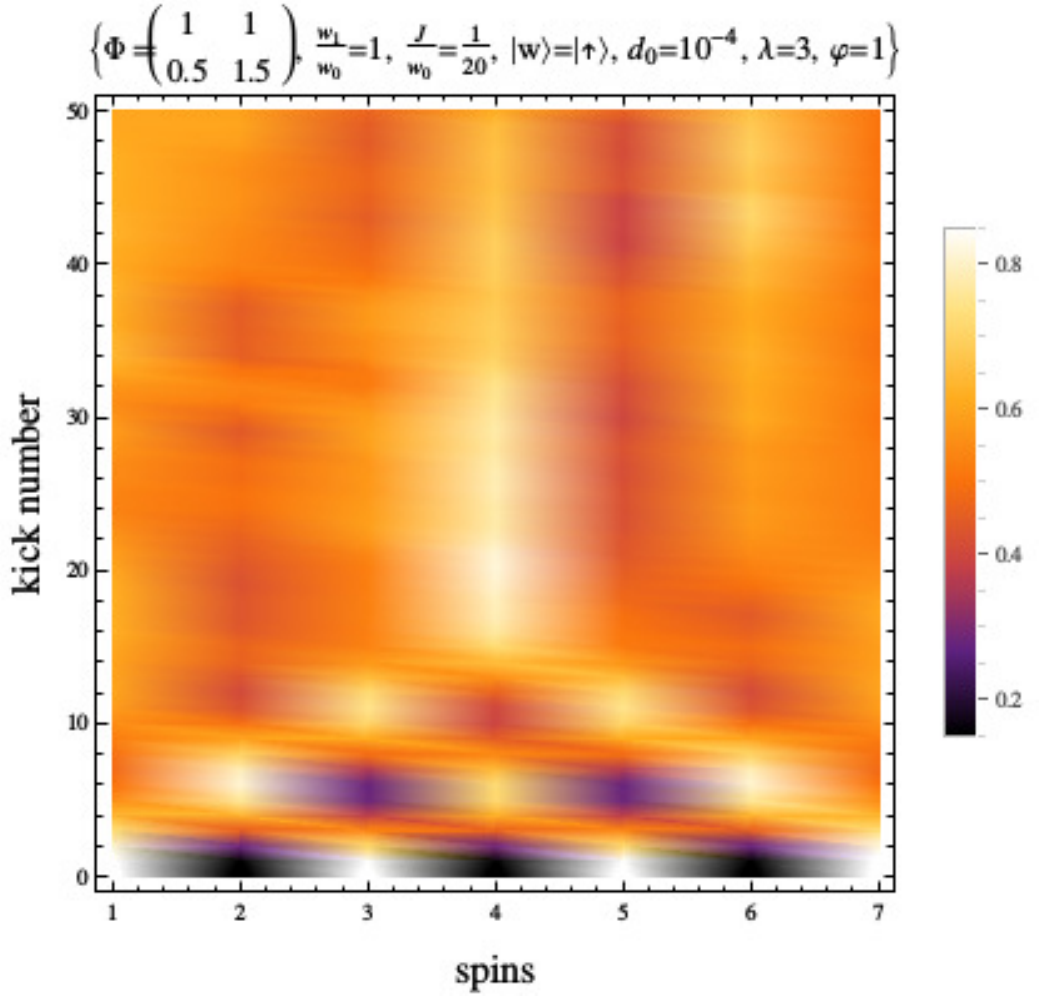}
\end{center}
\caption{\label{heisenberg} Density of the population $\langle \uparrow | \rho_n | \uparrow \rangle$  of 7 spins coupled by the Heisenberg interaction with respect to the spins and to the kick number. The chain is submitted to a chaotic kick bath where the kicks are in the direction of an eigenvector. The states of the spins are initially $|\psi_{2n+1}\rangle=\frac{1}{\sqrt{10}}(3 |\uparrow\rangle + |\downarrow \rangle)$, $|\psi_{2n}\rangle=\frac{1}{\sqrt{10}}( |\uparrow \rangle + 3 |\downarrow \rangle)$. $\Phi$ is the matrix defining the automorphism of the torus.}
\end{figure}

We can also observe another behaviour fig.~\ref{heisenberg}. In this one we have alternate the spin states. If the position of the spin in the chain is odd, then the spin state is $|\psi_{2n+1} \rangle =\frac{1}{\sqrt{10}}(3 |\uparrow\rangle + |\downarrow \rangle)$, and if it is even, $|\psi_{2n} \rangle=\frac{1}{\sqrt{10}}( |\uparrow\rangle + 3 |\downarrow \rangle)$. The horizon of coherence is about 13. Like previously we can know the states of the spins at the moment of the horizon of coherence with respect to $w_0$ and to the interaction parameter. There is also an other effect well seen in this graphic. There is a kind of state freezing. The upper state of the fourth spin at the horizon of coherence is conserved for a large number of kicks. We can also see this effect for the other density graphics. This phenomenon is explained on the next section.\\

Note :  For a sake of simplicity, we choose to not consider the case where the spins are almost all in the state $|\uparrow \rangle$ and/or $|\downarrow \rangle$, i.e. in the direction of an eigenvector. If the spin direction is initially near to an eigenvector, at $t=0$, there is no coherence between the spins because they are in a classical direction. So the effect of the horizon of coherence like we have described it in the third section (with a fall of the coherence and a large increase of the entropy) is not present. However, the results will be the same. From the time which corresponds to the horizon of coherence and at each kicks, the spins feel different kick strengths and delays. So the spins react to the kicks which induce that the information transmission is stopped. So the effect is the same.\\

We can now determinate the information transmission time between the spins coupled by a nearest-neighbour Heisenberg interaction. Using these analyses, it can be interesting to see if it is possible to realize a control experience. We have just observed what happened if we kick chaotically the system. Since the strength and the delay are modified all the time from one kick to another due to the automorphism of the torus, it may be interesting to considered other kind of kicks.

\subsection{Control of the information transmission}
We consider a closed spin chain where each spin is submitted to a nearest-neighbour Heisenberg interaction. This model requires a modification to complete the interaction Hamiltonian as follows 
\begin{equation}
H_{I} =  \sum_{n=1}^{N-1} H_{I_n} -J(S_x\otimes \mathrm{id}^{\otimes N-1} \otimes S_x + S_y \otimes \mathrm{id}^{\otimes N-1} \otimes S_y + S_z \otimes \mathrm{id}^{\otimes N-1} \otimes S_z)
\end{equation}
where we have just added the interaction term between the first and the last spin. \\

\begin{figure}
\begin{center}
\includegraphics[width=7.7cm]{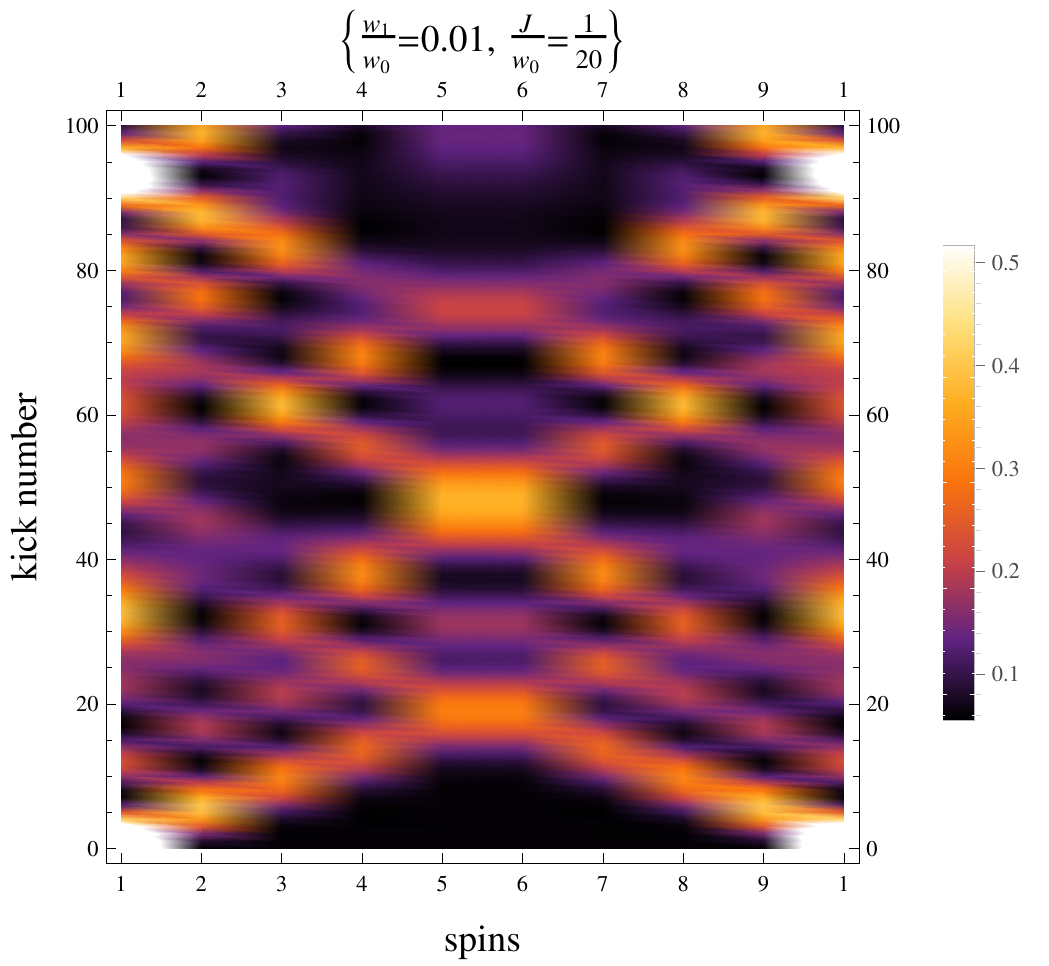}
\includegraphics[width=7.7cm]{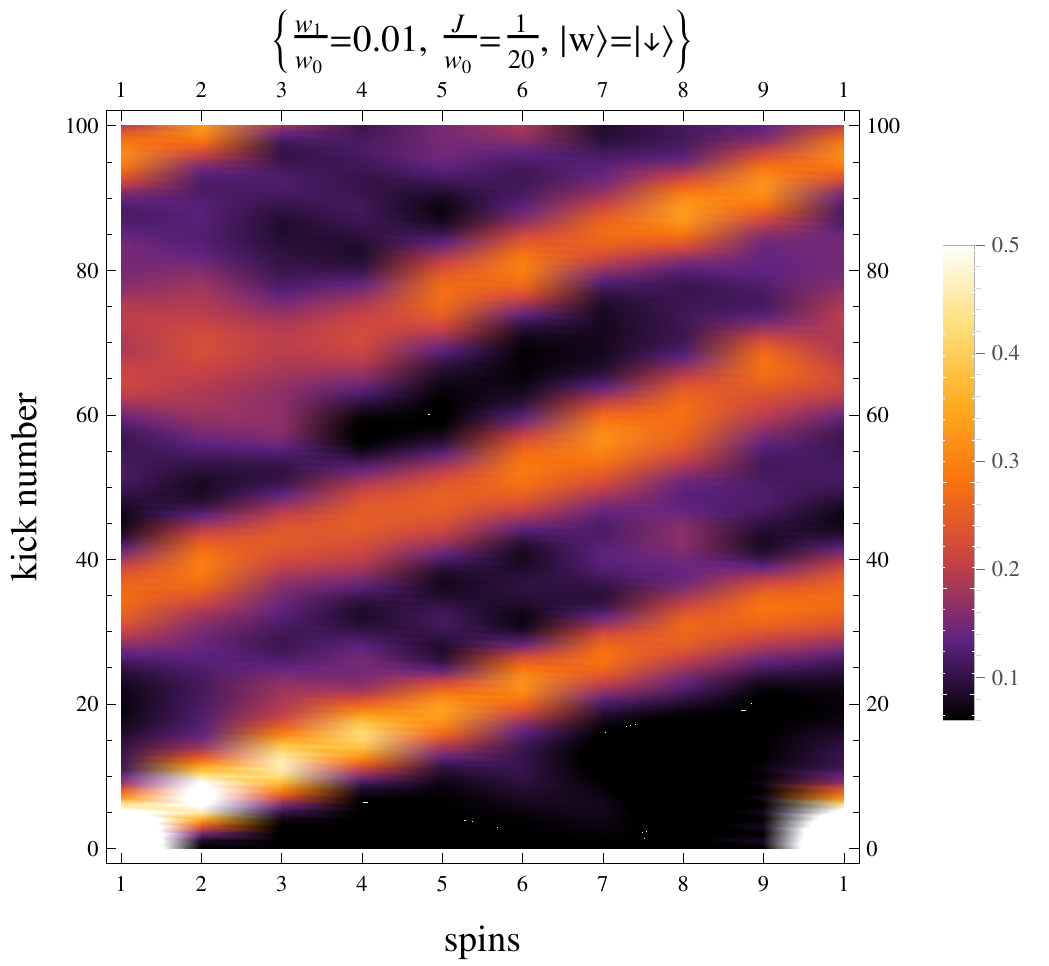}
\end{center}
\caption{\label{evolution_ss_frappe} Density of the population $\langle \uparrow | \rho_n | \uparrow \rangle$  of 9 spins coupled by a nearest-neighbour Heisenberg interaction with respect to the kick number and to the spins. For the left graphic, the chain is not submitted to kicks. However for the right one, only the ninth spin is kicked in the down direction ($w = |\downarrow>$) between the first and the tenth kick. All spins are initially in the state $\frac{1}{\sqrt{17}}(|\uparrow\rangle + 4 |\downarrow \rangle)$ except the first one which is in the up state.}
\end{figure}

Let all spins be in the initial state $\frac{1}{\sqrt{17}}(|\uparrow\rangle + 4 |\downarrow \rangle)$ except the first one which is in the state $|\uparrow>$. Without any kick we obtain a free information transmission between the spins as we can see on the left density graphic of fig.~\ref{evolution_ss_frappe}.
Since the chain is closed, the information of the first spin is transmitted to the second and to the ninth spin. For the control, we would like that the information goes only in one direction, toward the second spin. For this, we calculate the oscillation period of the first spin as if it has one neighbour, i.e. $T^{eff}_{edge} = \frac{\omega_0}{J \hbar}$. Here, we have the half of its oscillation, so $T^1 = \frac{\omega_0}{2J \hbar} = 10$. We choose to kick the ninth spin in order that it remains in a state near to $|\downarrow>$ during the first oscillation of the first spin, (approximately ten kicks). Thus, the first spin can only transmit its information to the second spin and behaves as if it has only one neighbour (this explains the calculation of the oscillation period of the first spin). For this control model, it is not interesting to use the dynamic of the chaotic kicks before the horizon of coherence. If we kick a spin chaotically, there is a modification of the strength and of the delay. Thus sometimes the spin is less kick and can oscillate more. If the kicks are stationary, they are all similar and with a large strength. This forces the spins to stay in the state near to the down one. This control is represented on the right density graphic of fig.~\ref{evolution_ss_frappe}, which is what we want to obtain.\\

This model is interesting because it does not present any interference between the spins. We force the initial to go toward one direction, toward the second spin. We introduce a way to transmit the information. In addition there is a quarter of period during which the spin $n+1$ has its information which decreases and the spin $n$ has no information. So the probability that the spin $n+1$ transmits again an information to the spin $n$ is really low.  \\

\begin{figure}
\begin{center}
\includegraphics[width=8cm]{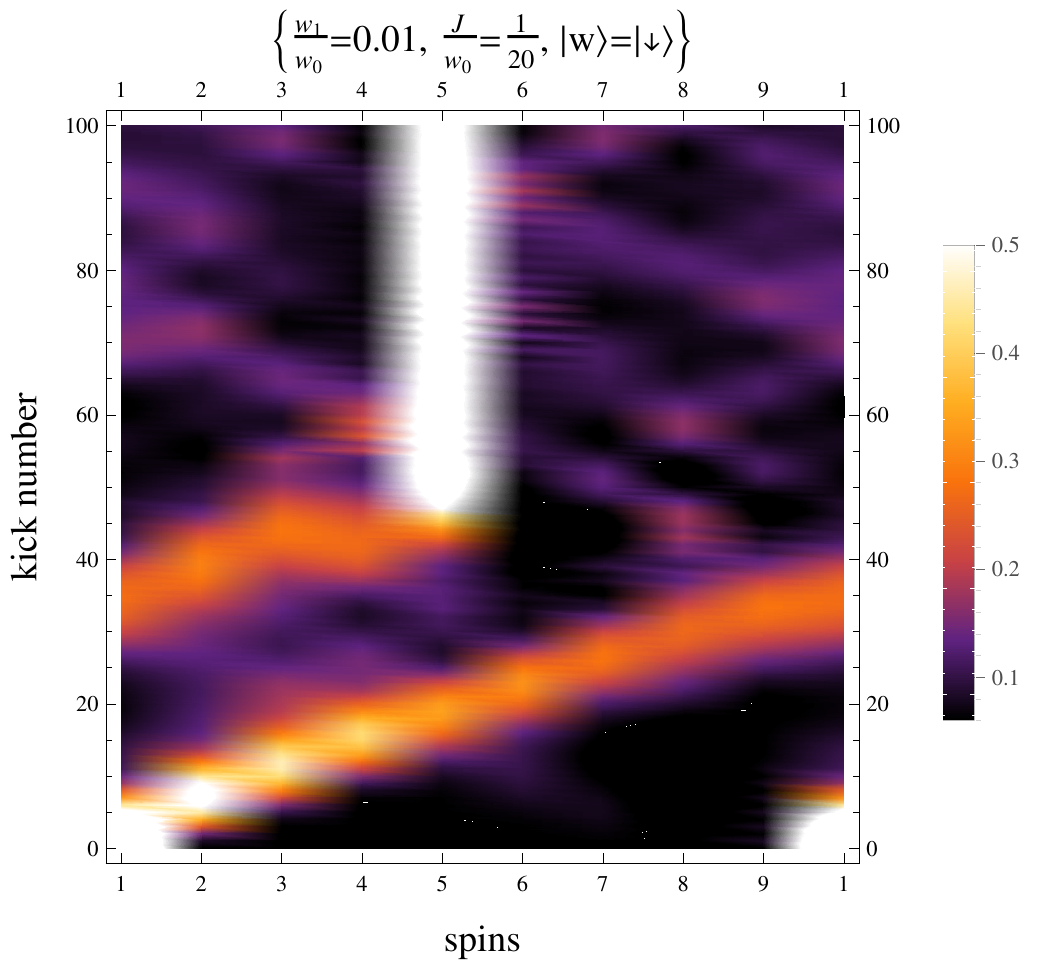}
\includegraphics[width=10cm]{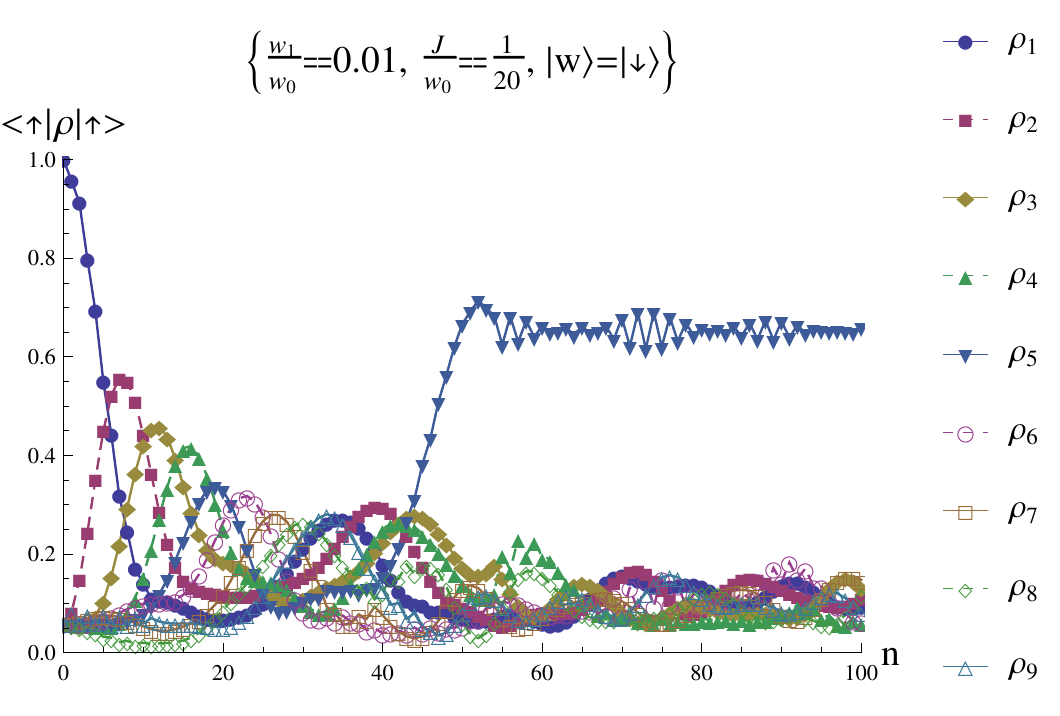}
\end{center}
\caption{\label{information_transmission} Up, density graphic of the population $\langle \uparrow | \rho_n | \uparrow \rangle$ of 9 spins coupled by a nearest-neighbour Heisenberg interaction, with respect to the kick number and to the spins. Down, population evolution as a function of the kicks. In a way that the fifth spin conserves all the information after that the information crosses it two times, the other spins are kicked in the down direction when the first oscillation is end for the spins 6, 7, 8 and 9 and after second oscillation is end for the spins 1, 2, 3 and 4. All spins are initially in the state $\frac{1}{\sqrt{17}}(|\uparrow\rangle + 4 |\downarrow \rangle)$ except the first one which is in the up state.}
\end{figure}

We can now stop the information transmission. We choose to stop the information when it is on the fifth spin and when it crosses it two times. We obtain fig.~\ref{information_transmission}. In order to obtain this graphic, we have to calculate the oscillation period of each spin and more precisely the period of two oscillations for the spin 1, 2, 3 and 4, and the period for only one oscillation for the spin 6, 7, 8 and 9. After that, we have to kick them stationary in the down direction when their information is the lowest. This allows to concentrate the information on the fifth spin. \\

\begin{figure}
\begin{center}
\includegraphics[width=7.7cm]{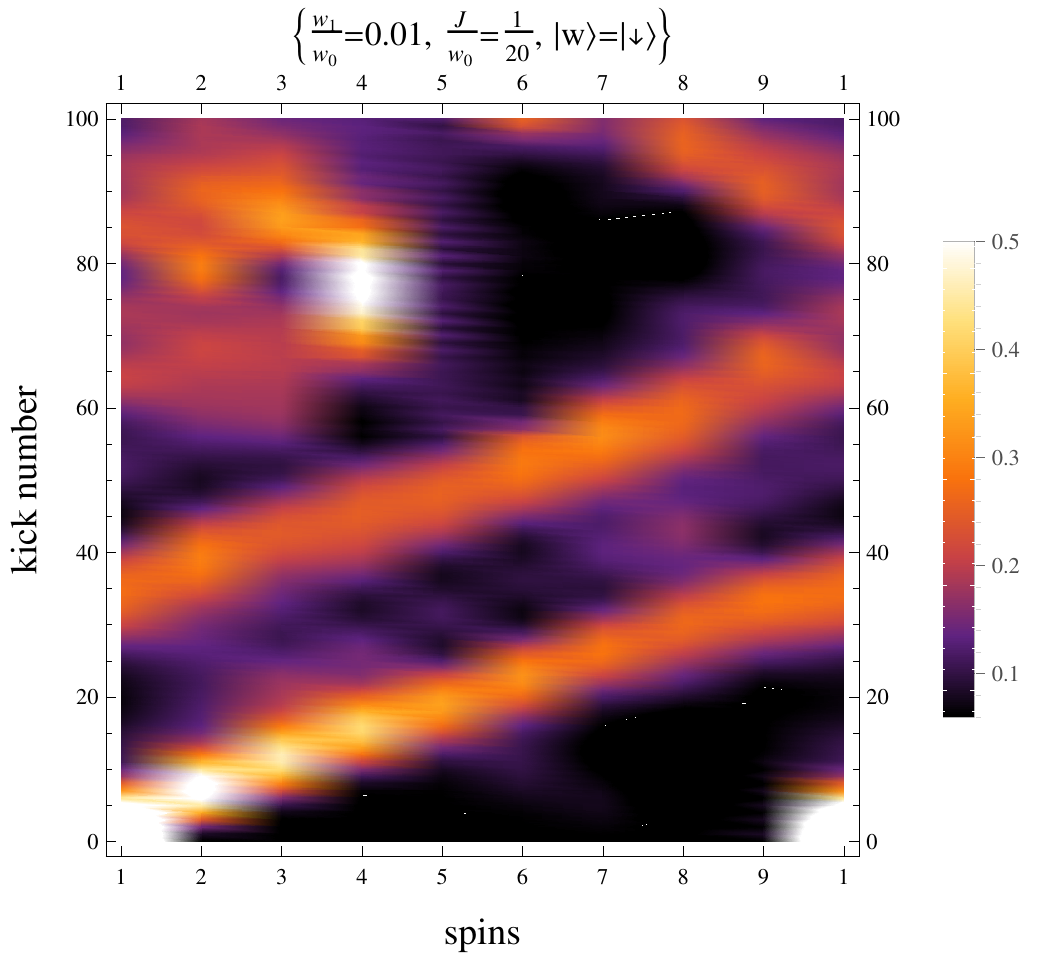}
\includegraphics[width=10cm]{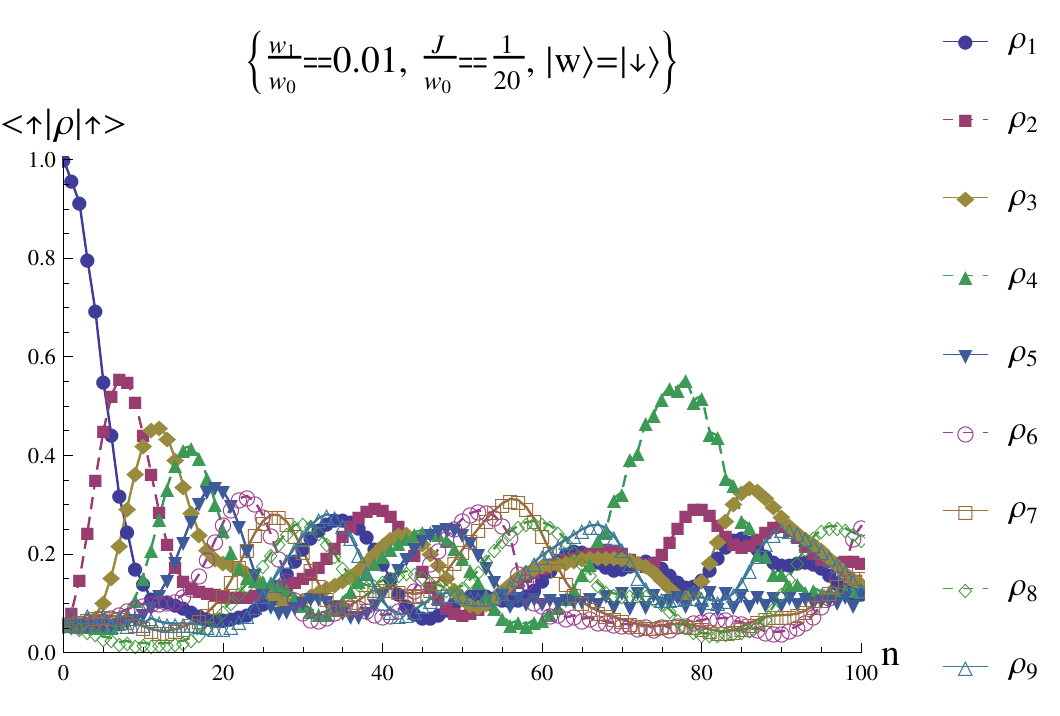}
\end{center}
\caption{\label{informationtransmissionmminstant} Up, density graphic of the population $\langle \uparrow | \rho_n | \uparrow \rangle$ of 9 spins coupled by a nearest-neighbour Heisenberg interaction, with respect to the kick number and to the spins. Down, population evolution with respect to the kicks. All spins are stationary kicked at the same time, i.e. when the second oscillation of the fourth and the sixth spin is down. The spins are initially in the state $\frac{1}{\sqrt{17}}(|\uparrow\rangle + 4 |\downarrow \rangle)$ except the first one which is in the up state.}
\end{figure}

Note : Here each spin is kicked at an appropriate time in order that each one is nearly in the state of the kick. However if we kick all spins (except the fifth one) at the same time, each spin can have a state near or different from the kick state. This induces an oscillation of the population, a lower or no information concentration. An example is given on fig.~\ref{informationtransmissionmminstant}. In order to perform a control, it is more efficient to kick the spin always in their state. It is also better to kick them in a direction of an eigenvector in order that they less oscillate.\\

This last analysis allows us to understand why there is a kind of freezing of the last spin  state in fig.~\ref{heisenbergdensitechaotiqueonde}, \ref{informationtransmission} and \ref{heisenberg}. We see just above that if we kick one spin at the appropriate time, we can force it to stay in its state kicking. Consider for example the right density graphic of fig.~\ref{heisenbergdensitechaotiqueonde}. At the beginning, there is a free variation of the spin oscillations. We only see the oscillation due to the interaction between the spin because all spins are nearly kicked similarly. But, after the horizon of coherence, all spins are kicked differently. So the spins react to the kicks. At the time of the horizon of coherence, all spins are nearly in the down state except the fourth one which is in a superposition $\alpha |\uparrow> + \beta |\downarrow>$ with $\alpha > \beta$. So all spins around the fourth one are forced to stay in their directions. This shows a kind of freezing. Thus we can conserve the information, like the control in this part, but with chaotic kicks.\\

\begin{figure}
\begin{center}
\includegraphics[width=8.5cm]{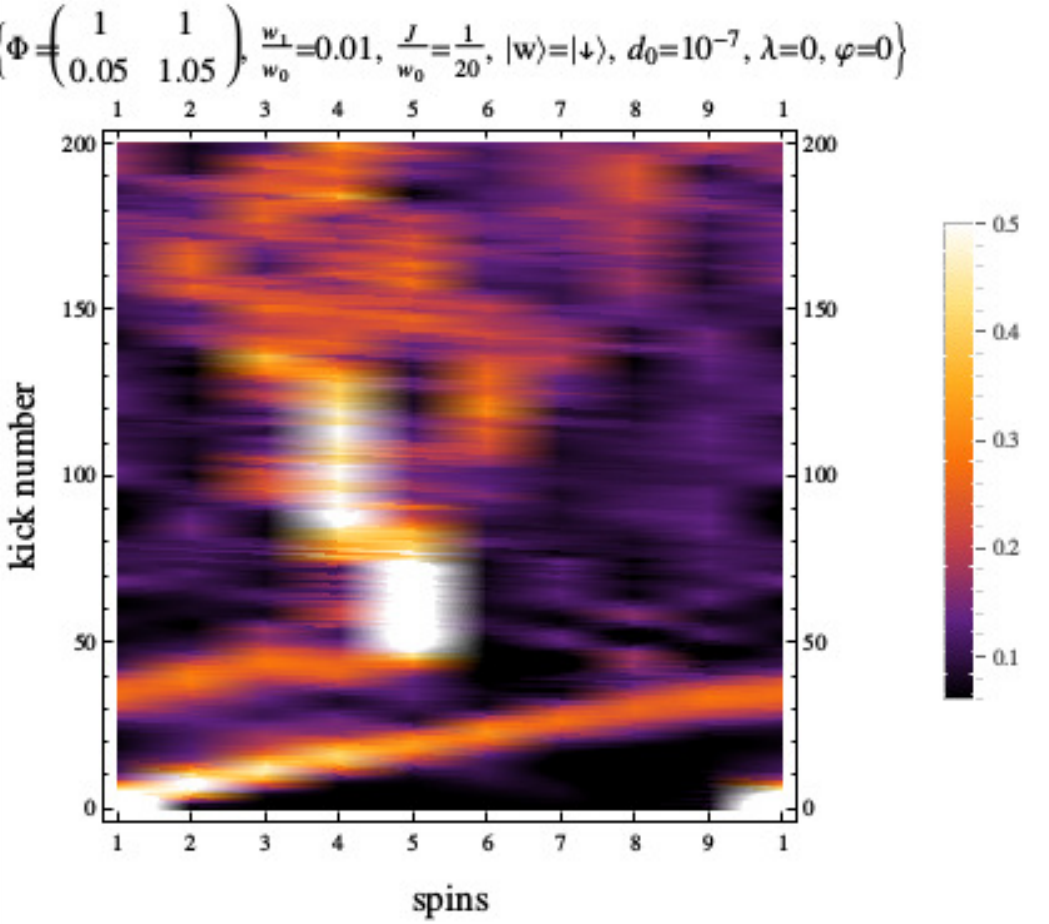}
\includegraphics[width=10cm]{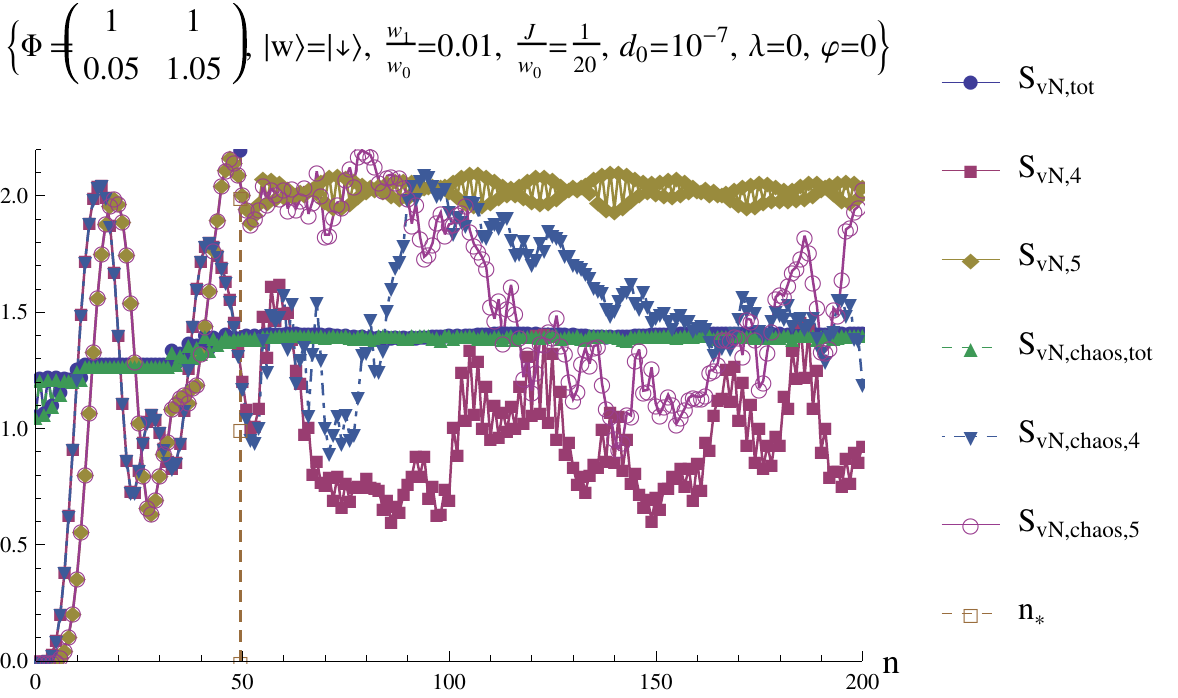}
\end{center}
\caption{\label{evolution_avec_frappe} Up density of the population $\langle \uparrow | \rho_n | \uparrow \rangle$ of 9 spins coupled by a nearest-neighbour Heisenberg interaction, with respect to the spin number of the chain and to the kicks. In order that the fifth spin conserve all the information after that the information crosses it two times, the other spins are kicked in the down direction when the first oscillation is end for the spins 6, 7, 8 and 9 and after second oscillation is end for the spins 1, 2, 3 and 4. All spin are initially in the state $\frac{1}{\sqrt{17}}(|\uparrow\rangle + 4 |\downarrow \rangle)$ except the first one which is in the up state. There is also a chaotic disruption of the kicks : $\Phi$ is the matrix defining the automorphism of the torus. The down graphic represents the evolution of the entropy with respect to the kicks for the average spin of the chain and for the fourth and the fifth spin when there is no disruption of the kick (respectively $S_{vN,tot}$, $S_{vN,4}$, $S_{vN,5}$) and when the kick are disturbed by a chaotic dynamics ($S_{vN,chaos,tot}$, $S_{vN,chaos,4}$, $S_{vN,chaos,5}$)}
\end{figure}

In this section, until now, we have just made a perfect control, i.e. nothing disturbed the kicks. We now introduce a chaotical disruption of the kick. Let the two kinds of data $(\lambda_n^i, \phi_n^i)$ be respectively the strength and the delay of the $i$-th kick on the $n$-th spin associated with the perfect control solution. We introduce another kick set $(\lambda_n^{dist.,i},\phi_n^{dist,i})$ which corresponds to the disruption induced by a chaotic dynamical process. The initial dispersion is chosen to be really small and the initial kick parameters are near to $0$ (we only want the chaotic effect and not the parameter propagations on the phase space induced by the automorphism on the torus). The new kicks are defined by $(\lambda_n^i, \phi_n^i) + (\lambda_n^{dist.,i},\phi_n^{dist,i})$. We obtain fig.~\ref{evolution_avec_frappe}. On the up graphic, we see that before 80 kicks, there is no modification of the control information. After it, the information stopped on the fifth spin begins to be scattered on the other spins. This is always seen on the down graphic which represents the entropy evolution with respect to the kick number. If there is no disruption of the control kicks, at the time where the information is stopped, the spin five has a large entropy whereas for all the others it is lower. When we add the chaotic disruption, the evolution is the same until the horizon of coherence where the entropy is large for all spins with a lot of oscillations (because the interaction is low).\\

This section allows us to perform a control. We have just seen that a free transmission of information during the horizon of coherence appears when the spins are chaotically kicks in the direction of an eigenvector. We have also realised a control by changing or concentrating a spin information. For this we have used some stationary kicks before the horizon of coherence. In this subsection, for the control the propagation of the kick parameters on the phase space before the horizon of predictability are considerably reduced because we choose to begin from $\lambda=0$ and $\phi=0$ (we only conserved the effects of the chaos). In this condition we can think that we can take other directions of kicks. However, other kick directions generally produce some population and coherence oscillations. For a kick on only one spin as for example what we have made on fig.~\ref{evolution_ss_frappe}, we can use another kick direction and the results are generally corrects, or better if the spin state is really different from an eigenvector. For example, if the spin states are $\frac{1}{\sqrt{2}}(|\uparrow\rangle + |\downarrow \rangle)$ except the first one which is in the up state, kicking in the direction of an eigenvector destroys all the spin information which stay near to $\frac{1}{2}$. In this case it is better to kick in the spin direction). But if we kick several spins, different oscillations appear which are transmitted to the other spins by the interaction. The spin information is then completely lost.  

\section{Conclusion}

In this paper, we have studied the behaviours of a spin chain submitted to a kick bath. The kicks are disturbed by chaotic dynamics which are given by the continuous automorphisms of the torus. The spins of the chain are coupled by a nearest-neighbour Heisenberg or Ising-Z interaction. With the Ising-Z coupling the system evolution is characterised by a coherence plateau and a horizon of coherence which are present when the kicks are in the direction of an eigenvector. The length of this plateau is well predicted by eq.~\ref{horizon}. 

The most interesting case in order to control the system is the Heisenberg coupling. This coupling presents the coherence plateau and the horizon of coherence which can be predicted by eq.~\ref{horizon} for the condition  $J \leq w_0$. The coherence conservation is present for all kick directions. We have seen that this coupling allows a conservation and a transmission of information during the horizon of coherence when the kicks are in an eigenvector direction. It is also possible to make some predictions concerning the spin chain evolution with respect to the interaction and/or to the kick frequency parameter and so to know the information evolution. We can speed up or slow down the transmission of information using these parameters. It is also possible to realise an interesting control of the spin information during the horizon of coherence using stationary kicks : we can stop the evolution of the information and concentrate it on only one spin during the horizon of coherence. 

If we can find a chaotic environment which presents a large horizon of coherence and that we can force the kicks to be in the direction of an eigenvector, it is possible to control freely the system. Other analyses will consist of finding an expression of the value of the horizon of coherence for the spins coupled by the Heisenberg interaction for the condition $J>w_0$ and so extending this work.

\appendix
\section{Coherence oscillation and stationary population of a spin chain coupled by an Ising-Z interaction}
\label{isingzdemo}
We have seen in section \ref{sectionisingz} that the spins of a chain coupled by an Ising-Z interaction and not submitted to kicks have a stationary population (at its initial value) and a coherence which oscillates. To understand and to prove it, consider two spins coupled by an Ising-Z interaction. Let all matrices be defined in the base $\{|\uparrow \uparrow\rangle, |\uparrow \downarrow\rangle, |\downarrow \uparrow\rangle, |\downarrow \downarrow\rangle \}$. In this case, the evolution operator becomes 
\begin{multline}
U^{(\mathrm{i})} = e^{-\imath \frac{ H_{0,I}}{\hbar w_0}(2 \pi - \varphi_2^{(\mathrm{i})})} \left[ \mathrm{id} \otimes \left(\mathrm{id} + \left(e^{- \imath \lambda_2^{(\mathrm{i})}}-1\right) W\right)\right]e^{-\imath \frac{ H_{0,I}}{\hbar w_0} (\varphi_2^{(\mathrm{i})} - \varphi_1^{(\mathrm{i})})}\\
\left[\left(\mathrm{id} + \left(e^{- \imath \lambda_1^{(\mathrm{i})}}-1\right) W\right)\otimes \mathrm{id} \right] e^{-\imath \frac{ H_{0,I}}{\hbar w_0}  \varphi_1^{(\mathrm{i})}}
\end{multline}
\begin{eqnarray}
U^{(\mathrm{i})} &=& e^{-\imath \frac{ H_{0,I}}{\hbar w_0}(2 \pi) } \left[ \mathrm{id} \otimes\mathrm{id} \right]e^{-\imath \frac{ H_{0,I}}{\hbar w_0} \times 0} \left[\mathrm{id} \otimes \mathrm{id} \right] e^{-\imath \frac{ H_{0,I}}{\hbar w_0} \times 0}\\
&=&e^{-\imath \frac{ H_{0,I}}{\hbar w_0}(2 \pi) } \begin{pmatrix}
1 & 0 & 0 & 0\\
0 & 1 & 0 & 0\\
0 & 0 & 1 & 0\\
0 & 0 & 0 & 1
\end{pmatrix}\\
&=&e^{-\imath \frac{ H_{0,I}}{\hbar w_0}(2 \pi) } 
\end{eqnarray}
With
\begin{eqnarray}
H_{0,I} &=& \begin{pmatrix}
0 & 0\\
0 & \frac{\hbar w_1}{2}
\end{pmatrix} \otimes \mathrm{id} + \mathrm{id}  \otimes\begin{pmatrix}
0 & 0\\
0 & \frac{\hbar w_1}{2}
\end{pmatrix} - J S_z \otimes S_z \\
&=& \begin{pmatrix}
0 & 0 & 0 & 0\\
0 & \frac{\hbar w_1}{2} & 0 & 0\\
0 & 0 & \frac{\hbar w_1}{2} & 0\\
0 & 0 & 0 & \hbar w_1
\end{pmatrix}-J \frac{\hbar^2}{4} \begin{pmatrix}
1 & 0 & 0 & 0\\
0 & -1 & 0 & 0\\
0 & 0 & -1 & 0\\
0 & 0 & 0 & 1 
\end{pmatrix}\\
&=&   \begin{pmatrix}
- J \frac{\hbar^2}{4} & 0 & 0 & 0\\
0 & J \frac{\hbar^2}{4}+ \frac{\hbar w_1}{2} & 0 & 0\\
0 & 0 & J \frac{\hbar^2}{4} + \frac{\hbar w_1}{2} & 0\\
0 & 0 & 0 & -J \frac{\hbar^2}{4}+ \hbar w_1
\end{pmatrix}
\end{eqnarray}
\begin{equation}
\Leftrightarrow U^{(\mathrm{i})} = e^{-\imath \frac{ H_{0,I}}{\hbar w_0}2 \pi } =  \begin{pmatrix}
e^{\imath \frac{\hbar J}{4 w_0}2 \pi} & 0 & 0 & 0\\
0 & e^{- \imath (\frac{\hbar J}{4 w_0} + \frac{ w_1}{2 w_0})2 \pi} & 0 & 0\\
0 & 0 & e^{- \imath( \frac{\hbar J}{4 w_0} + \frac{ w_1}{2 w_0})2 \pi} & 0\\
0 & 0 & 0 & e^{\imath (\frac{\hbar J}{4 w_0} - \frac{ w_1}{w_0})2 \pi}
\end{pmatrix}
\end{equation}
Let the complete wave function be
\begin{equation}
|\psi^{(0)}\rangle = |\Psi_1\rangle \otimes |\Psi_2 \rangle =  \begin{pmatrix}
\alpha \\
\beta \\
\gamma\\
\delta
\end{pmatrix}
\end{equation}
with $|\alpha|^2+|\beta|^2+|\gamma|^2+|\delta|^2=1$. The wave function evolution is
\begin{eqnarray}
|\psi^{(\mathrm{i})} \rangle &=& U^{(\mathrm{i})} |\psi^{(\mathrm{i}-1)}\rangle \\
&=& \begin{pmatrix}
\alpha e^{\mathrm{i}\imath \frac{\hbar J}{4 w_0}(2 \pi)}\\
\beta e^{-\mathrm{i} \imath (\frac{\hbar J}{4 w_0} + \frac{ w_1}{2 w_0})(2 \pi)} \\
\gamma e^{-\mathrm{i} \imath( \frac{\hbar J}{4 w_0} + \frac{ w_1}{2 w_0})(2 \pi)}\\
\delta e^{\mathrm{i} \imath (\frac{\hbar J}{4 w_0} - \frac{ w_1}{w_0})(2 \pi)}
\end{pmatrix} 
\end{eqnarray}
To obtain the coherence and the population of the first spin, we have to calculate the density matrix and the partial trace on the second spin. The density matrix is defined by
\begin{equation}
\rho^{(\mathrm{i})} = |\psi^{(\mathrm{i})} \rangle \langle \psi^{(\mathrm{i})} |
\end{equation}
\begin{equation}
\rho^{(\mathrm{i})} =  \begin{pmatrix}
\alpha e^{\mathrm{i}\imath \frac{\hbar J}{4 w_0}2 \pi}\\
\beta e^{-\mathrm{i} \imath (\frac{\hbar J}{4 w_0} + \frac{ w_1}{2 w_0})2 \pi} \\
\gamma e^{-\mathrm{i} \imath( \frac{\hbar J}{4 w_0} + \frac{ w_1}{2 w_0})2 \pi}\\
\delta e^{\mathrm{i}\imath (\frac{\hbar J}{4 w_0} - \frac{ w_1}{w_0})2 \pi}
\end{pmatrix} \begin{pmatrix}
\alpha^* e^{-\mathrm{i}\imath \frac{\hbar J}{2 w_0} \pi}&
\beta^* e^{\mathrm{i}\imath (\frac{\hbar J}{2 w_0} + \frac{ w_1}{ w_0}) \pi} &
\gamma^* e^{\mathrm{i}\imath( \frac{\hbar J}{2 w_0} + \frac{ w_1}{ w_0}) \pi}&
\delta^* e^{-\mathrm{i}\imath (\frac{\hbar J}{4 w_0} - \frac{ w_1}{w_0})2 \pi}
\end{pmatrix}
\end{equation}
\begin{equation}
\rho^{(\mathrm{i})}=  \begin{pmatrix}
\alpha\alpha^* & \alpha\beta^* e^{\mathrm{i} \imath (\frac{\hbar J}{2 w_0} + \frac{ w_1}{2 w_0})2 \pi} & \alpha  \gamma^* e^{\mathrm{i} \imath( \frac{\hbar J}{2 w_0} + \frac{ w_1}{2 w_0})2 \pi}& \alpha \delta^* e^{\mathrm{i}\imath\frac{ w_1}{w_0} 2 \pi}\\
\beta \alpha^* e^{-\mathrm{i} \imath (\frac{\hbar J}{2 w_0} + \frac{ w_1}{2 w_0})2 \pi} & \beta \beta^* & \beta  \gamma^* & \beta \delta^* e^{-\mathrm{i} \imath (\frac{\hbar J}{2 w_0} - \frac{ w_1}{2 w_0})2 \pi}\\
\gamma \alpha^* e^{-\mathrm{i} \imath( \frac{\hbar J}{2 w_0} + \frac{ w_1}{2 w_0})2 \pi} & \gamma \beta^* & \gamma \gamma^* & \gamma \delta^* e^{-\mathrm{i} \imath( \frac{\hbar J}{2 w_0} - \frac{ w_1}{2 w_0})2 \pi}\\
\delta \alpha^* e^{- \mathrm{i}\imath \frac{ w_1}{w_0}2 \pi} & \delta\beta^* e^{\mathrm{i}\imath (\frac{\hbar J}{2 w_0} - \frac{ w_1}{2 w_0})2 \pi} & \delta  \gamma^* e^{\mathrm{i}\imath (\frac{\hbar J}{2 w_0} - \frac{ w_1}{2 w_0})2 \pi} & \delta \delta^*
\end{pmatrix}
\end{equation}
The coherence of the first spin is
\begin{eqnarray}
\rho^{cohe,(\mathrm{i})}_1 &=& |\langle \uparrow \uparrow |\rho^\mathrm{i} | \downarrow \uparrow \rangle + \langle \uparrow \downarrow|\rho^\mathrm{i} | \downarrow \downarrow \rangle|\\
&=& |\left( \gamma \alpha^* e^{- \mathrm{i}\imath (\frac{\hbar J}{2 w_0} + \frac{ w_1}{2 w_0})(2 \pi)} + \delta\beta^* e^{\mathrm{i}\imath (\frac{\hbar J}{2 w_0} - \frac{ w_1}{2w_0})(2 \pi)} \right)|
\end{eqnarray}
We see that the coherence only depends on the exponential which is $2  \pi$ periodic. The up population of the first spin is given by
\begin{eqnarray}
\rho^{pop,(\mathrm{i})}_1 &=& \langle \uparrow \uparrow |\rho^\mathrm{i} | \uparrow \uparrow \rangle + \langle \uparrow \downarrow|\rho^\mathrm{i} | \uparrow \downarrow \rangle\\
&=& \left( \alpha \alpha^* + \beta \beta^*  \right)
\end{eqnarray}
The population is not modified with the time if there is no kick.\\

The extension of these analyses to $N$ coupled spins give the same results.

\section{Effect of the kick strength on an uncoupled spin kicked in a direction of an eigenvector}
\label{withoutcoupling}
 If we consider a spin without any interaction with its neighbours, we have the following evolution operator for the $i$-th kick
\begin{equation}
U^{(\mathrm{i})} = e^{-\imath \frac{H_0}{\hbar w_0} (2 \pi - \varphi^{(\mathrm{i})})} \left[ \mathrm{id} + \left(e^{-\imath \lambda^{(\mathrm{i})}} -1 \right)W \right] e^{-\imath \frac{H_0}{\hbar w_0} \varphi^{(\mathrm{i})}}
\end{equation}
All matrices are defined in the base $\{|\uparrow \uparrow\rangle, |\uparrow \downarrow\rangle, |\downarrow \uparrow\rangle, |\downarrow \downarrow\rangle \}$. We suppose that the kicks are in the direction of an eigenvector $W = |w\rangle \langle w| = \begin{pmatrix}
1 \\
0
\end{pmatrix}\begin{pmatrix}
1 & 0
\end{pmatrix} = \begin{pmatrix}
1 & 0\\
0 & 0
\end{pmatrix}$.
If we calculate the evolution of the evolution operator until the $m$-th kick, we obtain
\begin{eqnarray}
     \prod_{j=1}^m U^{(\mathrm{j})} &=& \prod_{j=1}^m \begin{pmatrix}
1 & 0\\
0 & e^{-\imath \frac{w_1}{2 w_0}(2\pi- \varphi^{(\mathrm{j})})}
\end{pmatrix} \begin{pmatrix}
e^{-\imath \lambda^{(\mathrm{j})}} & 0\\
0 & 1
\end{pmatrix} \begin{pmatrix}
1 & 0\\
0 & e^{-\imath \frac{w_1}{2 w_0}\varphi^{(\mathrm{j})}}
\end{pmatrix} \\[.2cm]
      &=& \prod_{j=1}^m \begin{pmatrix}
e^{-\imath \lambda^{(\mathrm{j})}} & 0\\
0 & e^{- \imath \frac{w_1}{w_0} \pi}
\end{pmatrix} \\[.4cm]
      &=&\begin{pmatrix}
e^{-\imath \sum_{j=1}^m \lambda^{(\mathrm{j})}} & 0\\
0 & e^{- \imath \sum_{j=1}^m \frac{w_1}{w_0} \pi }
\end{pmatrix}
\end{eqnarray}
The initial state is chosen to be $|\psi^{(0)} \rangle =\begin{pmatrix}
\alpha \\
\beta
\end{pmatrix}$ with $|\alpha|^2 + |\beta|^2=1$. The wave function at the $i$-th kick is
\begin{equation}
|\psi^{(i)} \rangle  = \begin{pmatrix}
\alpha e^{-\imath \sum_{j=1}^i \lambda^{(j)}}\\
\beta e^{- \imath \sum_{j=1}^i \frac{w_1}{w_0} \pi}
\end{pmatrix}
\end{equation}
 The density matrix is then
\begin{equation}
\rho^{(\mathrm{i})}  = |\psi^{(\mathrm{i})} \rangle \langle \psi^{(\mathrm{i})}| = \begin{pmatrix}
\alpha \alpha^* &  \alpha \beta^* e^{-\imath \sum_{j=1}^i \lambda^{(\mathrm{i})}} e^{\imath \sum_{k=1}^i \frac{w_1}{w_0} \pi}\\
\alpha^* \beta e^{\imath \sum_{j=1}^i \lambda^{(j)}} e^{-\imath \sum_{k=1}^i \frac{w_1}{w_0} \pi} & \beta \beta^*
\end{pmatrix}
\end{equation}
We see that there is no effect of the strength or of the delay on the population. The strength only induces a pure dephasing.

\section{Effect of the coupling on the kick strength when the kick is in the direction of an eigenvector}
We have seen in section \ref{sectionisingz} that if two spins are kicked with the same strength in the direction of an eigenvector, the coherence and the population are not modified. However, if the spins are kicked with various strengths, a modification appears. In order to understand mathematically what happens, we choose to make the calculation for two coupled spins. All matrices are defined in the base $\{|\uparrow \uparrow\rangle, |\uparrow \downarrow\rangle, |\downarrow \uparrow\rangle, |\downarrow \downarrow\rangle \}$.

\subsection{When the spins are coupled by an Ising-Z interaction}
\label{isingzdemo2}
The evolution operator is characterised at the $i$-th kick by 
\begin{multline}
U^{(\mathrm{i})} = e^{-\imath \frac{ H_{0,I}}{\hbar w_0}(2 \pi - \varphi_2^{(\mathrm{i})})} \left[ \mathrm{id} \otimes \left(\mathrm{id} + \left(e^{- \imath \lambda_2^{(\mathrm{i})}}-1\right) W\right)\right]e^{-\imath \frac{ H_{0,I}}{\hbar w_0} (\varphi_2^{(\mathrm{i})} - \varphi_1^{(\mathrm{i})})}\\
\left[\left(\mathrm{id} + \left(e^{- \imath \lambda_1^{(\mathrm{i})}}-1\right) W\right)\otimes \mathrm{id} \right] e^{-\imath \frac{ H_{0,I}}{\hbar w_0}  \varphi_1^{(\mathrm{i})}}
\end{multline}
In order to simplify the calculation, this demonstration does not take into account the possible variation of the strength and of the delay from one kick to another, i.e. $U^{(\mathrm{i})}=U$,  $\varphi_1^{(\mathrm{i})}=\varphi_1$, $\varphi_2^{(\mathrm{i})}=\varphi_2$, $\lambda_1^{(\mathrm{i})}=\lambda_1$ and $\lambda_2^{(\mathrm{i})}=\lambda_2$.

Let the kicks be in the direction of an eigenvector, so
\begin{center}
 $W = |w\rangle \langle w| = \begin{pmatrix}
1 \\
0
\end{pmatrix}\begin{pmatrix}
1 & 0
\end{pmatrix} = \begin{pmatrix}
1 & 0\\
0 & 0
\end{pmatrix}$
\end{center}
We have obtained in \ref{isingzdemo} the exponential of the Hamiltonian $H_{0,I}$
\begin{equation}
e^{-\imath \frac{ H_{0,I}}{\hbar w_0} } =  \begin{pmatrix}
e^{\imath \frac{\hbar J}{4 w_0}} & 0 & 0 & 0\\
0 & e^{- \imath (\frac{\hbar J}{4 w_0} + \frac{ w_1}{2 w_0})} & 0 & 0\\
0 & 0 & e^{- \imath( \frac{\hbar J}{4 w_0} + \frac{ w_1}{2 w_0})} & 0\\
0 & 0 & 0 & e^{\imath (\frac{\hbar J}{4 w_0} - \frac{ w_1}{w_0})}
\end{pmatrix}
\end{equation}
Let $\alpha = \frac{\hbar J}{4 w_0}$ and $\beta=\frac{ w_1}{2 w_0}$. The Hamiltonian becomes
\begin{equation}
e^{-\imath \frac{ H_{0,I}}{\hbar w_0}} =  \begin{pmatrix}
e^{\imath \alpha} & 0 & 0 & 0\\
0 & e^{- \imath (\alpha + \beta)} & 0 & 0\\
0 & 0 & e^{- \imath(\alpha + \beta)} & 0\\
0 & 0 & 0 & e^{\imath (\alpha - 2\beta)}
\end{pmatrix}
\end{equation}

\begin{multline}
\Leftrightarrow U=  \begin{pmatrix}
e^{\imath \alpha (2\pi - \varphi_2)} & 0 & 0 & 0\\
0 & e^{- \imath (\alpha + \beta)(2\pi - \varphi_2)} & 0 & 0\\
0 & 0 & e^{- \imath(\alpha + \beta)(2\pi - \varphi_2)} & 0\\
0 & 0 & 0 & e^{\imath (\alpha - 2\beta)(2\pi - \varphi_2)}
\end{pmatrix} \\
 \left[\begin{pmatrix}
1 & 0\\
0 & 1
\end{pmatrix} \otimes \begin{pmatrix}
e^{- \imath \lambda_2} & 0\\
0 & 1
\end{pmatrix}\right] \begin{pmatrix}
e^{\imath \alpha (\varphi_2 - \varphi_1)} & 0 & 0 & 0\\
0 & e^{- \imath (\alpha + \beta)(\varphi_2 - \varphi_1)} & 0 & 0\\
0 & 0 & e^{- \imath(\alpha + \beta)(\varphi_2 - \varphi_1)} & 0\\
0 & 0 & 0 & e^{\imath (\alpha - 2\beta)(\varphi_2 - \varphi_1)}
\end{pmatrix} \\
\left[ \begin{pmatrix}
e^{- \imath \lambda_1} & 0\\
0 & 1
\end{pmatrix}\otimes \begin{pmatrix}
1 & 0\\
0 & 1
\end{pmatrix} \right]\begin{pmatrix}
e^{\imath \alpha  \varphi_1} & 0 & 0 & 0\\
0 & e^{- \imath (\alpha + \beta) \varphi_1} & 0 & 0\\
0 & 0 & e^{- \imath(\alpha + \beta)\varphi_1} & 0\\
0 & 0 & 0 & e^{\imath (\alpha - 2\beta)\varphi_1}
\end{pmatrix}
\end{multline}
\begin{equation}
\Leftrightarrow U=\begin{pmatrix}
e^{\imath \alpha 2\pi}e^{-\imath (\lambda_2 + \lambda_1)} & 0 & 0 & 0 \\
0 & e^{-\imath (\alpha +\beta)2\pi}e^{-\imath  \lambda_1} & 0 & 0\\
0 & 0 & e^{-\imath (\alpha +\beta)2\pi}e^{-\imath  \lambda_2} & 0\\
0 & 0 & 0 & e^{\imath (\alpha - 2\beta) 2\pi}
\end{pmatrix}
\end{equation}
We choose the initial state to be : $|\Psi_1\rangle = \begin{pmatrix}
\chi \\
\zeta
\end{pmatrix}$ and $|\Psi_2\rangle = \begin{pmatrix}
\gamma\\
\delta
\end{pmatrix}$
\begin{equation}
|\psi^{(0)}\rangle = |\Psi_1\rangle \otimes |\Psi_2\rangle =\begin{pmatrix}
\chi \gamma\\
\chi \delta\\
\zeta \gamma\\
\zeta \delta
\end{pmatrix}
\end{equation}
The evolution of wave function at the $i$-th kick is
\begin{equation}
|\psi^{(\mathrm{i})}\rangle = U |\psi^{(\mathrm{i}-1)}\rangle = \begin{pmatrix}
\chi \gamma e^{\mathrm{i} \imath \alpha 2\pi}e^{-\mathrm{i} \imath (\lambda_2 + \lambda_1)}\\
\chi \delta e^{-\mathrm{i} \imath (\alpha +\beta)2\pi}e^{-\mathrm{i} \imath  \lambda_1} \\
\zeta \gamma e^{-\mathrm{i} \imath (\alpha +\beta)2\pi}e^{-\mathrm{i} \imath  \lambda_2}\\
\zeta \delta e^{\mathrm{i} \imath (\alpha - 2\beta) 2\pi}
\end{pmatrix}
\end{equation}
The density matrix is then
\begin{equation}
\rho^{(\mathrm{i})} = |\psi^{(\mathrm{i})}\rangle \langle \psi^{(\mathrm{i})}|
\end{equation}
The coherence of the first spin is
\begin{eqnarray}
\rho^{cohe,(\mathrm{i})}_1 &=& |\langle \uparrow \uparrow |\rho^{(\mathrm{i})} | \downarrow \uparrow \rangle + \langle \uparrow \downarrow|\rho^{(\mathrm{i})} | \downarrow \downarrow \rangle|\\
&=& |\chi \gamma \zeta^* \gamma^* e^{\mathrm{i} \imath (2 \alpha +\beta) 2\pi}e^{-\mathrm{i} \imath \lambda_1} + \chi \delta \zeta^* \delta^* e^{-\mathrm{i} \imath (2\alpha - \beta) 2\pi} e^{-\mathrm{i} \imath  \lambda_1}|  \\
&=& | \chi \gamma \zeta^* \gamma^* e^{\mathrm{i} \imath (2 \alpha +\beta) 2\pi} + \chi \delta \zeta^* \delta^* e^{-\mathrm{i} \imath (2\alpha - \beta) 2\pi}|
\end{eqnarray}
and the first spin up population
\begin{eqnarray}
\rho^{pop,(\mathrm{i})}_1 &=& \langle \uparrow \uparrow |\rho^{(\mathrm{i})} | \uparrow \uparrow \rangle + \langle \uparrow \downarrow|\rho^{(\mathrm{i})} | \uparrow \downarrow \rangle\\
&=& \chi \chi^* \gamma \gamma^* + \chi \chi^* \delta \delta^*
\end{eqnarray}
 The coherence and the population of the first spin does not change with respect to the kick number for a kick in a direction of an eigenvector as we have seen in section \ref{sectionisingz}. For the coherence of the average spin we have 
\begin{equation}
\rho^{cohe,(\mathrm{i})}_{tot} = \frac{1}{2} |\langle \uparrow \uparrow |\rho^{(\mathrm{i})} | \downarrow \uparrow \rangle + \langle \uparrow \downarrow|\rho^{(\mathrm{i})} | \downarrow \downarrow \rangle + \langle \uparrow \uparrow |\rho^{(\mathrm{i})} | \uparrow \downarrow \rangle + \langle \downarrow \uparrow |\rho^{(\mathrm{i})} | \downarrow \downarrow \rangle|
\end{equation}
\begin{multline}
\rho^{cohe,(\mathrm{i})}_{tot} = \frac{1}{2} \left| \chi \gamma\zeta^* \gamma^* e^{\mathrm{i} \imath (2\alpha +\beta)2\pi}e^{-\mathrm{i} \imath  \lambda_1} + \chi \delta \zeta^* \delta^* e^{-\mathrm{i} \imath  \lambda_1} e^{-\mathrm{i} \imath (2\alpha - \beta) 2\pi}\right.\\
\left.+ \chi \gamma \chi^* \delta^* e^{-\mathrm{i} \imath \lambda_2 }  e^{\mathrm{i} \imath (2\alpha +\beta)2\pi} + \zeta \gamma \zeta^* \delta^* e^{-\mathrm{i} \imath (2\alpha -\beta)2\pi}e^{-\mathrm{i} \imath  \lambda_2}  \right|
\end{multline}
In the case where $\lambda = \lambda_1 = \lambda_2$, we obtain
\begin{multline}
\rho^{cohe,(\mathrm{i})}_{tot} = \frac{1}{2} \left| \chi \gamma\zeta^* \gamma^* e^{\mathrm{i} \imath (2\alpha +\beta)2\pi} + \chi \delta \zeta^* \delta^* e^{-\mathrm{i} \imath (2\alpha - \beta) 2\pi}\right.\\
\left.+ \chi \gamma \chi^* \delta^* e^{\mathrm{i} \imath (2\alpha +\beta)2\pi} + \zeta \gamma \zeta^* \delta^* e^{-\mathrm{i} \imath (2\alpha -\beta)2\pi}\right|
\end{multline}
We see that if the spins are similarly kicked (same strength) in a direction of an eigenvector, there is no modification of the coherence of the average spin of the chain. However, when the strengths are different, the spins feel the kicks. The kick delay never influences the coherence and the population.

The extension to a large number of spins and to a variation of the strength and of the delay from one kick to another one gives the same results.

\subsection{When the spins are coupled by a Heisenberg interaction}
\label{force}
The evolution operator is characterised by 
\begin{multline}
U^{(\mathrm{i})} = e^{-\imath \frac{ H_{0,I}}{\hbar w_0}(2 \pi - \varphi_2^{(\mathrm{i})})} \left[ \mathrm{id} \otimes \left(\mathrm{id} + \left(e^{- \imath \lambda_2^{(\mathrm{i})}}-1\right) W\right)\right]e^{-\imath \frac{ H_{0,I}}{\hbar w_0} (\varphi_2^{(\mathrm{i})} - \varphi_1^{(\mathrm{i})})}\\
\left[\left(\mathrm{id} + \left(e^{- \imath \lambda_1^{(\mathrm{i})}}-1\right) W\right)\otimes \mathrm{id} \right] e^{-\imath \frac{ H_{0,I}}{\hbar w_0}  \varphi_1^{(\mathrm{i})}}
\end{multline}
As previously we simplify the calculation in not taking into account the possible variation of the strength and of the delay from one kick to another, i.e. $U^{(\mathrm{i})}=U$,  $\varphi_1^{(\mathrm{i})}=\varphi_1$, $\varphi_2^{(\mathrm{i})}=\varphi_2$, $\lambda_1^{(\mathrm{i})}=\lambda_1$ and $\lambda_2^{(\mathrm{i})}=\lambda_2$.

We choose to kick in the direction of an eigenvector, so
\begin{center}
 $W = |w\rangle \langle w| = \begin{pmatrix}
1 \\
0
\end{pmatrix}\begin{pmatrix}
1 & 0
\end{pmatrix} = \begin{pmatrix}
1 & 0\\
0 & 0
\end{pmatrix}$
\end{center}
We are only interested by the variation of the strength between the first and the second spin. For a sake of simplicity, we suppose that $\varphi_1 = \varphi_2 = 0$. The evolution operator is modified as follows 

\begin{multline}
U = e^{-\imath \frac{ H_{0,I}}{\hbar w_0}2 \pi} \left[\begin{pmatrix}
1 & 0\\
0 & 1
\end{pmatrix} \otimes \left[ \begin{pmatrix}
1 & 0\\
0 & 1
\end{pmatrix}  + \left( e^{- \imath \lambda_2}-1 \right) \begin{pmatrix}
1 & 0\\
0 & 0
\end{pmatrix}  \right] \right] e^{H_{0,I} \times 0} \\
 \left[ \left[ \begin{pmatrix}
1 & 0\\
0 & 1
\end{pmatrix}  + \left( e^{- \imath \lambda_1}-1 \right) \begin{pmatrix}
1 & 0\\
0 & 0
\end{pmatrix}  \right] \otimes \begin{pmatrix}
1 & 0\\
0 & 1
\end{pmatrix}\right]e^{H_{0,I} \times 0}
\end{multline}
with $ e^{H_{0,I} \times 0} = \begin{pmatrix}
1 & 0 & 0 & 0\\
0 & 1 & 0 & 0\\
0 & 0 & 1 & 0\\
0 & 0 & 0 & 1
\end{pmatrix}$.

\begin{equation}
\Leftrightarrow U = e^{-\imath \frac{ H_{0,I}}{\hbar w_0}2 \pi} \begin{pmatrix}
e^{-\imath (\lambda_1 + \lambda_2)} & 0 & 0 & 0\\
0 & e^{- \imath \lambda_1} & 0 & 0\\
0 & 0 & e^{- \imath \lambda_2} & 0 \\
0 & 0 & 0 & 1
\end{pmatrix}
\end{equation}
$H_{0,I} = H_0 + H_I$ with $H_0$ a diagonal matrix and $H_I$  a matrix with non-diagonal terms associated with the coupling. The matrix $H_{0,I}$ can be written as

\begin{equation}
H_{0,I} = \begin{pmatrix}
a & 0 & 0 & 0\\
0 & b & c & 0\\
0 & c & b & 0\\
0 & 0 & 0 & d
\end{pmatrix}
\end{equation} 
with $a \neq d$ because of the shape of $H_0$. The exponential of such a matrix becomes

\begin{equation}
e^{-\imath \frac{H_{0,I}}{\hbar w_0} 2 \pi} = \begin{pmatrix}
u & 0 & 0 & 0\\
0 & v & w & 0\\
0 & w & v & 0\\
0 & 0 & 0 & x
\end{pmatrix}
\end{equation}
and the evolution operator becomes 

\begin{equation}
U =  \begin{pmatrix}
u e^{-\imath (\lambda_1 + \lambda_2)} & 0 & 0 & 0\\
0 & v e^{- \imath \lambda_1} & w e^{- \imath \lambda_2} & 0\\
0 & w e^{- \imath \lambda_1} & v e^{- \imath \lambda_2} & 0 \\
0 & 0 & 0 & x
\end{pmatrix}
\end{equation}
Let the spin state be at $t=0$ 
\begin{center}
$|\Psi_1\rangle = \begin{pmatrix}
1\\
0
\end{pmatrix}$ \\
 $|\Psi_2\rangle =  \frac{1}{\sqrt{2}} \begin{pmatrix}
1\\
1
\end{pmatrix}$.
\end{center}
\begin{equation}
\Leftrightarrow |\psi^{(0)}\rangle = | \Psi_1 \rangle\otimes |\Psi_2 \rangle =  \frac{1}{\sqrt{2}} \begin{pmatrix}
1 \\
1\\
0\\
0
\end{pmatrix}
\end{equation}
The evolution with respect to the kick number is given by
\begin{equation}
|\psi^{(i+1)}\rangle = U |\psi^{(i)} \rangle
\end{equation}
In order to know the effect of the kick on the population, we calculate the three first up populations of the first spin ($\rho_1^{pop,(i)}$). For this, we have to calculate the complete wave function $|\psi^{(i)}\rangle$ and the density matrix $\rho^{(i)}$.

\begin{center}
$|\psi^{(0)}\rangle = | \Psi_1 \rangle\otimes |\Psi_2 \rangle =  \frac{1}{\sqrt{2}} \begin{pmatrix}
1 \\
1\\
0\\
0
\end{pmatrix}$
\end{center}
\begin{equation}
\Leftrightarrow \rho^{(0)} = |\psi^{(0)}\rangle \langle \psi^{(0)}| = \frac{1}{2} . \begin{pmatrix}
1 & 1 & 0 & 0\\
1 & 1 & 0 & 0\\
0 & 0 & 0 & 0\\
0 & 0 & 0 & 0
\end{pmatrix}
\end{equation}
The up population of the first spin is given by the partial trace on the second spin.
\begin{equation}
\Leftrightarrow  \psi_1^{(0)}  = \langle \uparrow \downarrow | \rho^{(0)} |\uparrow \downarrow \rangle +  \langle \uparrow \uparrow | \rho^{(0)} |\uparrow \uparrow \rangle= 1
\end{equation}

\begin{equation}
|\psi^{(1)}\rangle = U |\psi^{(0)} \rangle = \frac{1}{\sqrt{2}} \begin{pmatrix}
u e^{- \imath (\lambda_1 + \lambda_2)} \\
v e^{- \imath \lambda_1}\\
w e^{- \imath \lambda_1} \\
0
\end{pmatrix}
\end{equation}
\begin{equation}
\Leftrightarrow \rho^{(1)} = |\psi^{(1)}\rangle \langle \psi^{(1)}| = \frac{1}{2} \begin{pmatrix}
uu^* & uv^* e^{- \imath \lambda_2} & uw^* e^{-\imath \lambda_2} & 0\\
vu^* e^{\imath \lambda_2} & vv^* & vw^* & 0\\
wu^* e^{\imath \lambda_2} & wv^* & ww^* & 0\\
0 & 0 & 0 & 0
\end{pmatrix}
\end{equation}
\begin{equation}
\Leftrightarrow  \rho_1^{pop,(1)}  = \langle \uparrow \downarrow | \rho^{(1)} |\uparrow \downarrow \rangle +  \langle \uparrow \uparrow | \rho^{(1)} |\uparrow \uparrow \rangle= \frac{1}{2} (uu^* + vv^*)
\end{equation}

\begin{equation}
|\psi^{(2)}\rangle = U |\psi^{(1)} \rangle = \frac{1}{\sqrt{2}} \begin{pmatrix}
u^2 e^{- 2 \imath (\lambda_1 + \lambda_2)} \\
v^2 e^{- 2 \imath \lambda_1} + w^2 e^{-\imath (\lambda_1 + \lambda_2)}\\
vw e^{- 2 \imath \lambda_1} + wv e^{-\imath (\lambda_1 + \lambda_2)} \\
0
\end{pmatrix}
\end{equation}
\begin{multline}
\Leftrightarrow  \rho_1^{pop,(2)}  = \langle \uparrow \downarrow | \rho^{(2)} |\uparrow \downarrow \rangle +  \langle \uparrow \uparrow | \rho^{(2)} |\uparrow \uparrow \rangle\\
= \frac{1}{2} (u^2(u^2)^* + v^2(v^2)^* + w^2(w^2)^* + v^2(w^2)^* e^{-\imath (\lambda_1 -  \lambda_2)}+ w^2(v^2)^* e^{\imath (\lambda_1 - \lambda_2)})
\end{multline}
In the same way, we obtain 
\begin{multline}
\Leftrightarrow  \rho_1^{pop,(3)}  = \langle \uparrow \downarrow | \rho^{(3)} |\uparrow \downarrow \rangle +  \langle \uparrow \uparrow | \rho^{(3)} |\uparrow \uparrow \rangle\\
= \frac{1}{2} \left(u^3(u^3)^* + vv^* \left[ v^2(v^2)^* + 5 w^2(w^2)^* + v^2(w^2)^* e^{2 \imath (\lambda_2 - \lambda_1)} + 2 v^2 (w^2)^* e^{-\imath (\lambda_1-\lambda_2)} \right. \right.\\
\left. \left.+ w^2(v^2)^* e^{2 \imath (\lambda_1 - \lambda_2)} + 2 w^2 (w^2)^* e^{\imath (\lambda_1 - \lambda_2)}+ 2 w^2 (v^2)^* e^{\imath(\lambda_1- \lambda_2)} + 2 w^2 (w^2)^* e^{-\imath (\lambda_1 - \lambda_2)}\right] \right)
\end{multline}

Across the three up state of the first spin, we see that the strength can affect the population only if it is different for two coupled spins. If the strength is the same, we can easily see that it disappears and the spin evolution is only due to the coupling.

This demonstration can also be made for a kick strength which is not the same for every kicks and for more coupled spins. The conclusion will be the same.

\end{document}